\documentclass[journal]{IEEEtran}
\usepackage{amsmath}
\usepackage{mathtools, cuted}
\usepackage{setspace}
\usepackage{flushend, cuted}
\usepackage{lipsum}
\usepackage{lingmacros}
\usepackage{tree-dvips}
\usepackage{varioref}
\usepackage{colortbl}
\usepackage{url}
\usepackage{hyperref}
\usepackage{balance}

\usepackage{graphicx}
\usepackage{epstopdf}
\usepackage{color,soul}
\usepackage{amsfonts}
\usepackage{subfigure}
\usepackage{multicol}
\usepackage{multirow}
\usepackage{hhline}
\usepackage{amssymb}
\usepackage{newlfont}
\usepackage{mathrsfs}
\usepackage{mathtools}
\usepackage{mathrsfs}
\usepackage{cite}
\usepackage{multirow}
\usepackage{nomencl}   
\usepackage{fancyhdr}
\usepackage{caption}
\usepackage{slashbox}
\usepackage{float}
\usepackage{placeins}
\usepackage{tabularx}
\usepackage[table]{xcolor}
\usepackage[linesnumbered,ruled]{algorithm2e}
\SetKwRepeat{Do}{do}{while}%

\DeclareRobustCommand{\frac}[3][0pt]{%
	{\begingroup\hspace{#1}#2\hspace{#1}\endgroup\over\hspace{#1}#3\hspace{#1}}}

\let\norm\undefined 
\DeclarePairedDelimiter\norm{\lVert}{\rVert}

\ifCLASSINFOpdf
\else
\fi
\begin{document}

\title{Second-Order Ultrasound Elastography with $L1$-norm Spatial Regularization}
%
%
\author{Md~Ashikuzzaman, \textit{Student Member}, \textit{IEEE}
        and~Hassan~Rivaz, \textit{Senior Member}, \textit{IEEE}
\thanks{Md Ashikuzzaman and Hassan Rivaz are with the Department
of Electrical and Computer Engineering, Concordia University, Montreal,
QC, H3G 1M8, Canada.
 Email: m\_ashiku@encs.concordia.ca~and~hrivaz@ece.concordia.ca}
}

%

\maketitle
\setlength{\abovedisplayskip}{1pt}
\setlength{\belowdisplayskip}{1pt}
\begin{abstract}
Time delay estimation (TDE) between two radio-frequency (RF) frames is one of the major steps of quasi-static ultrasound elastography, which detects tissue pathology by estimating its mechanical properties. Regularized optimization-based techniques, a prominent class of TDE algorithms, optimize a non-linear energy functional consisting of data constancy and spatial continuity constraints to obtain the displacement and strain maps between the time-series frames under consideration. The existing optimization-based TDE methods often consider the $L2$-norm of displacement derivatives to construct the regularizer. However, such a formulation over-penalizes the displacement irregularity and poses two major issues to the estimated strain field. First, the boundaries between different tissues are blurred. Second, the visual contrast between the target and the background is suboptimal. To resolve these issues, herein, we propose a novel TDE algorithm where instead of $L2$-, $L1$-norms of both first- and second-order displacement derivatives are taken into account to devise the continuity functional. We handle the non-differentiability of $L1$-norm by smoothing the absolute value function's sharp corner and optimize the resulting cost function in an iterative manner. We call our technique Second-Order Ultrasound eLastography with $L1$-norm spatial regularization ($L1$-SOUL). In terms of both sharpness and visual contrast, $L1$-SOUL substantially outperforms GLUE, OVERWIND, and SOUL, three recently published TDE algorithms in all validation experiments performed in this study. In cases of simulated, phantom, and \textit{in vivo} datasets, respectively, $L1$-SOUL achieves $67.8\%$, $46.81\%$, and $117.35\%$ improvements of contrast-to-noise ratio (CNR) over SOUL. The $L1$-SOUL code can be downloaded from \url{ http://code.sonography.ai}.                                          
\end{abstract}


\IEEEpeerreviewmaketitle
                  
\section{Introduction}
Tissue pathologies such as tumor, cancer, benign lesion, and cyst often alter the elastic properties of tissue. Ultrasound elastography~\cite{ophir_91, mohammadi2020statistical} is an emerging medical imaging technique that aims at detecting such abnormalities of tissue by mapping its mechanical properties. Being a non-invasive and easy-to-use modality, over the last few decades, ultrasound elastography has been successfully employed in liver tissue classification~\cite{frulio2013ultrasound,DPAM}, assessment of breast~\cite{hall2003vivo,rglue_ius}, kidney~\cite{peride2016value, grenier2013renal} and prostate~\cite{cochlin2002elastography,mousavi2014towards}, ablation monitoring~\cite{rivaz2008ablation,bharat2005monitoring, varghese2003elastographic}, cardiac~\cite{chen2009ultrasound, strachinaru2017cardiac, konofagou2002myocardial} and vascular health~\cite{de2000characterization, li2019two, maurice2004noninvasive} assessment, surgical planning~\cite{della2021predicting, sastry2017applications} and numerous other clinical applications. Among different ultrasound elastography techniques, quasi-static elastography~\cite{varghese2009quasi} has gained popularity because of its low cost and ease of use. The ultimate goal of quasi-static elastography is to estimate the distribution of elasticity over the tissue region of interest (ROI) using the boundary information and the tissue strain induced by some external force. The first step of strain estimation is to slowly compress the tissue and collect time-series radio-frequency (RF) data with the same ultrasound probe, which is hand-held or attached to a mechanical arm. Then some time-delay estimation (TDE) technique calculates the displacement field between the pre- and the post-deformed RF frames, which is spatially differentiated to obtain the strain map. The strain map often reveals tissue pathology by displaying a color contrast between the healthy and diseased tissues.

Among the steps of ultrasound strain imaging, TDE is a crucial one which is performed by one of the three established techniques: window-based~\cite{luo2010fast, hall2003vivo, kibria2015class, al2020locally}, machine learning-based~\cite{ali_2020, kibria2018gluenet, gao2019learning} and regularized optimization- or energy-based~\cite{rpca_glue, intro14, pan2014two, islam2018new, mglue}. The window-based or block-matching techniques consider a data window around the target sample in the pre-deformed RF frame and track that window in the post-deformed frame by finding the position of maximum normalized cross-correlation (NCC)~\cite{Zahiri_2006, hybrid} or zero-phase crossing~\cite{Xunchang_2004, sharmin_2013}. These techniques' performance is highly controlled by the proper choice of window size and the inter-window overlap. A large data window provides a smoother displacement map sacrificing accuracy, whereas a small data window improves the accuracy and resolution but leads to a noisy displacement estimate. They also can fail if the level of noise in the window is large. Machine learning-based techniques exploit the power of data and train a model such as convolutional neural network (CNN) to estimate the displacement field between the RF frames under consideration. Although the machine learning-based algorithms are an interesting addition to the ultrasound elastography literature, their clinical adoption is subject to the availability of sufficient training data.       
 
\begin{figure}
	\centering
	\subfigure[Algorithm illustration]{{\includegraphics[width=.5\textwidth]{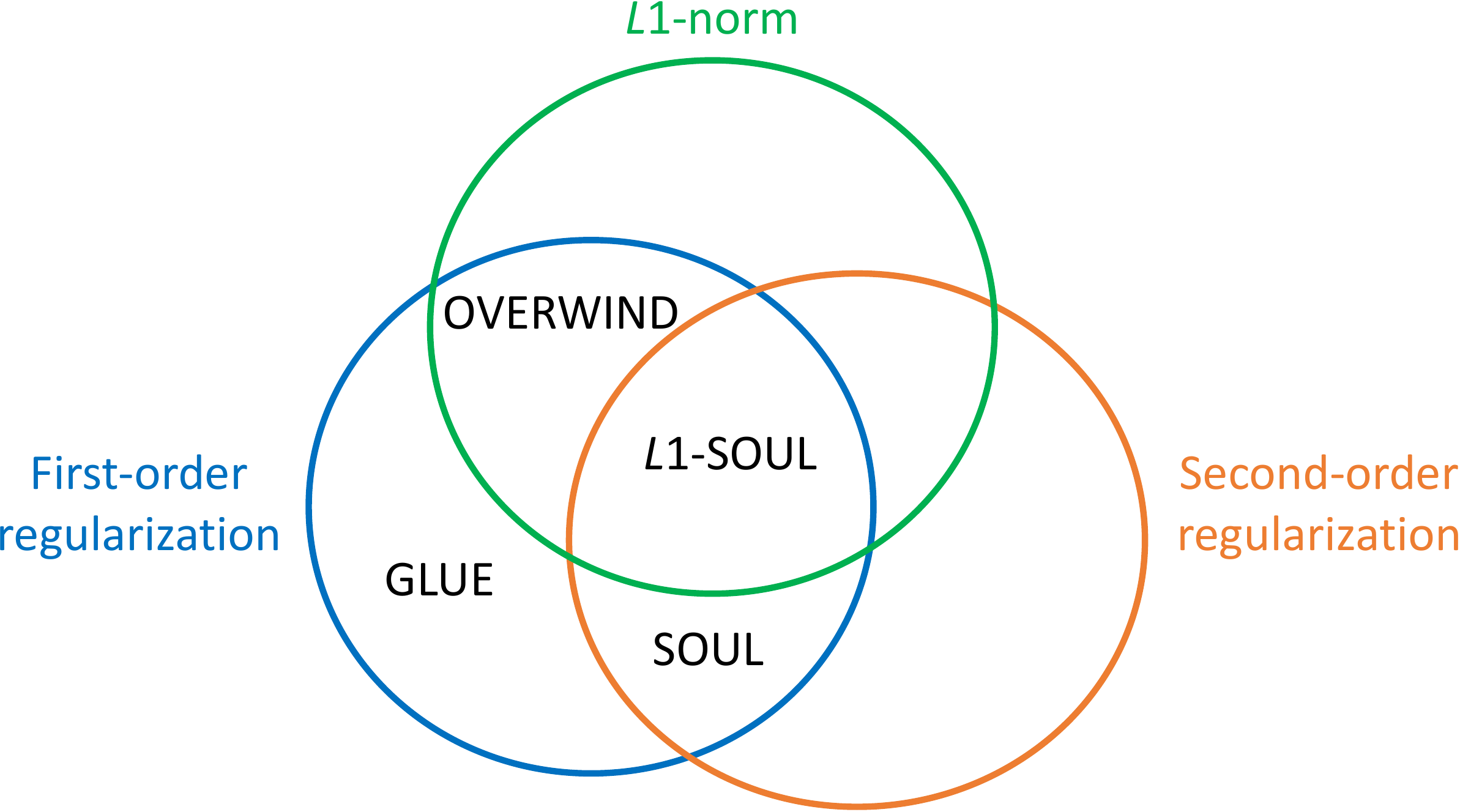}}}
	\subfigure[Ground truth]{{\includegraphics[width=.105\textwidth]{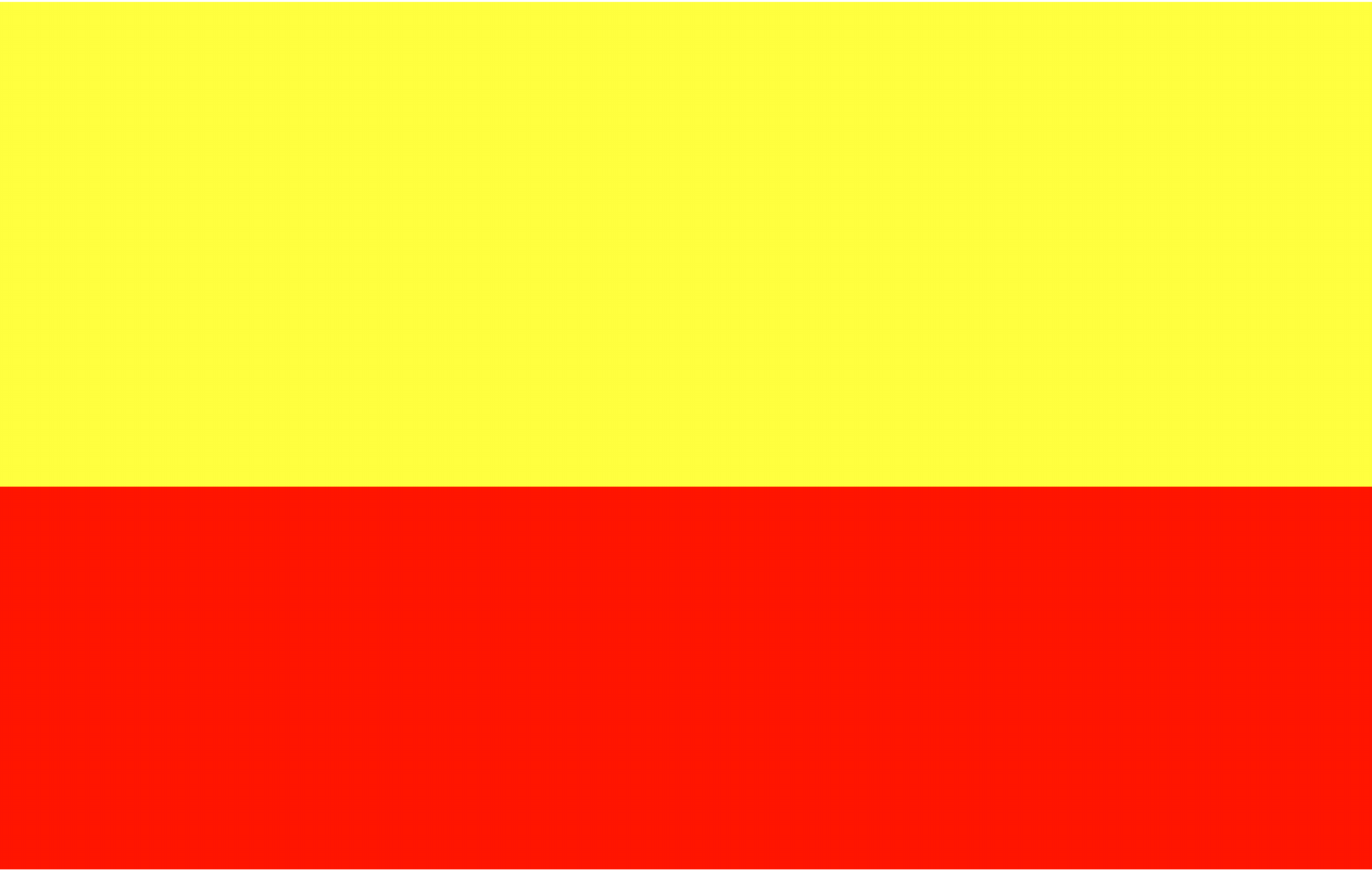}}}%
	\quad
    \subfigure[OVERWIND]{{\includegraphics[width=.105\textwidth]{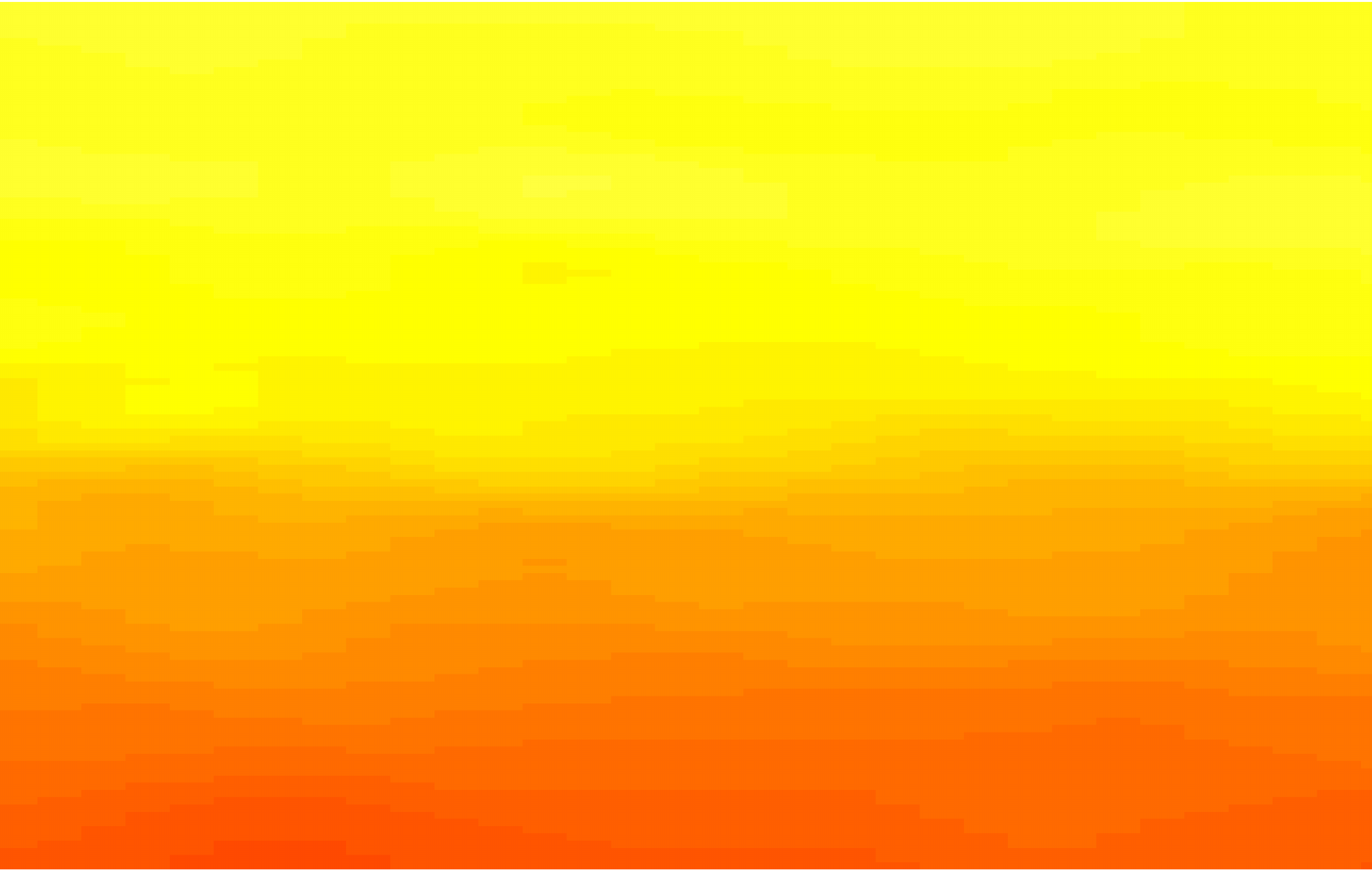}}}%
    \quad
    \subfigure[SOUL]{{\includegraphics[width=.105\textwidth]{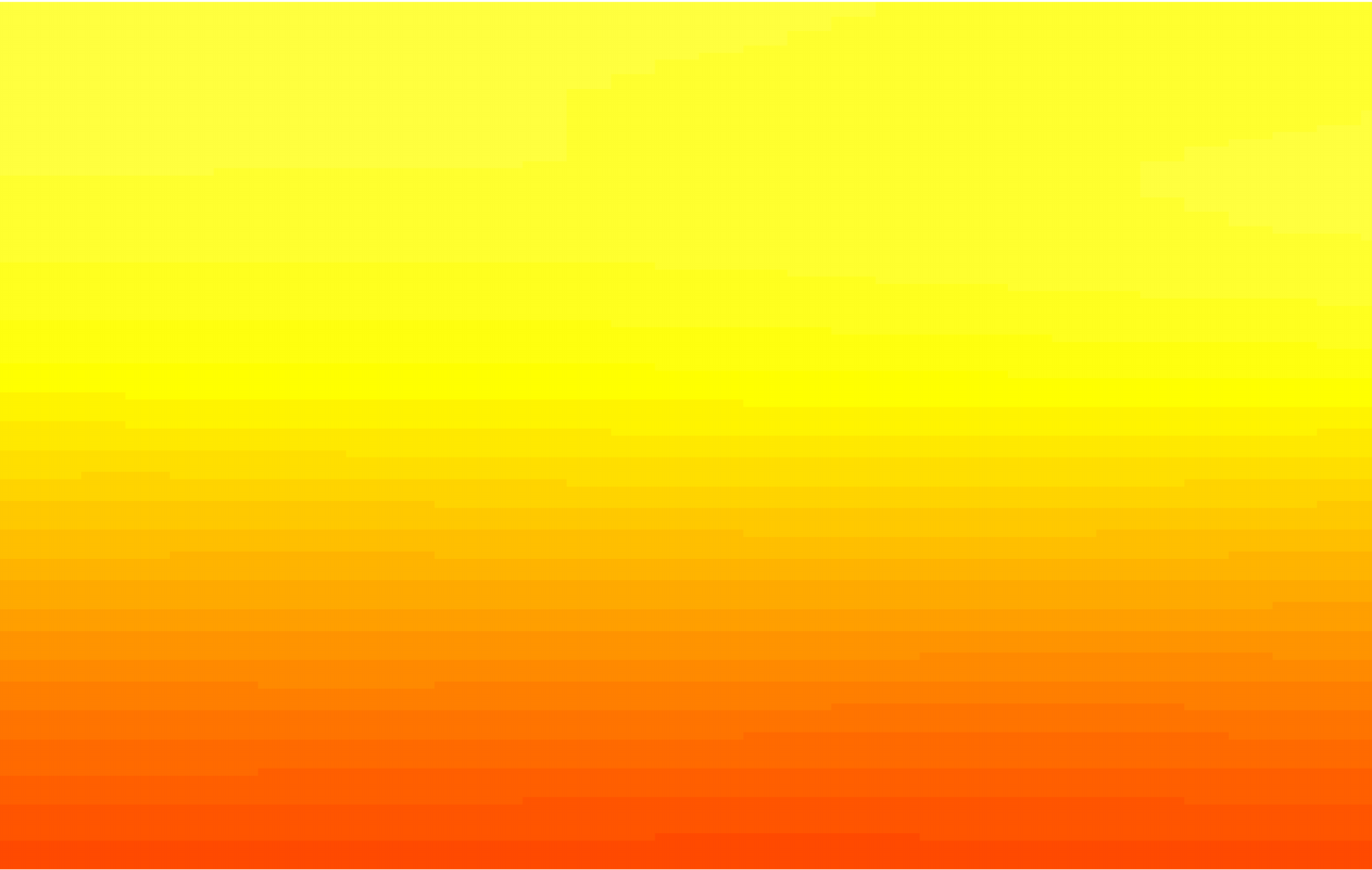}}}%
    \quad
    \subfigure[$L1$-SOUL]{{\includegraphics[width=.105\textwidth]{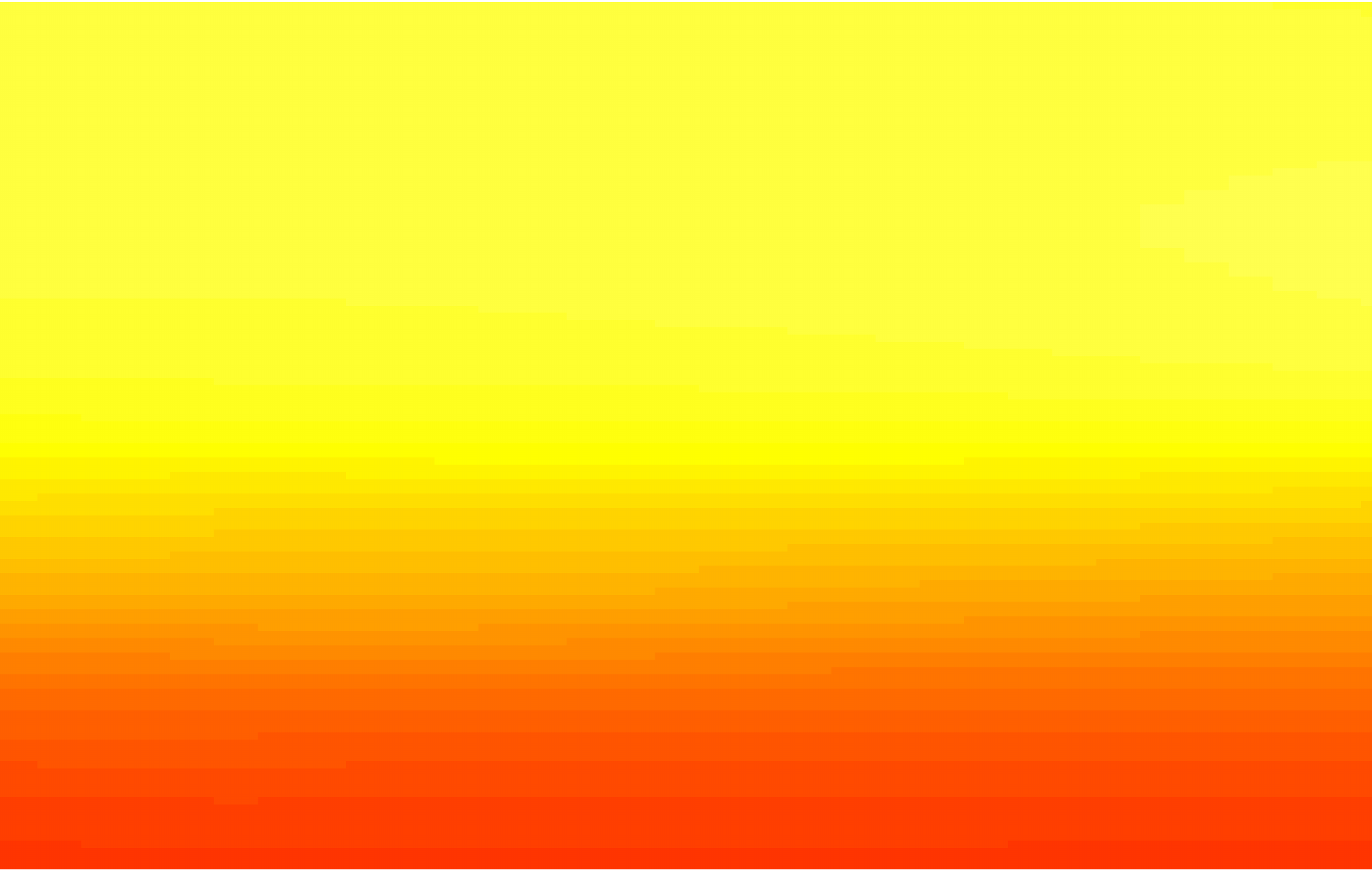}}}
	\caption{Illustration of different TDE algorithms. The Venn diagram in (a) shows the differences among GLUE, OVERWIND, SOUL, and $L1$-SOUL. (b)-(e) demonstrate the edge-preserving abilities of OVERWIND, SOUL, and $L1$-SOUL.}
	\label{venn_methods}
\end{figure}

This work is focused on regularized optimization-based approach, the last mainstream TDE class which optimizes a non-linear cost function consisting of data fidelity, spatial~\cite{glue,soul} and temporal~\cite{guest} continuity constraints to obtain accurate and spatially smooth displacement and strain maps. The downside of many optimization-based techniques is that they are computationally demanding and take a long time to execute. However, Dynamic Programming (DP)~\cite{DP,intro14} and analytic optimization-based techniques such as Dynamic Programming Analytic Minimization (DPAM)~\cite{DPAM} and GLobal Ultrasound Elastography (GLUE)~\cite{glue} have resolved the issue of computational expense. GLUE penalizes the first derivatives of the axial and lateral displacements to construct the spatial regularizer. This first-order regularization scheme enforces the displacement field to be constant. But biological tissue yields a smooth and linearly increasing displacement profile over depth in response to a quasi-static compression. Therefore, GLUE's regularizer does not represent tissue-deformation physics accurately. To resolve this limitation of GLUE, Second-Order Ultrasound eLastography (SOUL)~\cite{soul} has been proposed, which takes both the first- and second-order spatial continuity constraints into account and provides substantially higher quality displacement and strain maps. Like GLUE, SOUL uses the $L2$-norm as the penalty term. This $L2$-norm regularization over-penalizes sharp transitions in the displacement field due to the square operation. As a consequence, the resulting strain map suffers from two major issues. First, the edges of the target structures (i.e., tumor) are not sharp. Second, the visual contrast between the healthy and pathologic tissues is not satisfactory, especially if the target is small. In this work, we address these issues by penalizing the $L1$-norms of both first- and second-order displacement derivatives instead of $L2$-norms. Our contribution is twofold. First, we devise a novel cost function consisting of data similarity and $L1$-norm first- and second-order spatial continuity terms. Second, we analytically optimize the formulated penalty function by adopting an iterative scheme. We name our technique \textbf{$L1$-SOUL}: Second-Order Ultrasound eLastography with $L1$-norm spatial regularization. Although $L1$-norm continuity constraint was used in tOtal Variation rEgulaRization and WINDow-based time delay estimation (OVERWIND)~\cite{overwind} before, it considered $L1$-norms of first-order displacement derivatives only. In contrast, the $L1$-SOUL cost function takes $L1$-norms of both first- and second-order displacement derivatives into account and therefore uses the powers of both second-order regularization and $L1$-norm to devise a complete TDE technique. Fig.~\ref{venn_methods} highlights the similarities and differences among GLUE, OVERWIND, SOUL, and $L1$-SOUL. 

The proposed technique is validated against three simulated phantom, one experimental breast phantom, and three \textit{in vivo} liver cancer datasets. Similar to our previous work~\cite{RAPID_TMI,guest,soul}, we have published the $L1$-SOUL code at \url{ http://code.sonography.ai}.           

\section{Methods}
Let $I_{1}(i,j)$ and $I_{2}(i,j)$, $1 \leq i \leq m$, $1 \leq j \leq n$ denote two ultrasound RF frames acquired from a tissue before and after deformation, respectively. Our ultimate goal is to calculate the tissue strain resulting from the applied force. To that end, we estimate the axial and lateral displacement fields between $I_{1}$ and $I_{2}$.

This section first describes SOUL~\cite{soul}, a previously published displacement estimation technique. Then it presents the technical and mathematical details of $L1$-SOUL, the proposed technique.   

\subsection{Second-Order Ultrasound eLastography (SOUL)}
SOUL estimates the integer axial and lateral displacement fields $a_{i,j}$ and $l_{i,j}$ from DP~\cite{DP}. The DP integer estimates are fine-tuned by optimizing a non-linear cost function consisting of data fidelity, first- and second-order spatial continuity terms:  

\begin{equation}
\begin{aligned}
&C (\Delta a_{1,1},...,\Delta a_{m,n},\Delta l_{1,1},...,\Delta l_{m,n}) = \norm{\textrm{Data fidelity}}_{2}^{2}+ \\
&\gamma \norm{\partial_{y}a_{f}}_{2}^{2}+
\alpha_{1} \norm{\partial_{y}a}_{2}^{2} + \alpha_{2} \norm{\partial_{x}a}_{2}^{2} + \beta_{1} \norm{\partial_{y}l}_{2}^{2} + \beta_{2} \norm{\partial_{x}l}_{2}^{2}+\\ 
& \theta_{1} \norm{\partial_{y}^{2}a}_{2}^{2} + \theta_{2} \norm{\partial_{x}^{2}a}_{2}^{2} + \lambda_{1} \norm{\partial_{y}^{2}l}_{2}^{2} + \lambda_{2} \norm{\partial_{x}^{2}l}_{2}^{2}
\end{aligned}
\label{eq:SOUL_c}
\end{equation}



\noindent
where $(\partial_{y}a_{f})_{1,j}$ denotes the first-order derivative of the axial displacement of $j$-th RF line's first sample, and is defined as:

\begin{equation}
\begin{aligned}
(\partial_{y}a_{f})_{1,j} = a_{1,j}+\Delta a_{1,j}
\end{aligned}
\label{eq:dela_af}
\end{equation}

\noindent
Note that the displacement of the imaginary sample prior to the first sample of an RF line is zero~\cite{soul}. $(\partial_{y}a)_{i,j}$, $(\partial_{x}a)_{i,j}$, $(\partial_{y}l)_{i,j}$, and $(\partial_{x}l)_{i,j}$ stand for the biased derivatives of the axial and lateral displacements, and are given by:     

\begin{equation}
\begin{aligned}
(\partial_{y}a)_{i,j} = a_{i,j}+\Delta a_{i,j}-a_{i-1,j}- \Delta a_{i-1,j}-\epsilon_{a}
\end{aligned}
\label{eq:dela1}
\end{equation}

\begin{equation}
\begin{aligned}
(\partial_{x}a)_{i,j} = a_{i,j}+\Delta a_{i,j}-a_{i,j-1}- \Delta a_{i,j-1}-\epsilon_{a}
\end{aligned}
\label{eq:dela2}
\end{equation}

\begin{equation}
\begin{aligned}
(\partial_{y}l)_{i,j} = l_{i,j}+\Delta l_{i,j}-l_{i-1,j}- \Delta l_{i-1,j}-\epsilon_{l}
\end{aligned}
\label{eq:dell1}
\end{equation}

\begin{equation}
\begin{aligned}
(\partial_{x}l)_{i,j} = l_{i,j}+\Delta l_{i,j}-l_{i,j-1}- \Delta l_{i,j-1}-\epsilon_{l}
\end{aligned}
\label{eq:dell2}
\end{equation}

$(\partial_{y}^{2}a)_{i,j}$, $(\partial_{x}^{2}a)_{i,j}$, $(\partial_{y}^{2}l)_{i,j}$, and $(\partial_{x}^{2}l)_{i,j}$ indicate the second-order derivatives of the axial and lateral displacement fields:     

\begin{equation}
\begin{aligned}
(\partial_{y}^{2}a)_{i,j} = &a_{i-1,j}+\Delta a_{i-1,j}+a_{i+1,j}+\Delta a_{i+1,j}-\\
&2a_{i,j}-2\Delta a_{i,j}
\end{aligned}
\label{eq:dela3}
\end{equation}

\begin{equation}
\begin{aligned}
(\partial_{x}^{2}a)_{i,j} = &a_{i,j-1}+\Delta a_{i,j-1}+a_{i,j+1}+\Delta a_{i,j+1}-\\
&2a_{i,j}-2\Delta a_{i,j}
\end{aligned}
\label{eq:dela4}
\end{equation}

\begin{equation}
\begin{aligned}
(\partial_{y}^{2}l)_{i,j} = &l_{i-1,j}+\Delta l_{i-1,j}+l_{i+1,j}+\Delta l_{i+1,j}-\\
&2l_{i,j}-2\Delta l_{i,j}
\end{aligned}
\label{eq:dell3}
\end{equation}

\begin{equation}
\begin{aligned}
(\partial_{x}^{2}l)_{i,j} = &l_{i,j-1}+\Delta l_{i,j-1}+l_{i,j+1}+\Delta l_{i,j+1}-\\
&2l_{i,j}-2\Delta l_{i,j}
\end{aligned}
\label{eq:dell4}
\end{equation}

$\Delta a_{i,j}$ and $\Delta l_{i,j}$ are the refinement displacements to be calculated. $\alpha_{1}$, $\alpha_{2}$ and $\beta_{1}$, $\beta_{2}$ denote the first-order spatial continuity weights in the axial and lateral directions, respectively. $\theta_{1}$, $\theta_{2}$ and $\lambda_{1}$, $\lambda_{2}$, respectively, are the axial and lateral second-order regularization parameters. $\gamma$ is the continuity weight for the first sample of each scan line. $\epsilon_{a}$ and $\epsilon_{l}$ indicate axial and lateral adaptive regularization parameters which prevent the underestimation of displacement fields~\cite{guest,soul} and are calculated as:

\begin{equation}
\begin{array}{lll}
\epsilon_{a} = \frac{a_{m}-a_{1}}{m-1}, \ \ \epsilon_{l} = \frac{l_{n}-l_{1}}{n-1}
\end{array}
\label{eq:epsilon1}
\end{equation}

The implementation details of SOUL can be found in~\cite{soul}.

\subsection{Second-Order Ultrasound eLastography with $L1$-norm spatial regularization ($L1$-SOUL)}
SOUL incorporates the $L2$-norm of displacement derivatives to formulate the penalty function. Although the combination of first- and second-order continuity constraints model tissue deformation accurately, the $L2$-norm often over-penalizes displacement and strain discontinuities, which results in undesired spreading of tumor edge and reduction of target-background contrast. To resolve this issue, instead of $L2$-norm, the regularization term of $L1$-SOUL employs the $L1$-norms of both first- and second-order displacement derivatives. However, $L1$-norm makes the cost function hard to optimize since $\lvert u \rvert$ is not differentiable around $u = 0$. Similar to~\cite{overwind}, we handle this situation by replacing $\lvert u \rvert$ with $\xi(u) = \sqrt{\eta^{2} + u^{2}}$ to introduce smoothness to the sharp corner of $\lvert u \rvert$ and make it differentiable where $\eta$ is a small-valued tunable parameter which controls the level of corner smoothness.

The cost function associated with $L1$-SOUL is given below.

\begin{equation}
\begin{aligned}
&C_{s} (\Delta a_{1,1},...,\Delta a_{m,n},\Delta l_{1,1},...,\Delta l_{m,n}) =\\ 
&\sum\limits_{j=1}^n \sum\limits_{i=1}^m [I_{1}(i,j)-I_{2}(i+a_{i,j}+\Delta a_{i,j},j+l_{i,j}+\Delta l_{i,j})]^{2} +\\
&2\gamma \eta_{0} \norm{\partial_{y}a_{f}}_{1}+
2\alpha_{1}\eta_{1} \norm{\partial_{y}a}_{1} + 2\alpha_{2}\eta_{1} \norm{\partial_{x}a}_{1} + 2\beta_{1}\eta_{1} \norm{\partial_{y}l}_{1}\\ 
& + 2\beta_{2}\eta_{1} \norm{\partial_{x}l}_{1}+2\theta_{1}\eta_{2} \norm{\partial_{y}^{2}a}_{1} + 2\theta_{2}\eta_{2} \norm{\partial_{x}^{2}a}_{1} + 2\lambda_{1}\eta_{2} \norm{\partial_{y}^{2}l}_{1}\\ 
&+ 2\lambda_{2}\eta_{2} \norm{\partial_{x}^{2}l}_{1} 
\end{aligned}
\label{eq:c_soulmate}
\end{equation}

\noindent
where $\norm{\partial_{y}a_{f}}_{1}$ is defined as:

\begin{equation}
\begin{aligned}
\norm{\partial_{y}a_{f}}_{1} = \sum\limits_{j=1}^n \sqrt{\eta_{0}^{2} + (\partial_{y}a_{f})_{1,j}^2}
\end{aligned}
\label{eq:r0}
\end{equation}

Here, $\eta_{0}$ is the sharpness-controlling parameter. $\norm{\partial_{y}a}_{1}$, $\norm{\partial_{x}a}_{1}$, $\norm{\partial_{y}l}_{1}$, and $\norm{\partial_{x}l}_{1}$ are defined as:

\begin{equation}
\begin{aligned} 
\norm{\cdot}_{1} = \sum\limits_{j=1}^{n} \sum\limits_{i=1}^{m} \sqrt{\eta_{1}^{2} + (\cdot)_{i,j}^2}
\end{aligned}
\label{eq:soulmate_r1}
\end{equation}

\noindent
where $\eta_{1}$ controls the level of smoothness introduced to the absolute first-order derivatives. $\norm{\partial_{y}^{2}a}_{1}$, $\norm{\partial_{x}^{2}a}_{1}$, $\norm{\partial_{y}^{2}l}_{1}$, and $\norm{\partial_{x}^{2}l}_{1}$ are defined as:

\begin{equation}
\begin{aligned} 
\norm{\cdot}_{1} = \sum\limits_{j=1}^{n} \sum\limits_{i=1}^{m} \sqrt{\eta_{2}^{2} + (\cdot)_{i,j}^2}
\end{aligned}
\label{eq:soulmate_r2}
\end{equation}

\noindent
where $\eta_{2}$ is the sharpness-controlling parameter for the second-order regularization.

Our data function is non-linear due to the appearance of $\Delta a$ and $\Delta l$ within $I_{2}(\cdot)$. We remove this non-linearity by approximating $I_{2}$ by its first-order Taylor series expansion:

\begin{equation}
\begin{aligned}
&I_{2}(i+a_{i,j}+\Delta a_{i,j},j+l_{i,j}+\Delta l_{i,j}) \approx \\
&I_{2}(i+a_{i,j},j+l_{i,j})+\Delta a_{i,j}I_{2,a}^{'}+\Delta l_{i,j}I_{2,l}^{'} 
\end{aligned}
\label{eq:i2_taylor}
\end{equation}

\noindent
where $I_{2,a}^{'}$ and $I_{2,l}^{'}$, respectively, denote the axial and lateral derivatives of $I_{2}$. We now optimize $C_{s}$ by setting $\frac{\partial C_{s,i,j}}{\partial \Delta a_{i,j}}=0$ and $\frac{\partial C_{s,i,j}}{\partial \Delta l_{i,j}}=0$. The derivatives of $\sqrt{\eta_{p}^{2} + deriv^{2}}$-like regularization terms introduce non-linearity to the equations and make optimization intractable. We overcome this issue by adopting an iterative approach where $\frac{1}{\sqrt{\eta_{p}^{2} + deriv^{2}}}$ is updated using the derivatives from previous iteration and kept intact within the current iteration. Finally, we obtain the following linear system of equations.

\begin{equation}
(H+D+D_{2})\Delta d = H_{1}\mu - (D+D_{2})d + b_{s}
\label{eq:soul_axb}
\end{equation}

\noindent
where $d=[a_{1,1},l_{1,1},a_{1,2},l_{1,2},\dots,a_{m,n},l_{m,n}]^T$ contains the DP initial displacement estimates. The fine-tuning axial and lateral displacement estimates are stacked in $\Delta d=[\Delta a_{1,1},\Delta l_{1,1},\Delta a_{1,2},\Delta l_{1,2},\dots,\Delta a_{m,n},\Delta l_{m,n}]^T$. $D$ and $D_{2}$ are sparse matrices of size $2mn \times 2mn$ which contain functions of first- and second-order continuity weights. $H$ and $H_{1}$, respectively, are symmetric tridiagonal and diagonal matrices of size $2mn \times 2mn$ which contain functions of data derivatives. $\mu \in \mathbb{R}^{2mn \times 1}$ consists of the data residuals. $b_{s} \in \mathbb{R}^{2mn \times 1}$ stands for the first-order adaptive regularization term which contains functions of $\epsilon_{a}$, $\epsilon_{l}$, $\alpha_{1}$, $\alpha_{2}$, $\beta_{1}$, $\beta_{2}$, and $\eta_{1}$.

\begin{figure*}
	\centering
	\subfigure[Ground truth]{{\includegraphics[width=0.2\textwidth,height=0.31\textwidth]{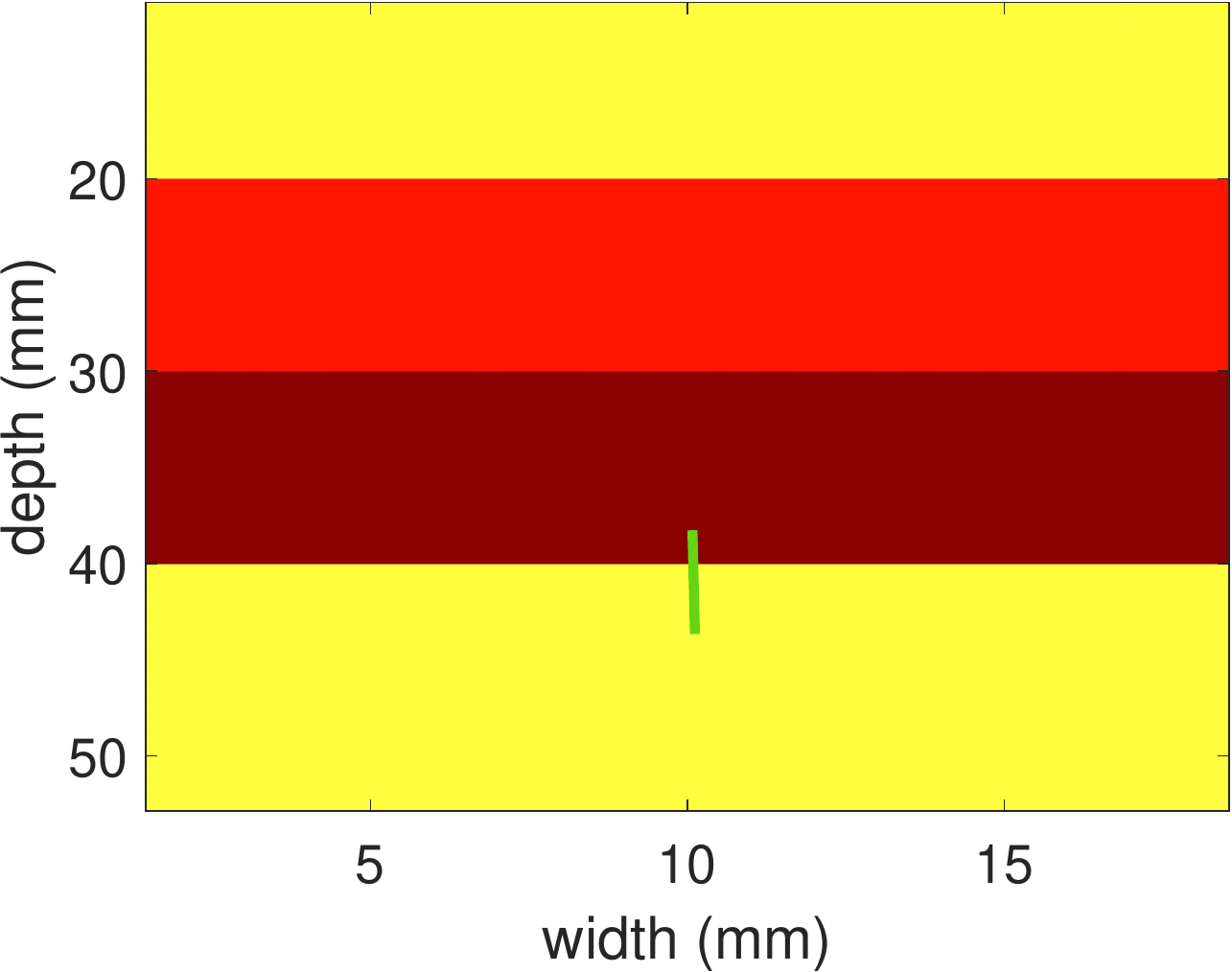}}}%
	\subfigure[GLUE]{{\includegraphics[width=0.2\textwidth,height=0.31\textwidth]{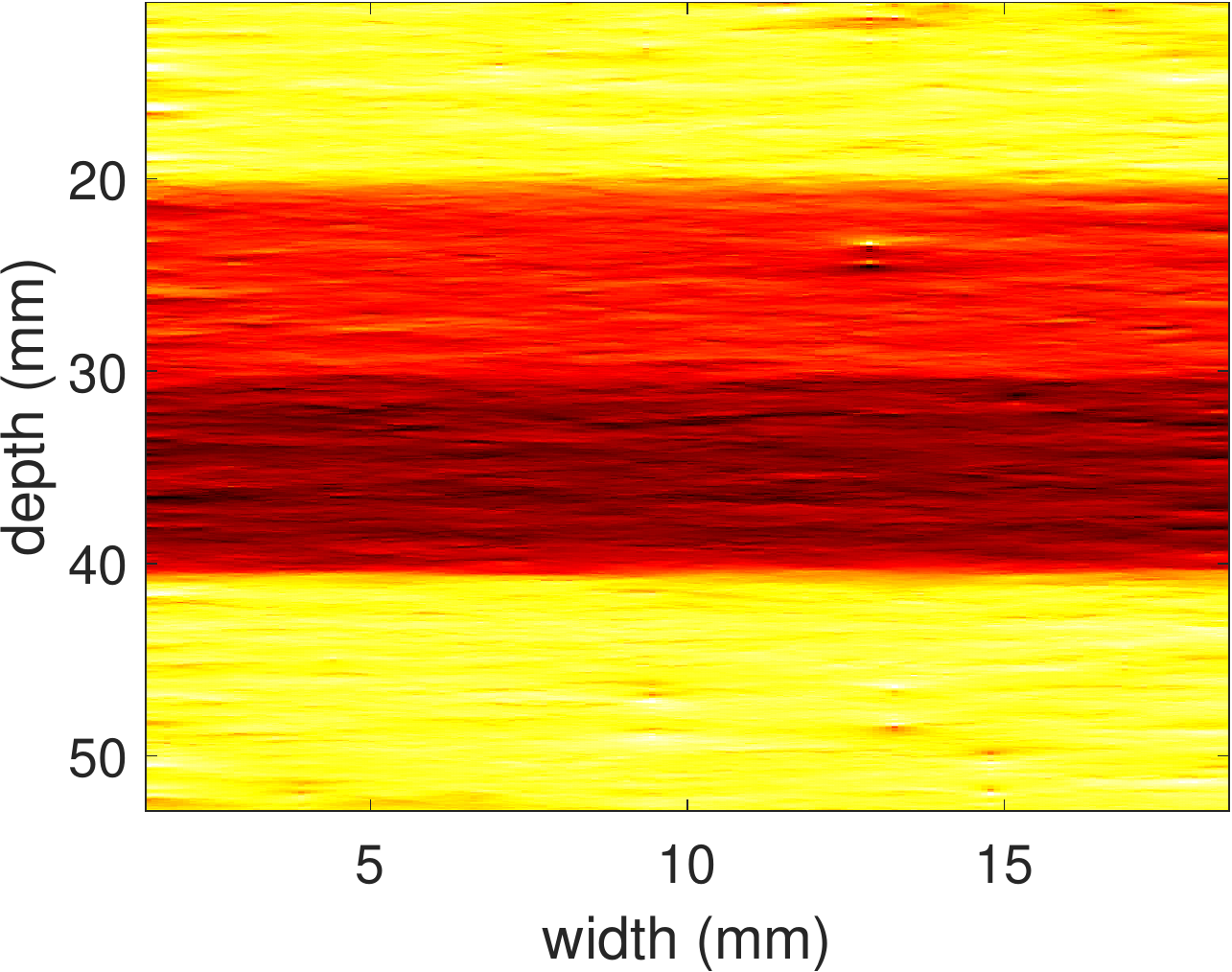}}}%
	\subfigure[OVERWIND]{{\includegraphics[width=0.2\textwidth,height=0.31\textwidth]{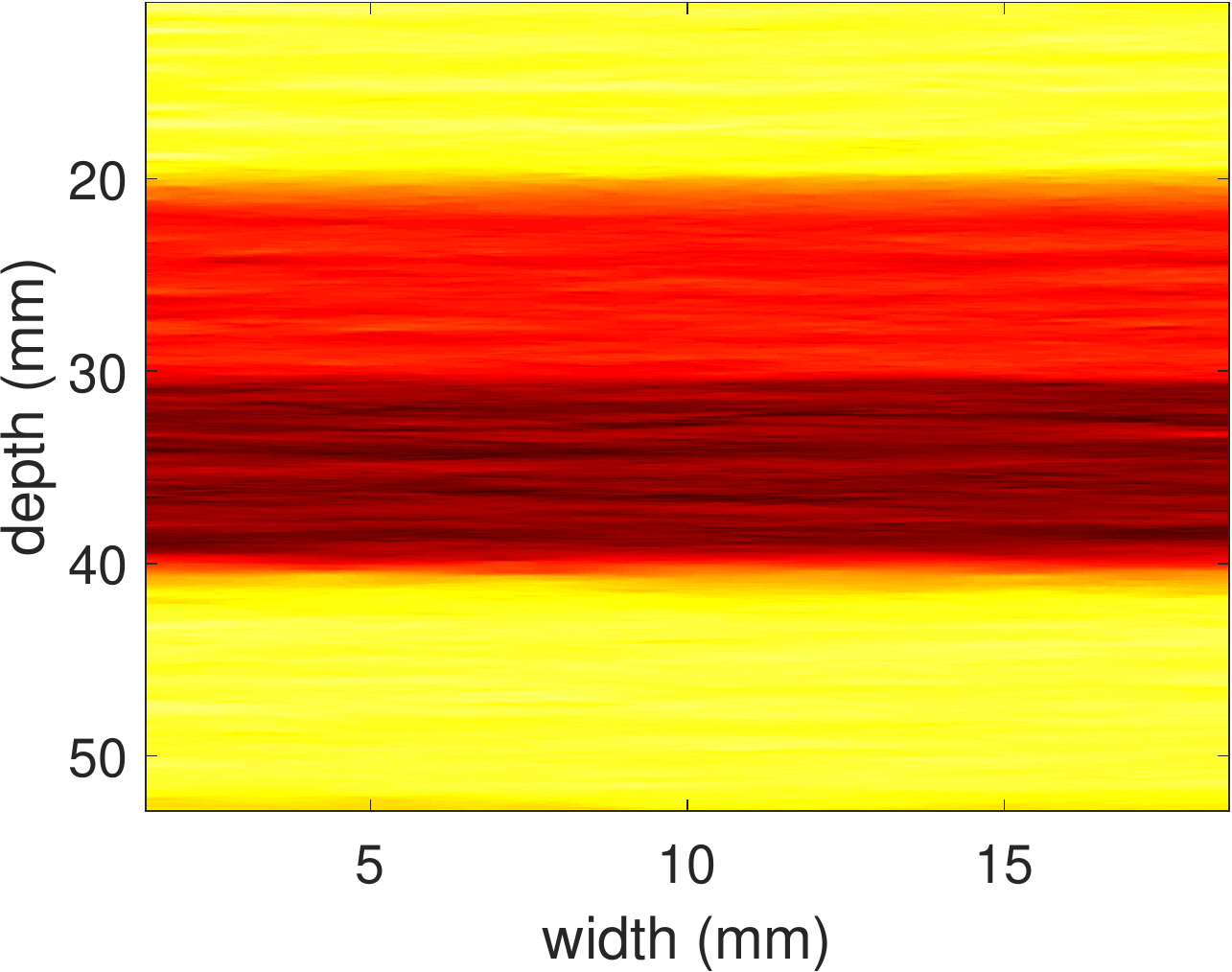} }}%
	\subfigure[SOUL]{{\includegraphics[width=0.2\textwidth,height=0.31\textwidth]{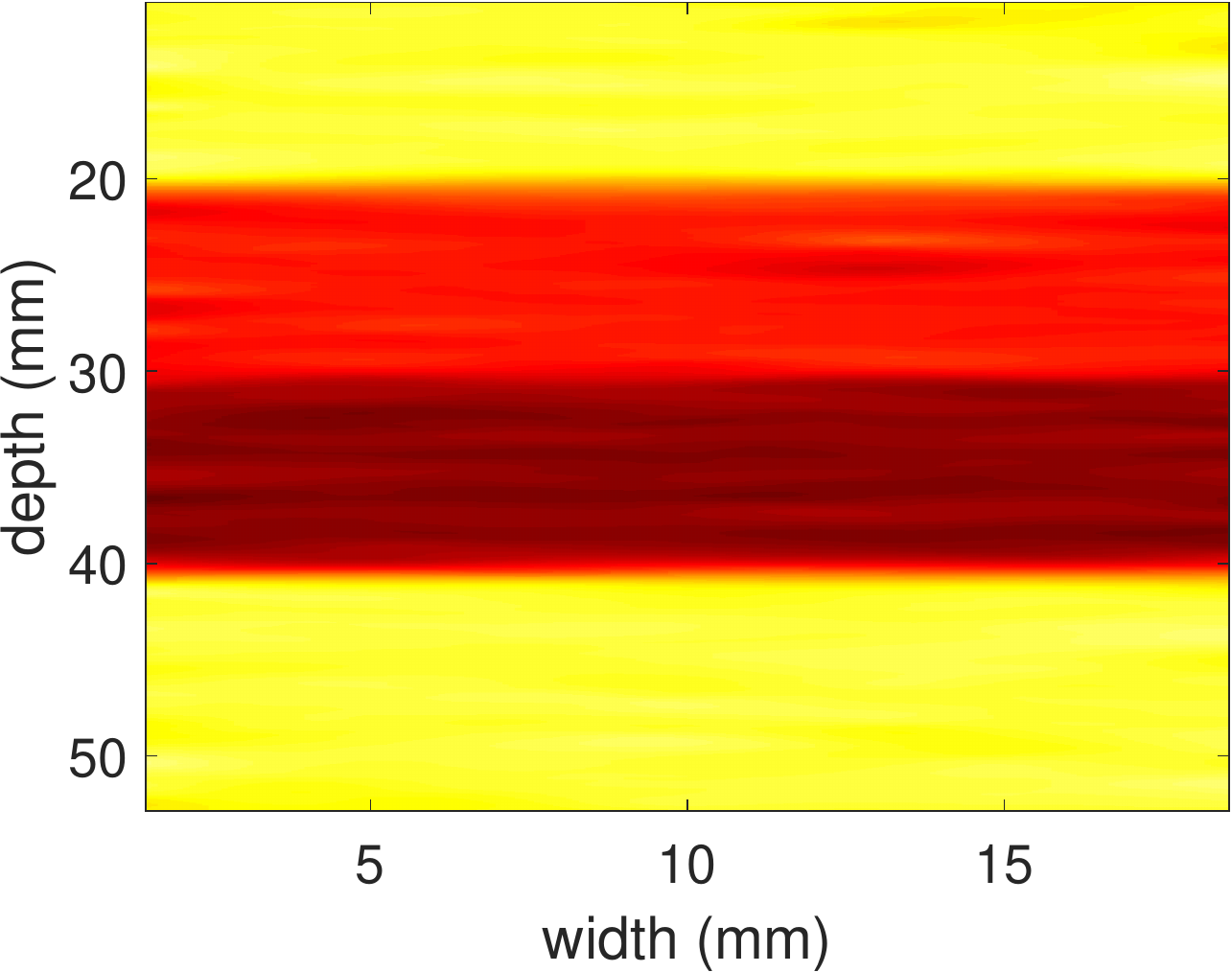} }}%
	\subfigure[$L1$-SOUL]{{\includegraphics[width=0.2\textwidth,height=0.31\textwidth]{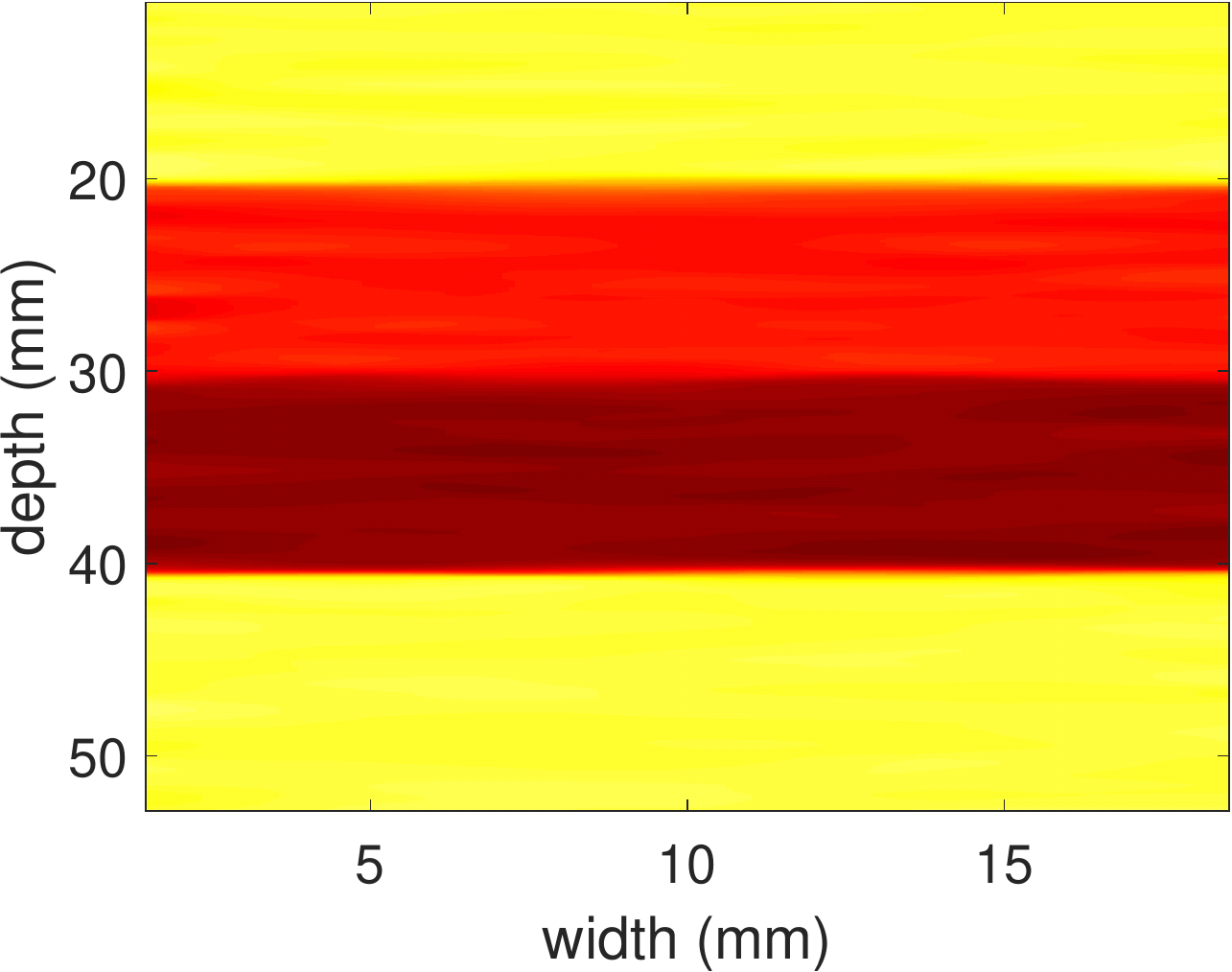} }}
	\subfigure[Ground truth]{{\includegraphics[width=0.2\textwidth,height=0.31\textwidth]{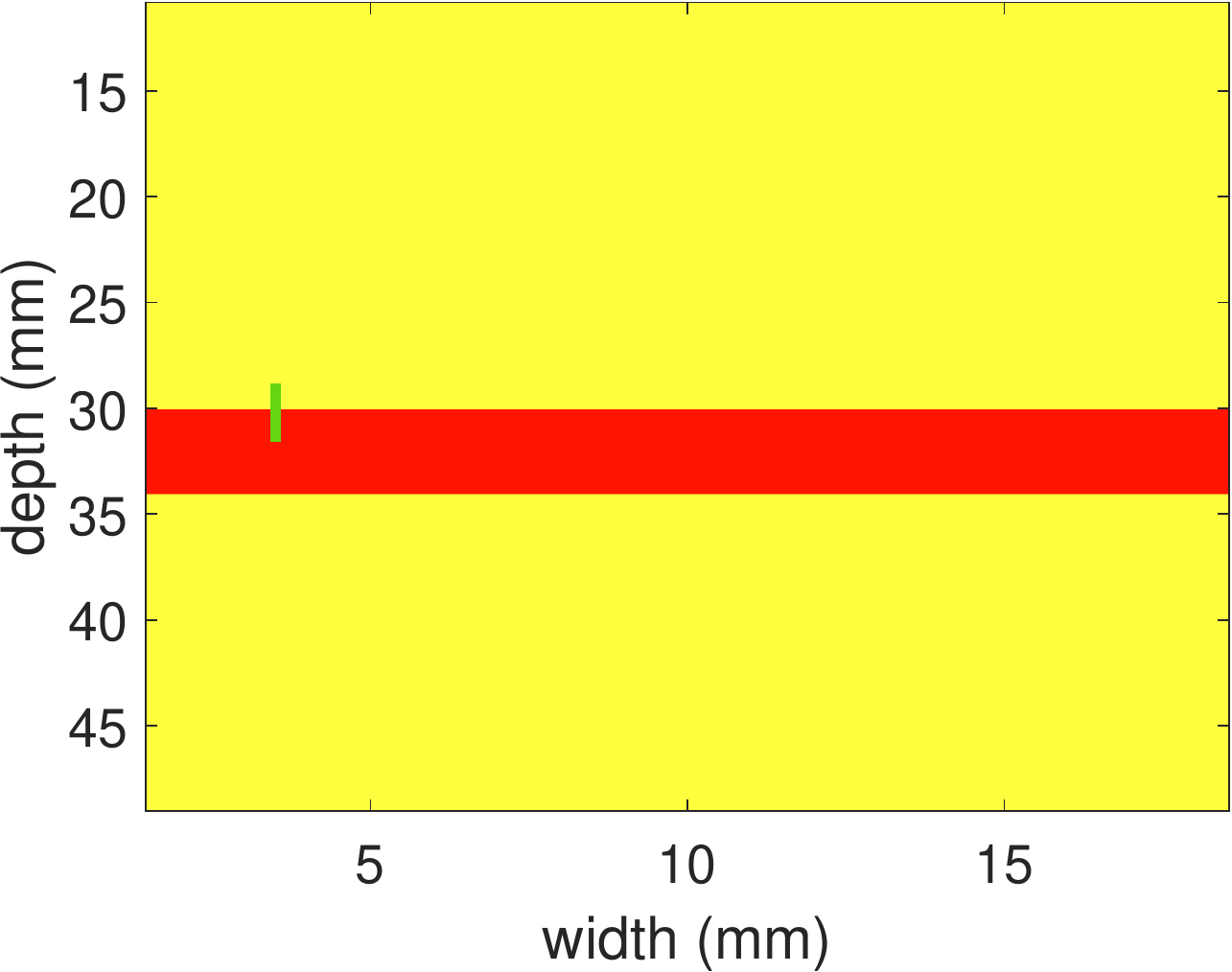}}}%
	\subfigure[GLUE]{{\includegraphics[width=0.2\textwidth,height=0.31\textwidth]{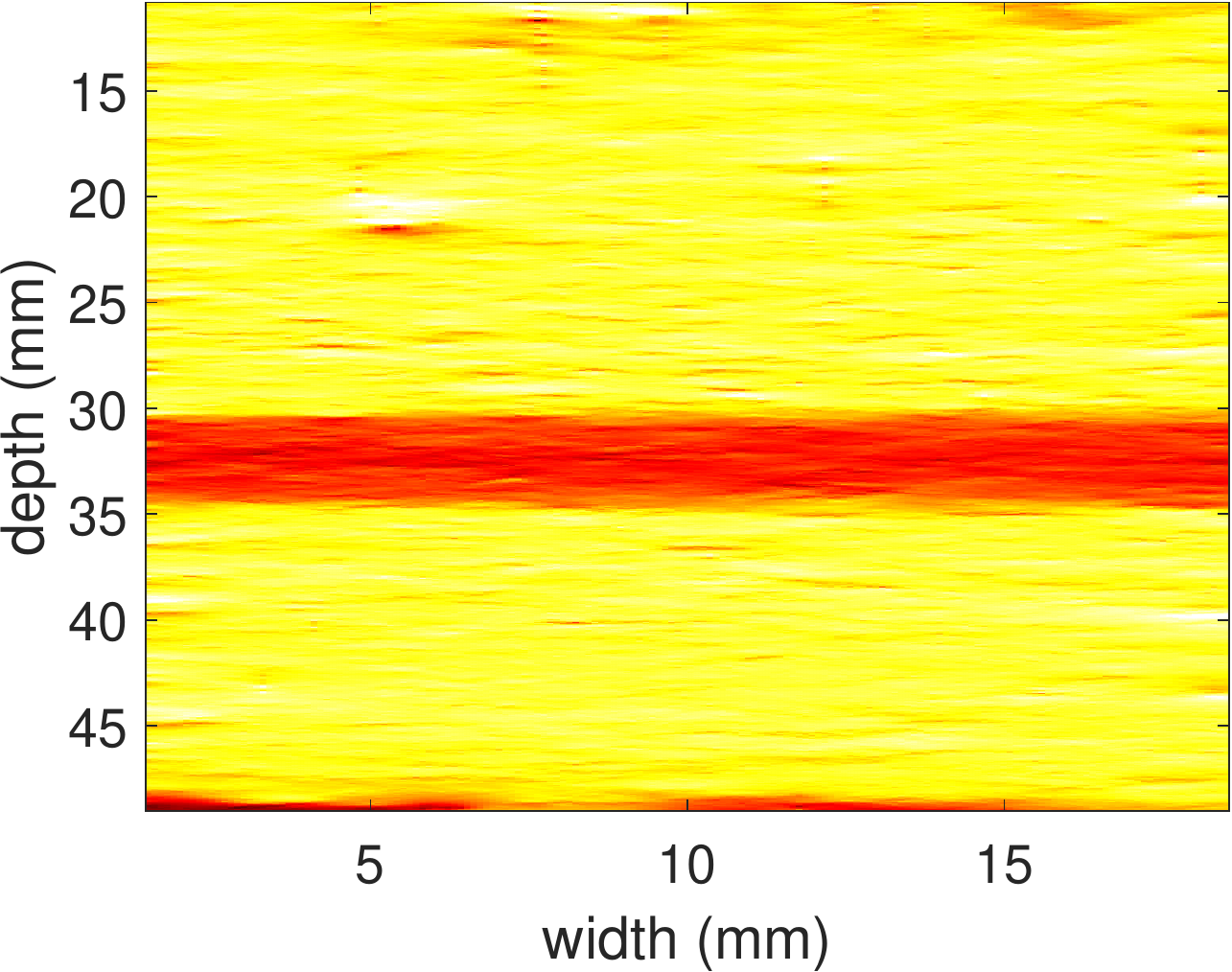}}}%
	\subfigure[OVERWIND]{{\includegraphics[width=0.2\textwidth,height=0.31\textwidth]{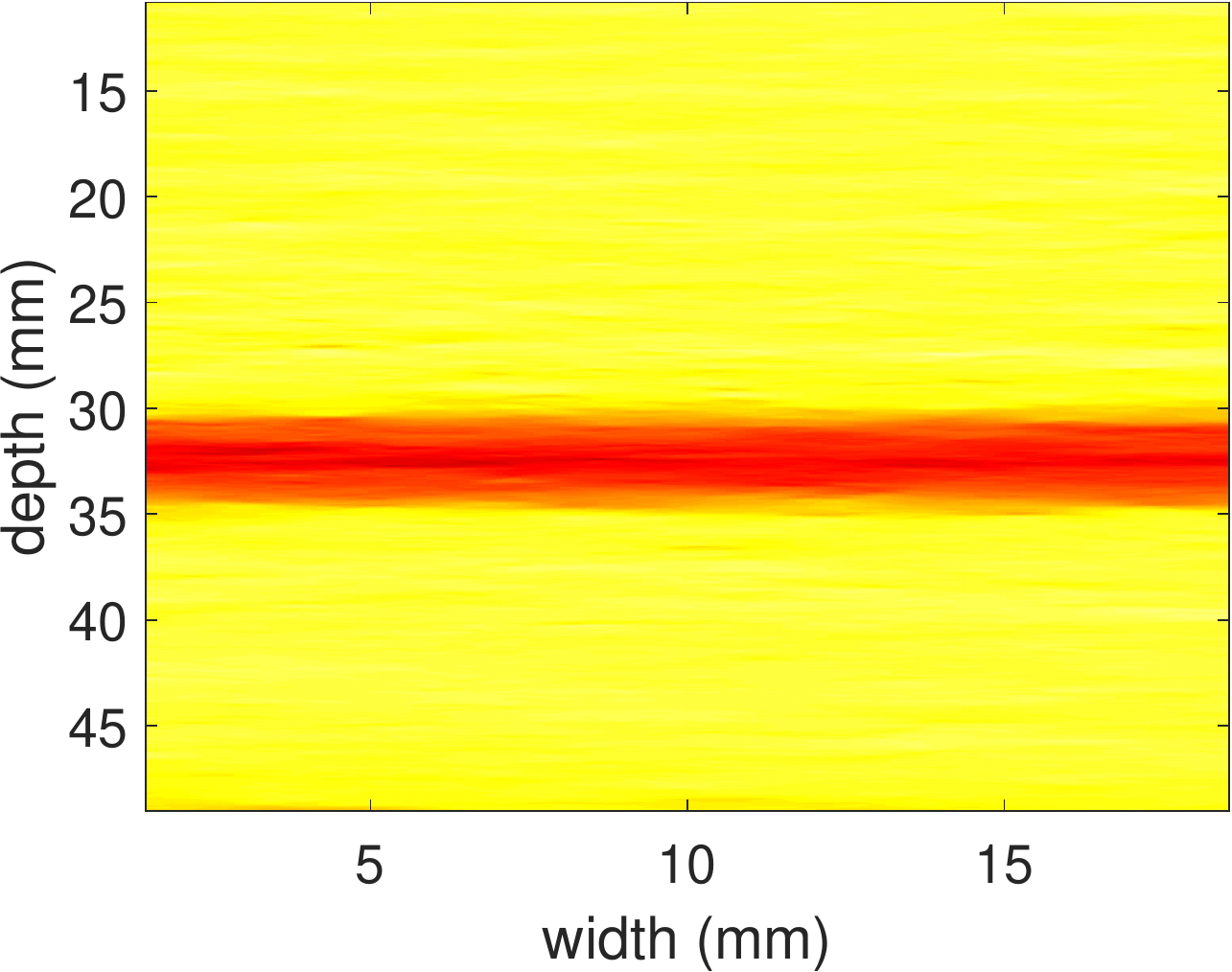} }}%
	\subfigure[SOUL]{{\includegraphics[width=0.2\textwidth,height=0.31\textwidth]{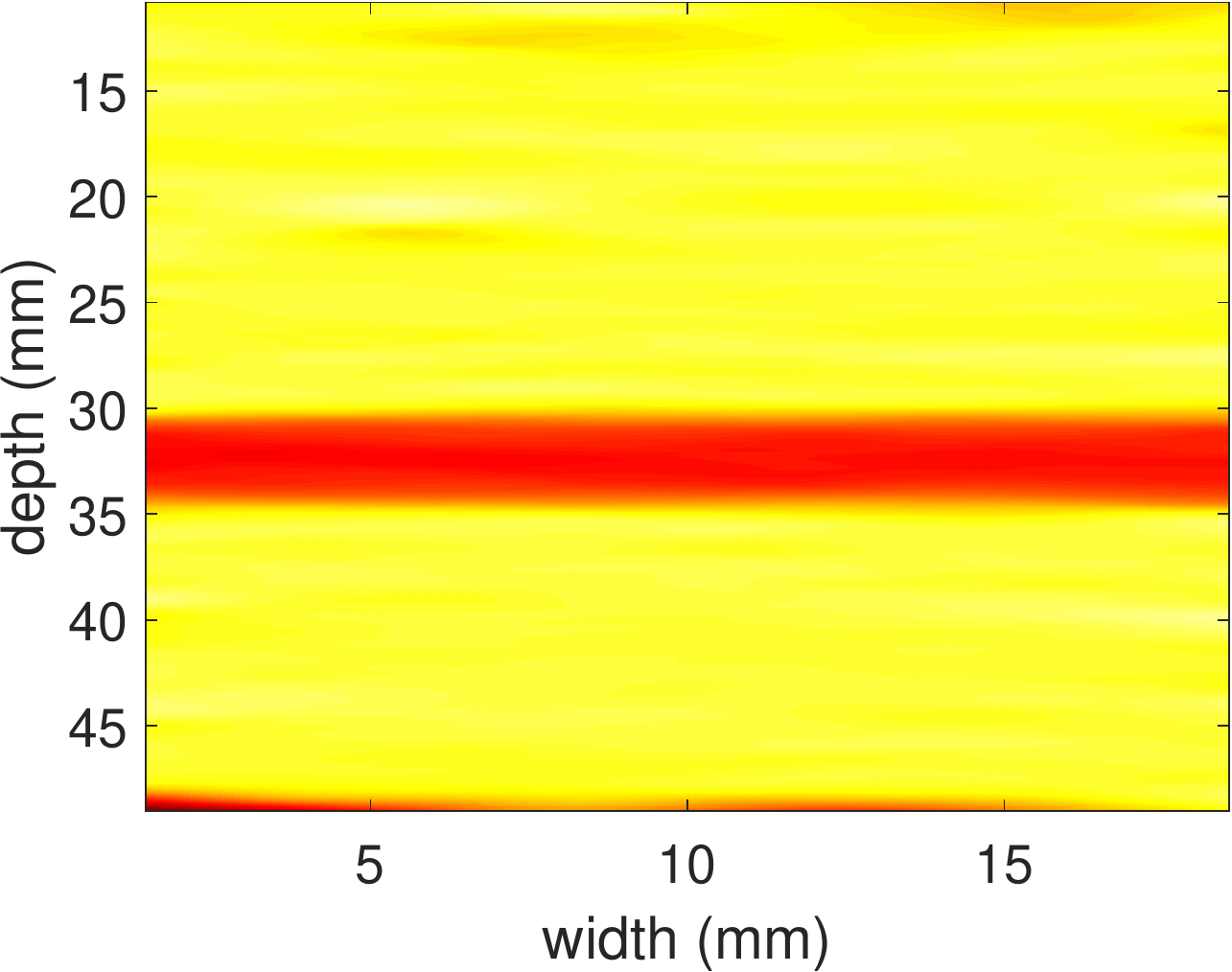} }}%
	\subfigure[$L1$-SOUL]{{\includegraphics[width=0.2\textwidth,height=0.31\textwidth]{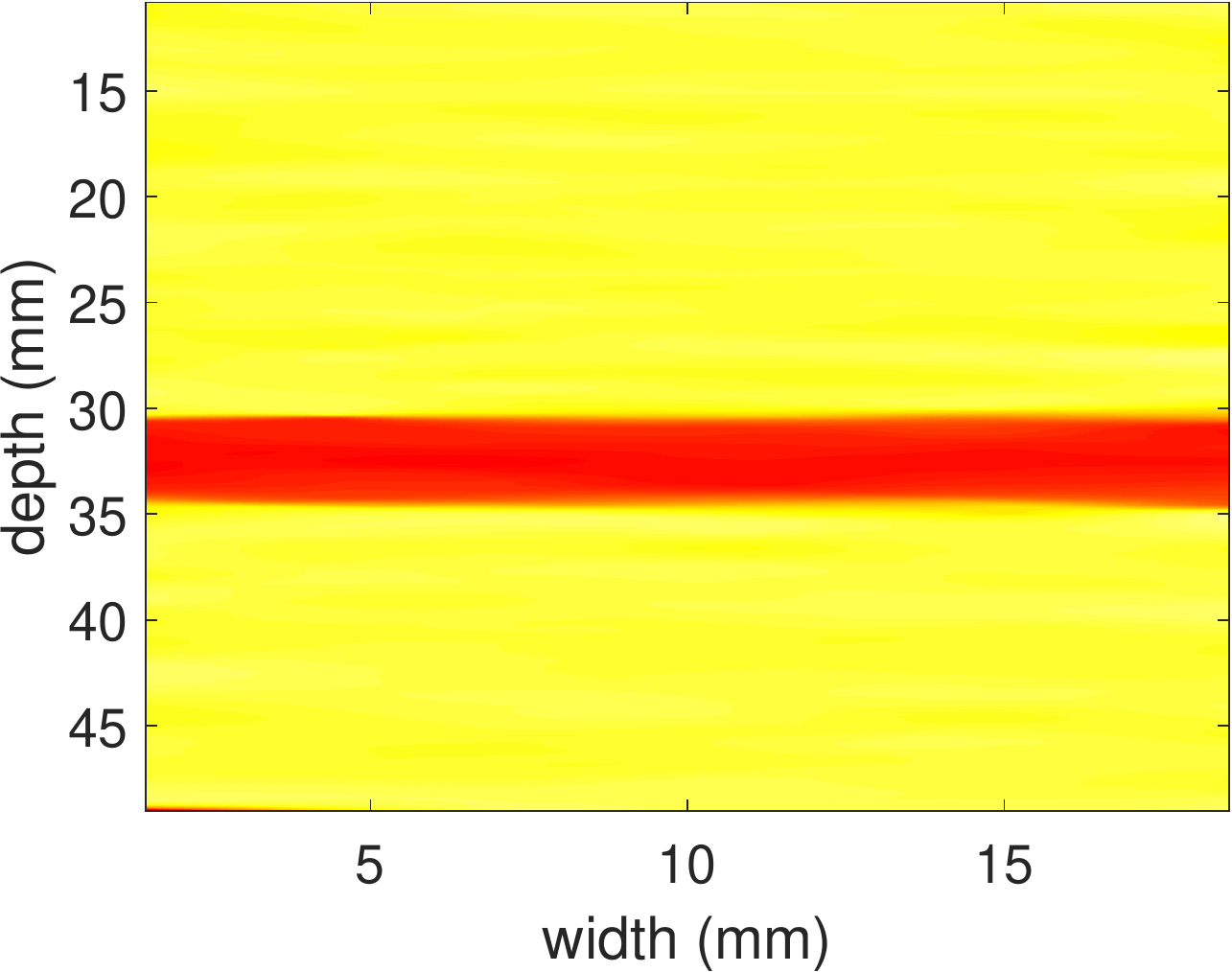} }}
	\subfigure[Strain, four-layer dataset]{{\includegraphics[width=.48\textwidth]{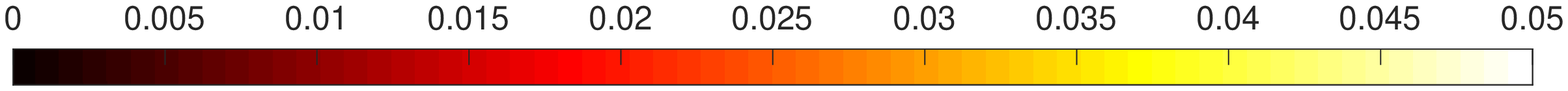}}}
	\subfigure[Strain, thin-layer dataset]{{\includegraphics[width=.48\textwidth]{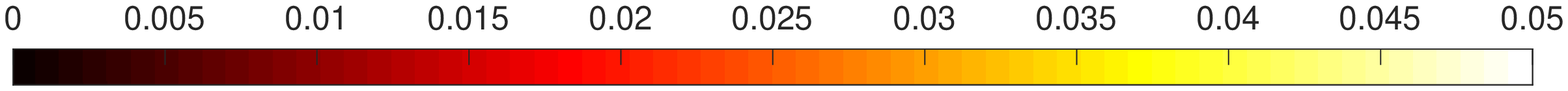}}}
	\caption{Axial strain images obtained from the simulated phantoms with four and thin layers. Rows 1 and 2 correspond to four- and thin-layers, respectively. Columns 1 to 5 correspond to the ground truth and axial strain maps estimated by GLUE, OVERWIND, SOUL, and $L1$-SOUL, respectively. The kernel length is set to 3 samples in all strain images for differentiating the displacement images. The vertical green bars in column 1 indicate the locations of ESFs shown in Fig.~\ref{esf}.}
	\label{layer}
\end{figure*}

\begin{figure}
	\centering
	\subfigure[Four-layer dataset]{{\includegraphics[width=.25\textwidth]{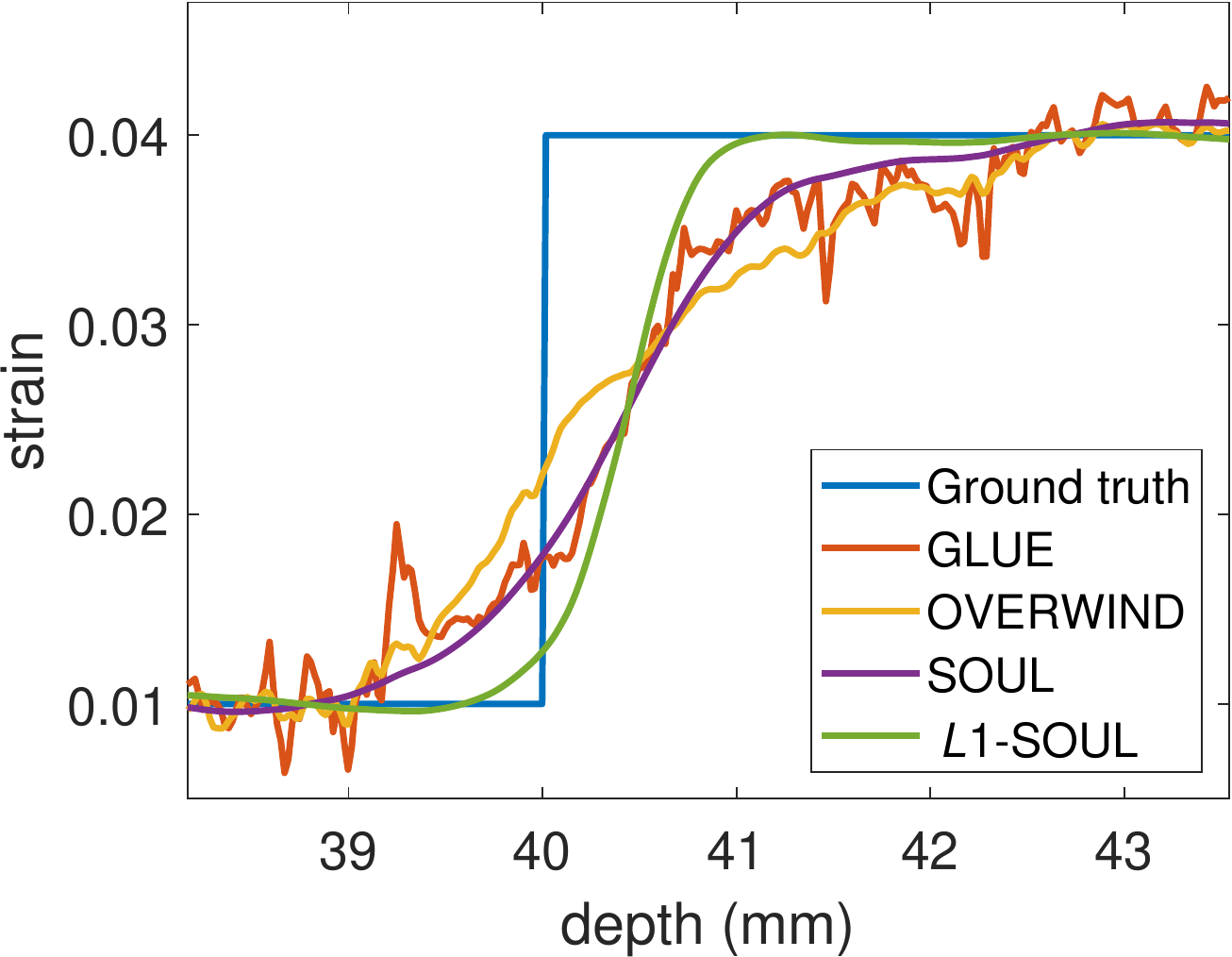}}}%
	\subfigure[Thin-layer dataset]{{\includegraphics[width=.25\textwidth]{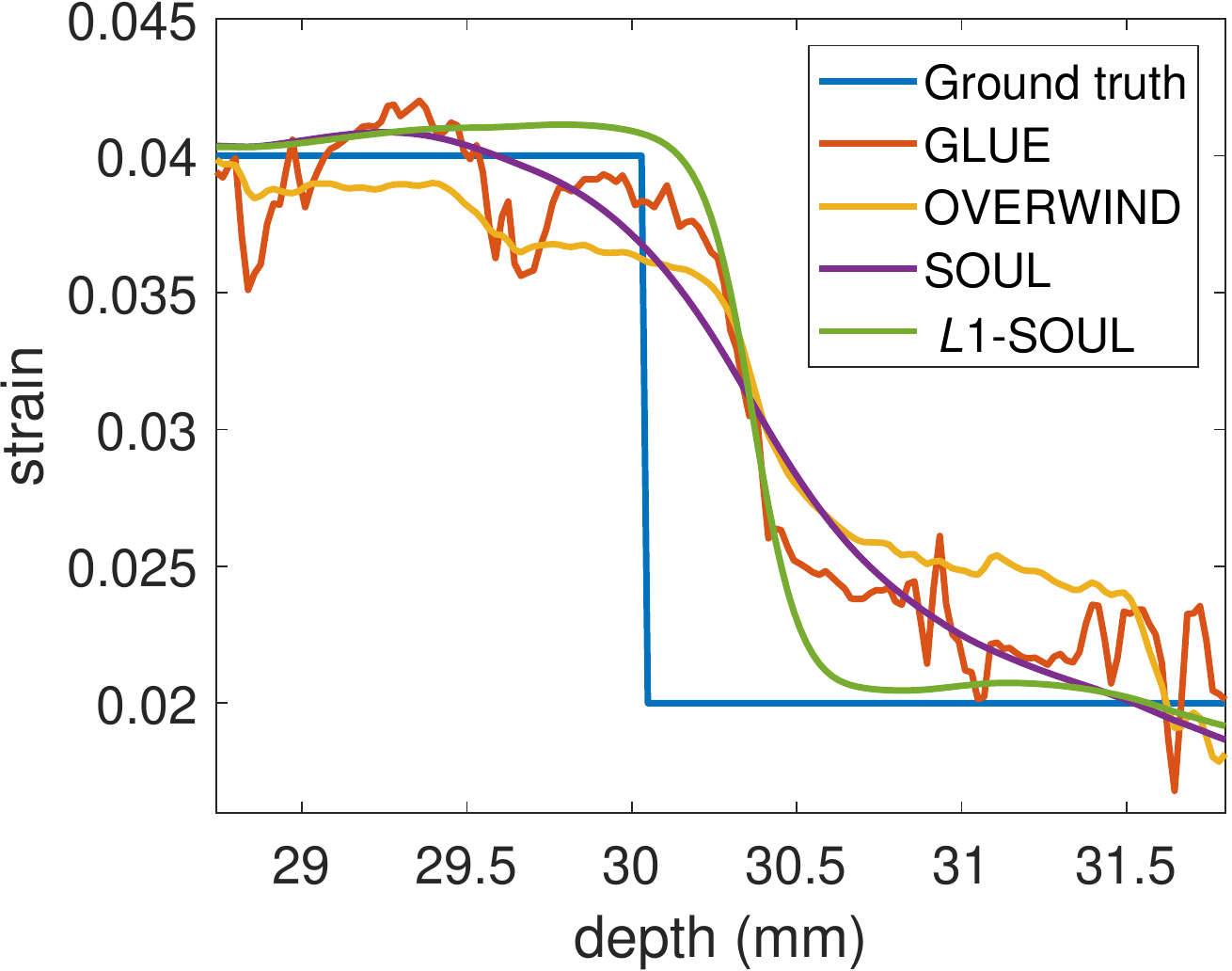}}}
	\caption{ESFs for the four- and thin-layer simulated datasets over the vertical lines shown in Figs.~\ref{layer}(a) and \ref{layer}(f), respectively. Columns 1 and 2 correspond to four- and thin-layer phantoms, respectively.}
	\label{esf}
\end{figure}

\begin{table}[tb]  
	\centering
	\caption{Edge-resolution (mm) for the four- and thin-layer simulated phantoms.} 
	\label{table_edge}
	\begin{tabular}{c c c c c c c} 
		\hline
		$ $  $ $&  Four-layer phantom & Thin-layer phantom\\
		\hline
		GLUE & 3.71 &  1.06\\
		OVERWIND & 3.71 &  1.06\\
		SOUL & 3.71 &  1.60\\
		$L1$-SOUL & \textbf{1.64}  & \textbf{0.73}\\
		\hline
	\end{tabular}
\end{table}

\begin{table}[tb]  
	\centering
	\caption{Mean SSIM values for the four- and thin-layer simulated phantoms.} 
	\label{table_mssim}
	\begin{tabular}{c c c c c c c} 
		\hline
		$ $  $ $&  Four-layer phantom & Thin-layer phantom\\
		\hline
		GLUE & 0.12 &  0.06\\
		OVERWIND & 0.51 &  0.47\\
		SOUL & 0.87 &  0.78\\
		$L1$-SOUL & \textbf{0.91}  & \textbf{0.87}\\
		\hline
	\end{tabular}
\end{table}

\begin{table}[tb]  
	\centering
	\caption{$L2$-errors for the four- and thin-layer simulated phantoms.} 
	\label{table_l2error}
	\begin{tabular}{c c c c c c c} 
		\hline
		$ $  $ $&  Four-layer phantom & Thin-layer phantom\\
		\hline
		GLUE & 2.32 &  2.36\\
		OVERWIND & 1.93 &  1.49\\
		SOUL & 1.86 &  1.79\\
		$L1$-SOUL & \textbf{1.85}  & \textbf{1.43}\\
		\hline
	\end{tabular}
\end{table}

\begin{figure*}
	\centering
	\subfigure[Ground truth]{{\includegraphics[width=.2\textwidth, height=.23\textwidth]{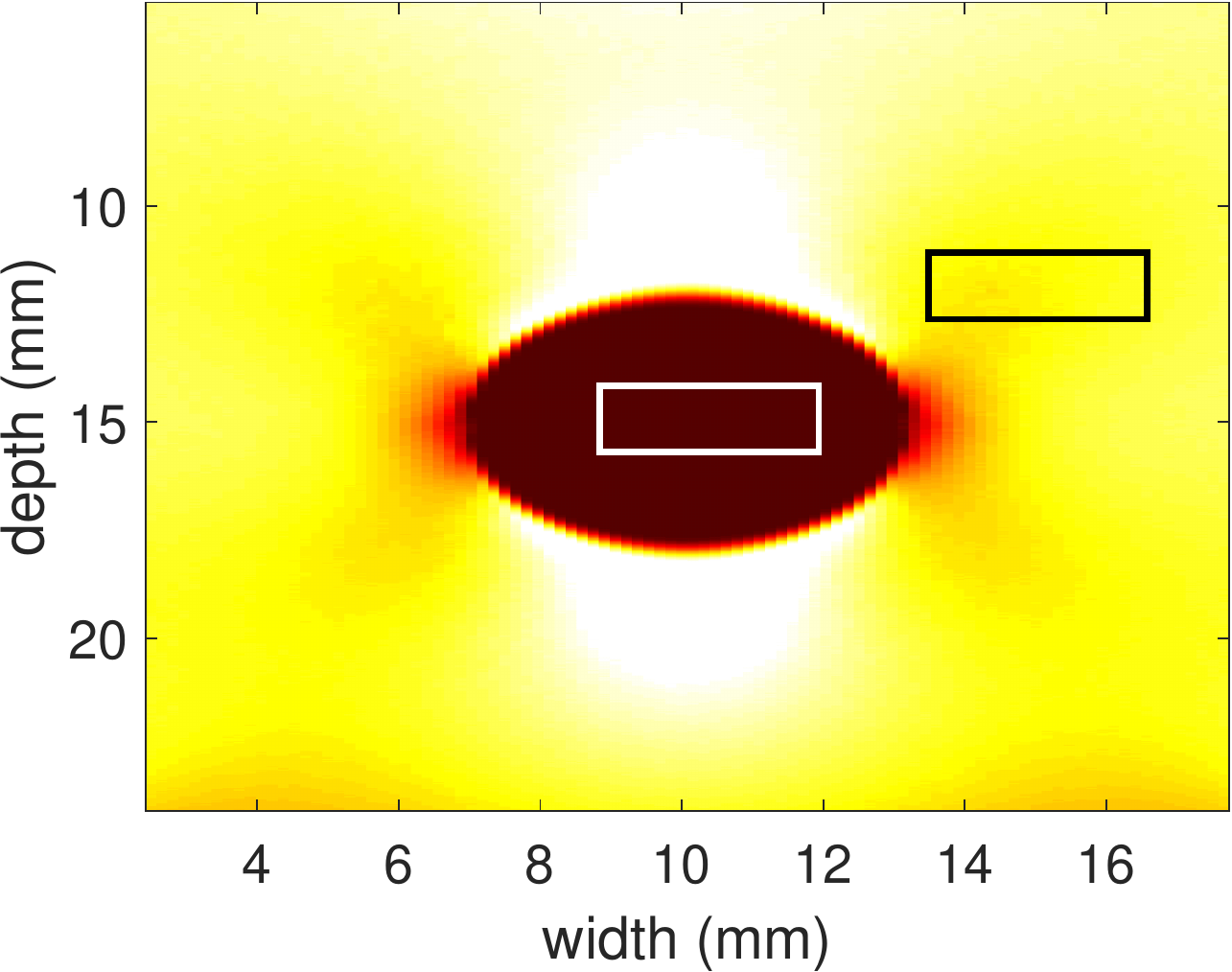}}}%
	\subfigure[GLUE]{{\includegraphics[width=.2\textwidth, height=.23\textwidth]{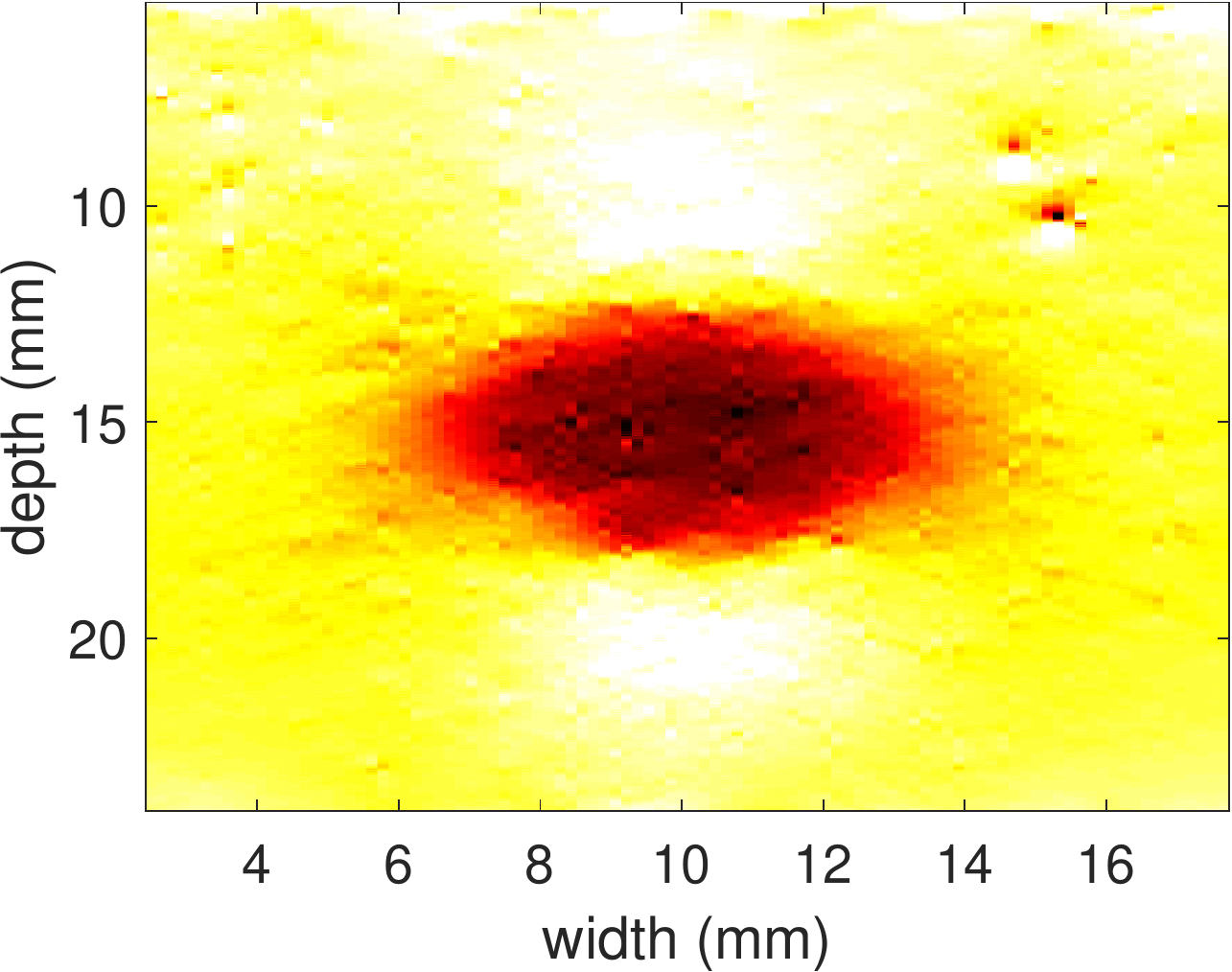}}}%
	\subfigure[OVERWIND]{{\includegraphics[width=.2\textwidth, height=.23\textwidth]{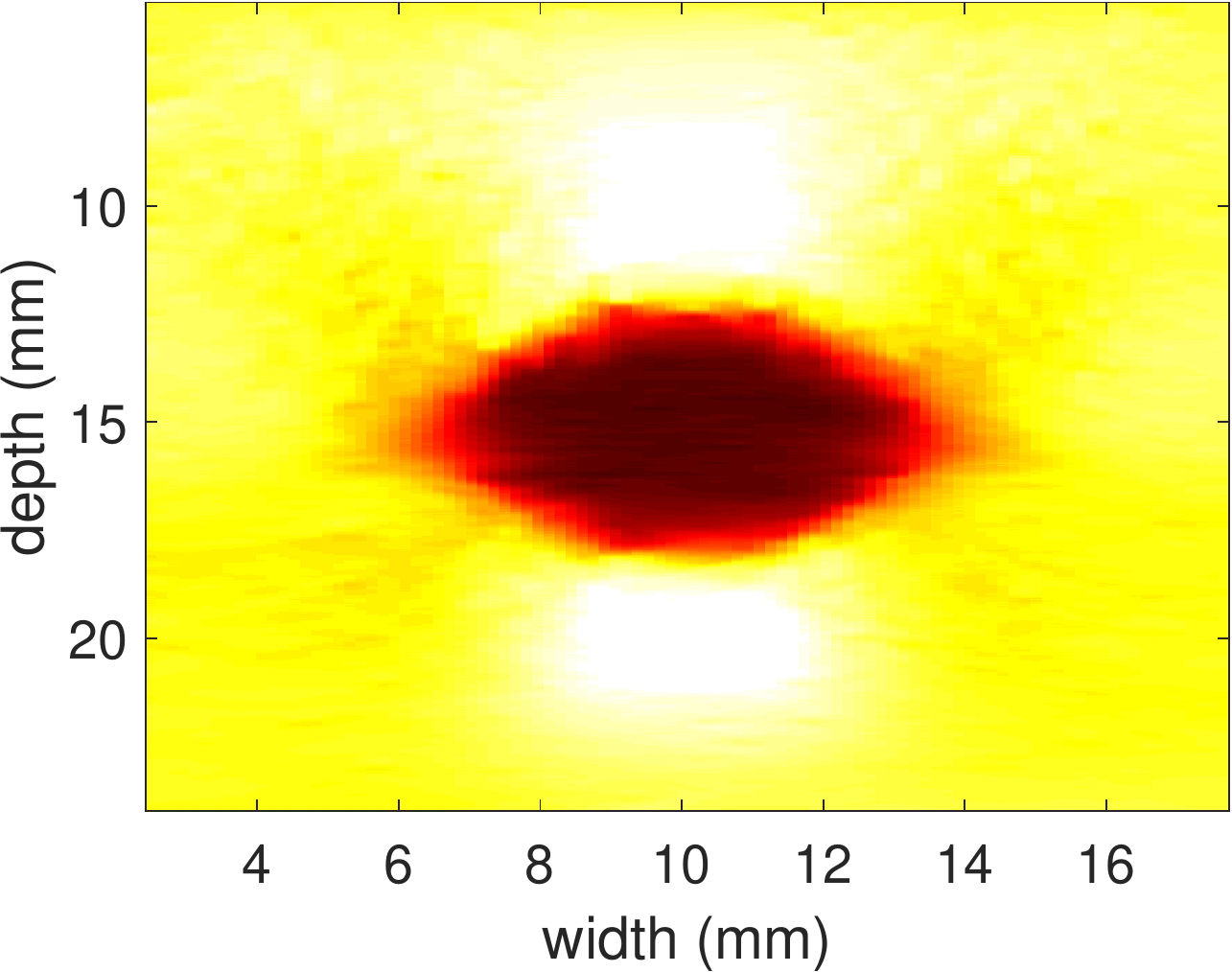} }}%
	\subfigure[SOUL]{{\includegraphics[width=.2\textwidth, height=.23\textwidth]{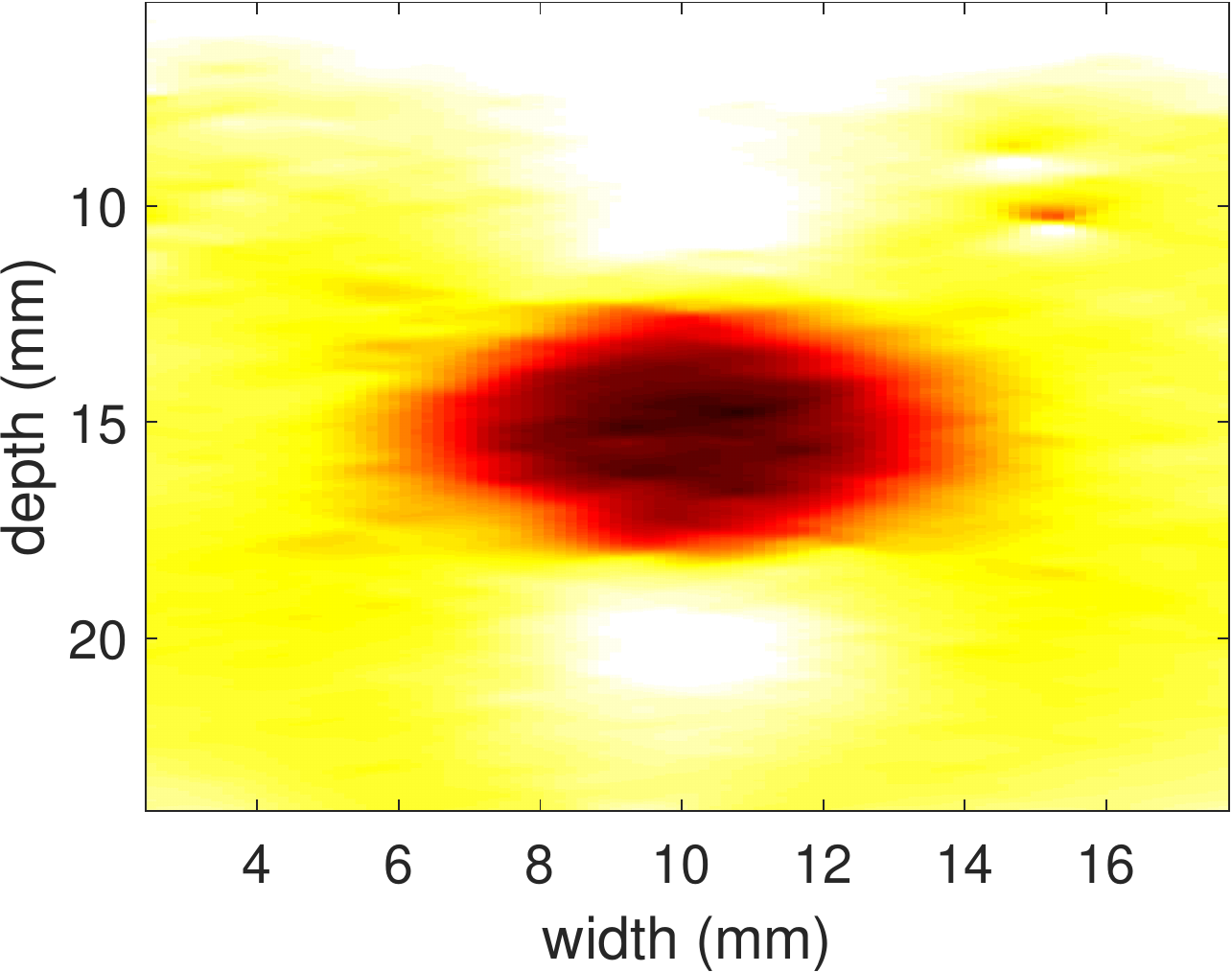} }}%
	\subfigure[$L1$-SOUL]{{\includegraphics[width=.2\textwidth, height=.23\textwidth]{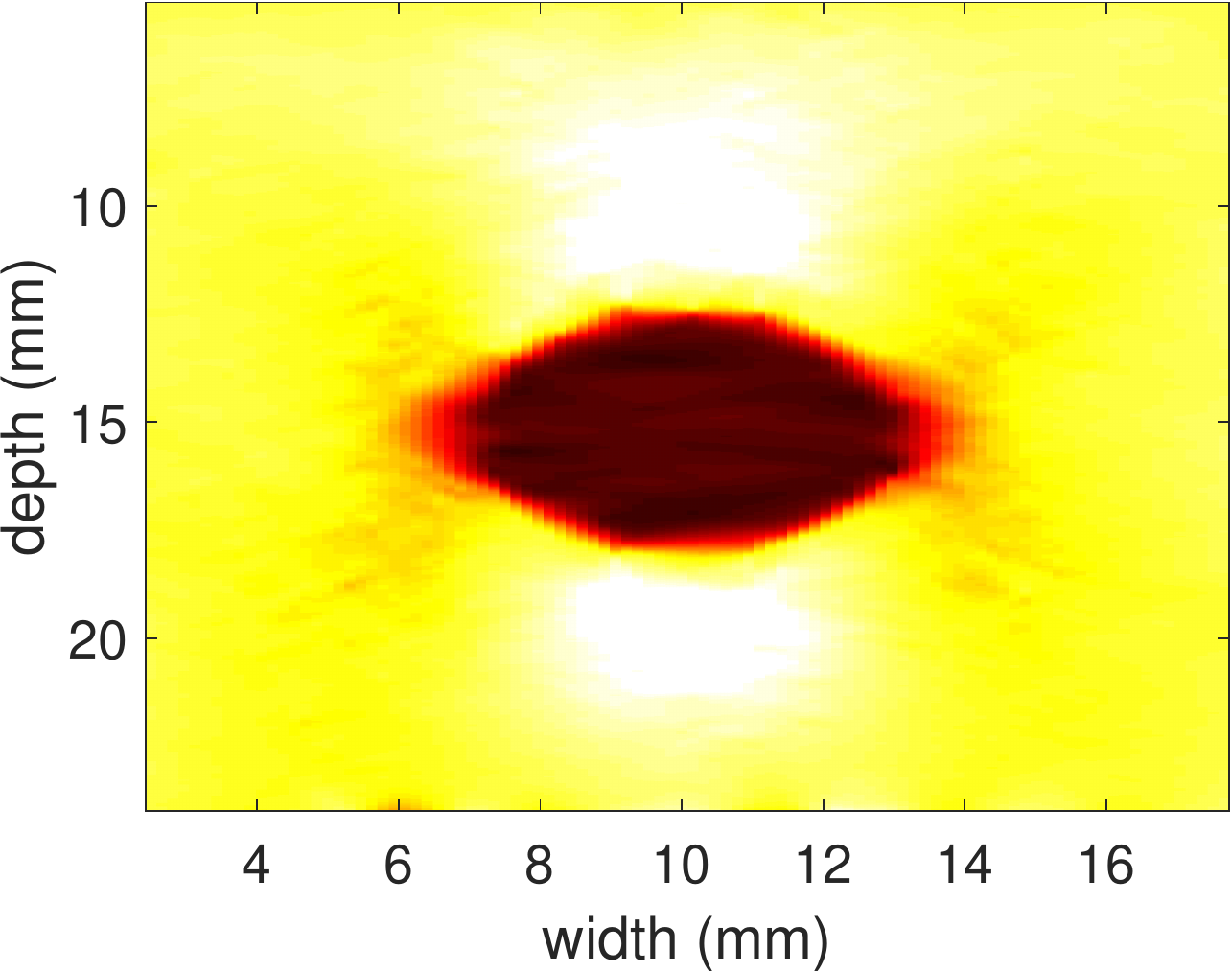} }}
	\subfigure[Strain]{{\includegraphics[width=.6\textwidth]{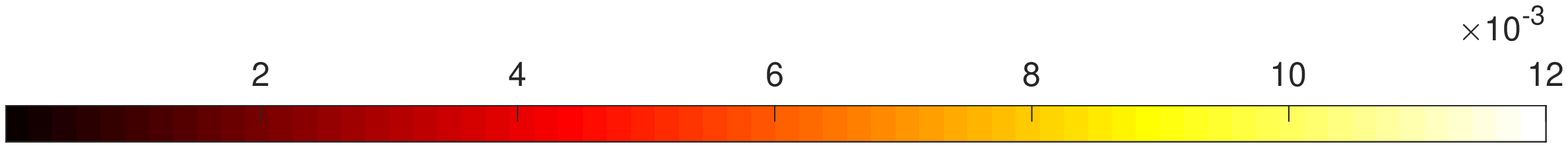}}}
	\caption{Axial strain images obtained from the simulated phantom with hard inclusion. Columns 1 to 5 show the strain images obtained using FEM, GLUE, OVERWIND, SOUL, and $L1$-SOUL, respectively. The kernel length is set to 3 samples in all strain images for differentiating the displacement images.}
	\label{hard_simu}
\end{figure*}

\begin{table}[tb]  
	\centering
	\caption{SNR, CNR, and SR corresponding to the simulated dataset with stiff inclusion. CNR and SR are obtained using the white target and black background windows shown in Fig.~\ref{hard_simu}(a), whereas SNR is calculated using the background window only.} 
	\label{table_hard}
	\begin{tabular}{c c c c c c c} 
		\hline
		$ $  $ $&    SNR & CNR & SR\\
		\hline
		GLUE &  19.72 &  14.97 & 0.20\\
		OVERWIND & 27.65 &  31.31 & \textbf{0.16}\\
		SOUL & 26.88 &  23.76 & 0.17\\
		$L1$-SOUL & \textbf{37.29}  & \textbf{39.87} & \textbf{0.16}\\
		\hline
	\end{tabular}
\end{table}

Once the refinement displacement fields are estimated, they are added to the initial ones to find the final axial and lateral displacement maps. The axial deformation field is spatially differentiated using a least-squares technique to obtain the axial strain map.

\begin{figure*}
	\centering
	\subfigure[B-mode]{{\includegraphics[width=.2\textwidth, height=.22\textwidth]{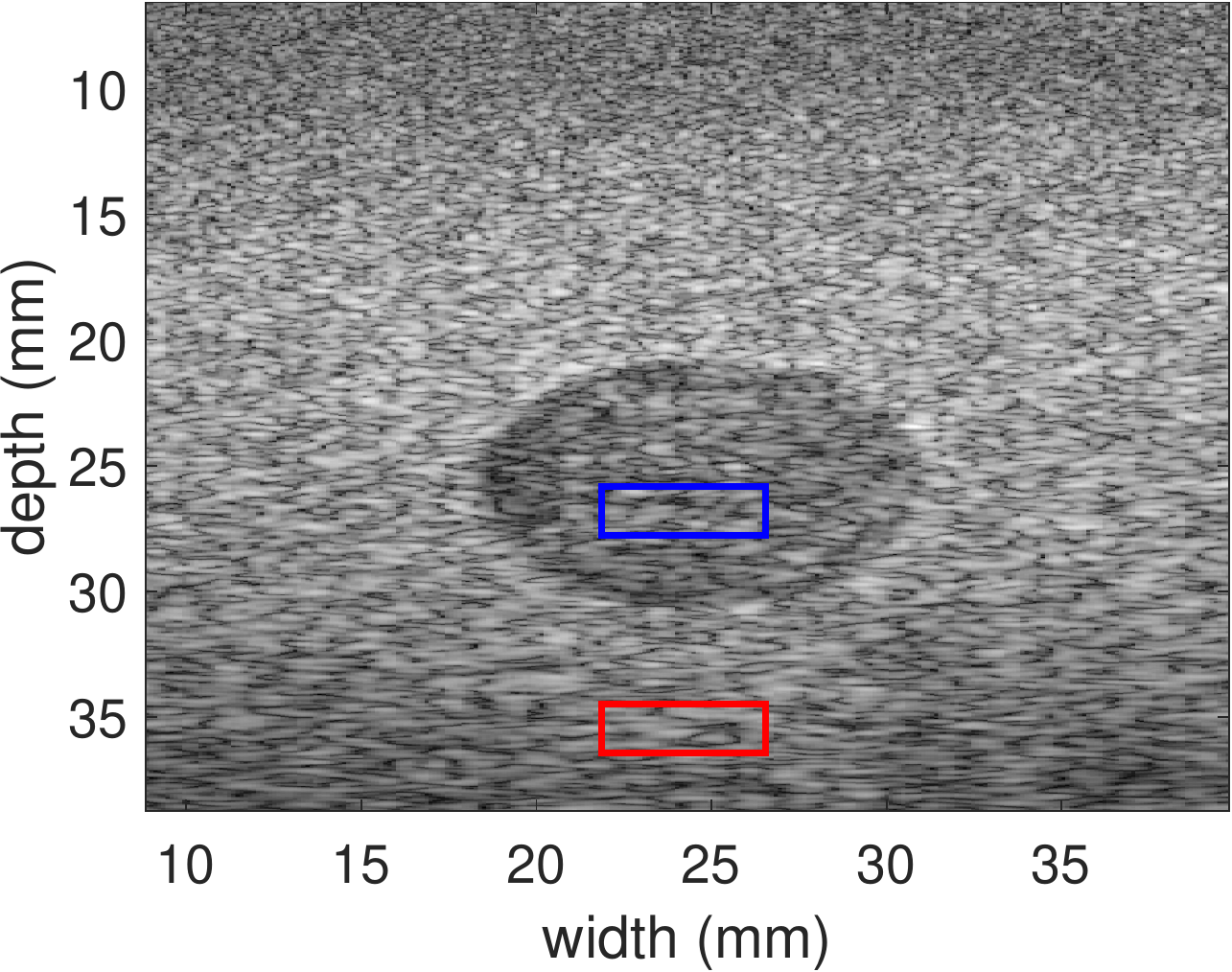}}}%
	\subfigure[GLUE]{{\includegraphics[width=.2\textwidth, height=.22\textwidth]{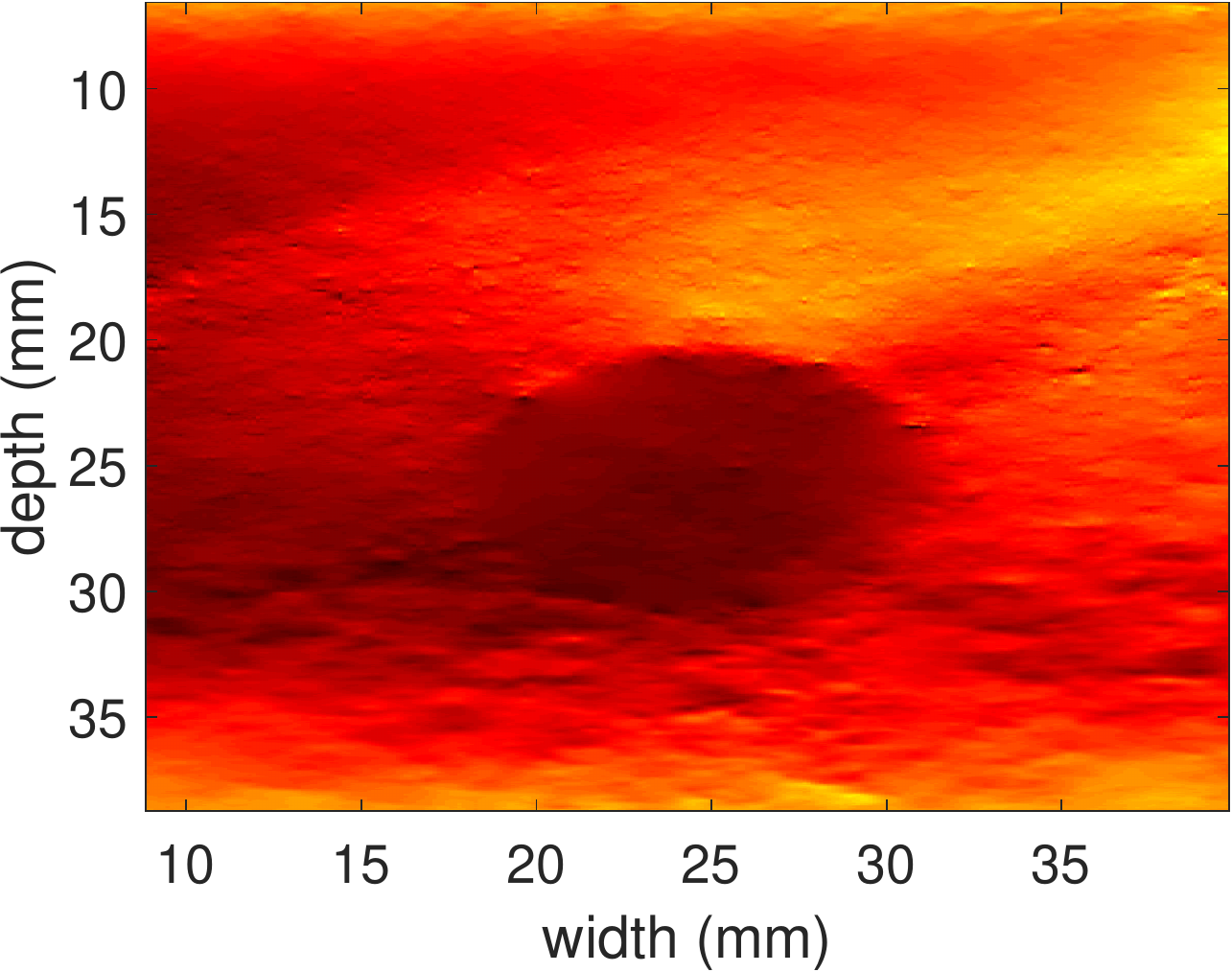}}}%
	\subfigure[OVERWIND]{{\includegraphics[width=.2\textwidth, height=.22\textwidth]{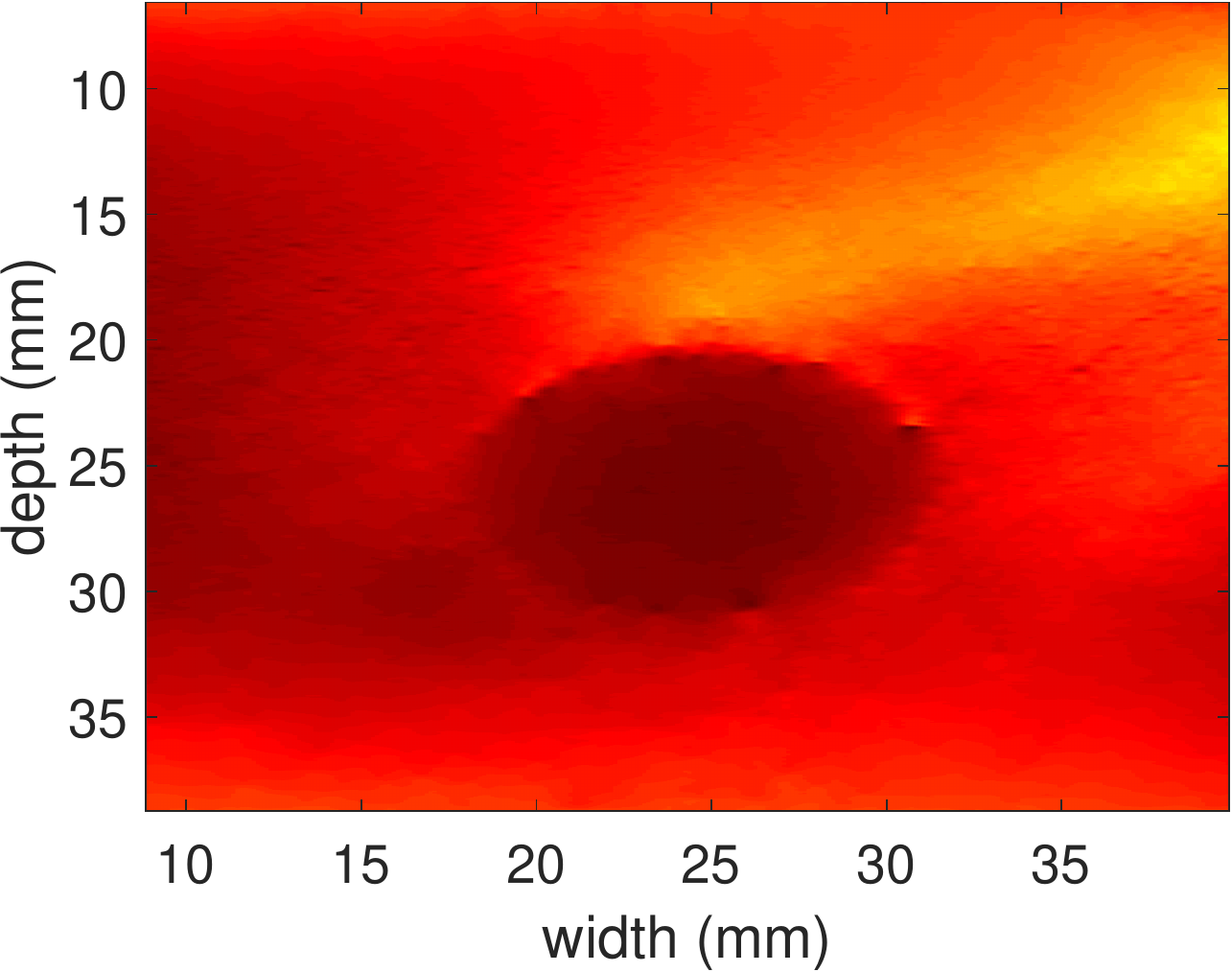} }}%
	\subfigure[SOUL]{{\includegraphics[width=.2\textwidth, height=.22\textwidth]{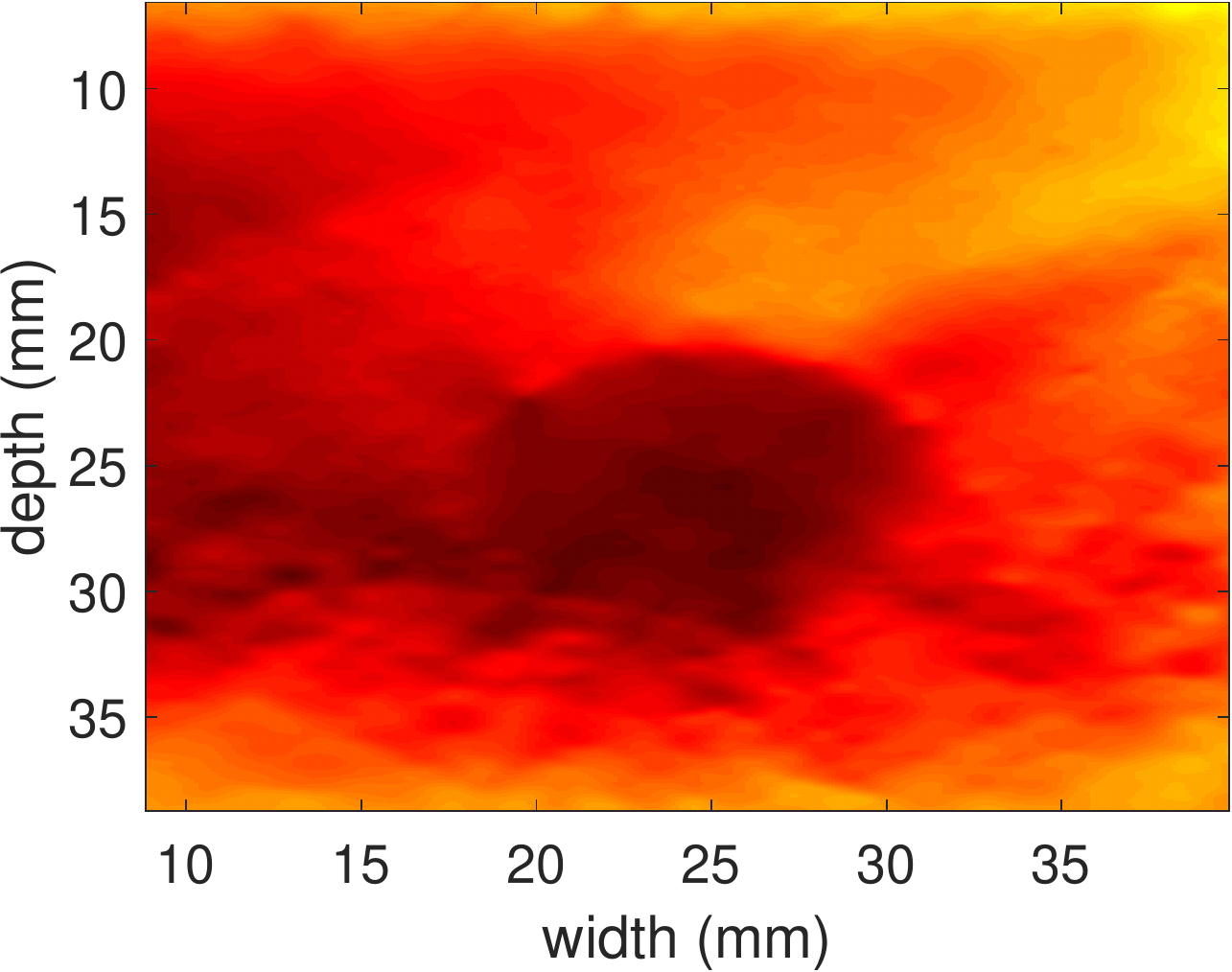} }}%
	\subfigure[$L1$-SOUL]{{\includegraphics[width=.2\textwidth, height=.22\textwidth]{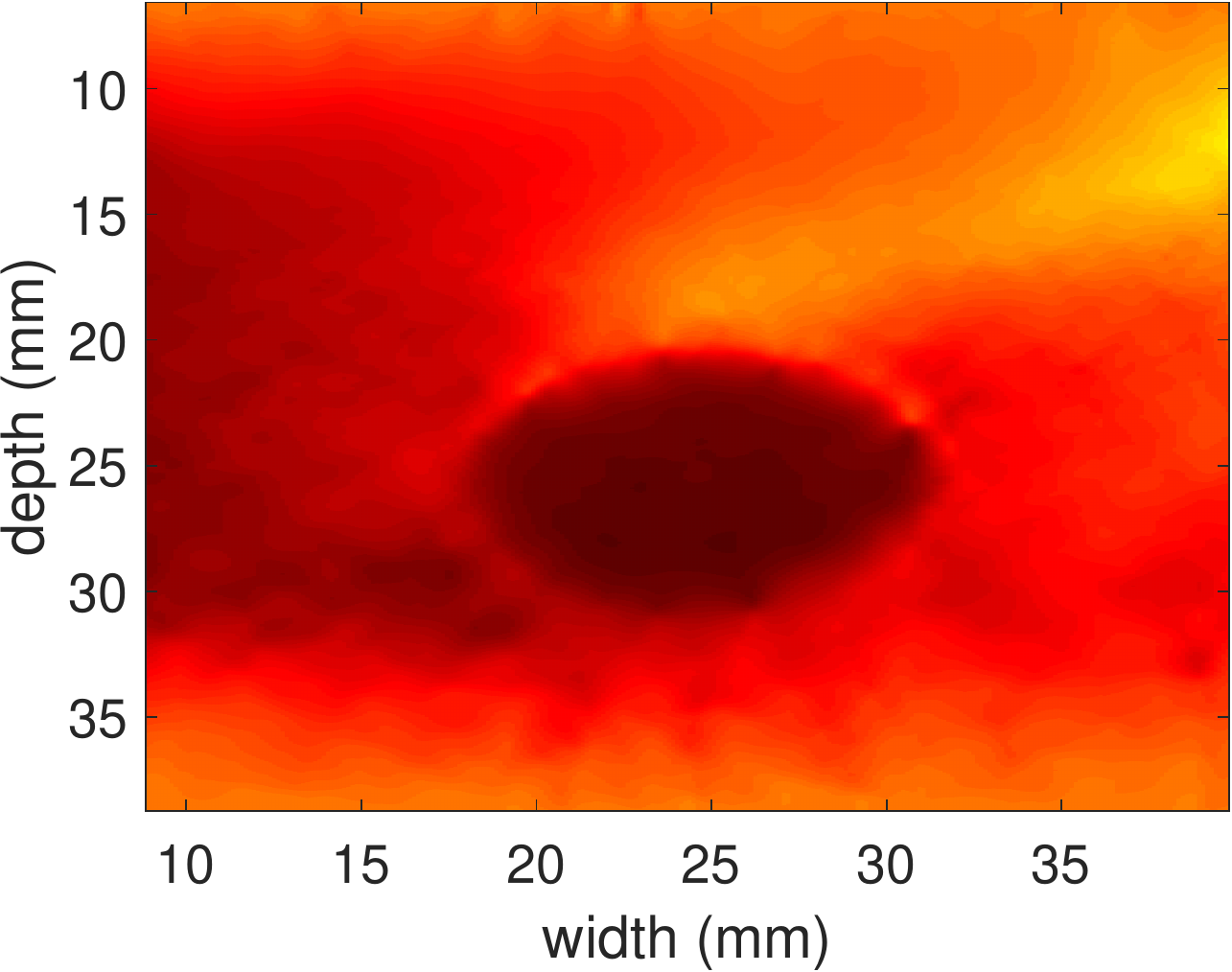} }}
	\subfigure[Strain]{{\includegraphics[width=.6\textwidth]{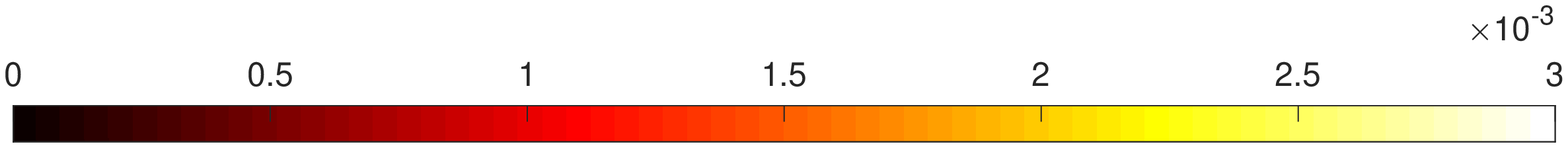}}}
	\caption{Axial strain results corresponding to the breast phantom. Columns 1 to 5 depict the B-mode image and the strain images generated by GLUE, OVERWIND, SOUL, and $L1$-SOUL, respectively. The kernel length is set to 3 samples in all strain images for differentiating the displacement images.}
	\label{perform_phan}
\end{figure*}

\begin{table}[tb]  
	\centering
	\caption{SNR, CNR, and SR for the breast phantom dataset. CNR and SR are calculated using the blue target and red background windows shown in Fig.~\ref{perform_phan}(a). SNR is obtained from the background window only.} 
	\label{table_phan}
	\begin{tabular}{c c c c c c c} 
		\hline
		$ $  $ $&    SNR & CNR & SR\\
		\hline
		GLUE &  7.27 &  6.30 & 0.38\\
		OVERWIND & \textbf{15.76} &  12.00 & 0.45\\
		SOUL & 11.06 &  9.87 & 0.35\\
		$L1$-SOUL & 14.83  & \textbf{14.49} & \textbf{0.30}\\
		\hline
	\end{tabular}
\end{table} 

\subsection{Ultrasound Simulation and Data Acquisition}
Three validation phantoms were simulated where the first phantom contains four tissue layers with inter-varying elasticities, and the second one contains a thin layer with higher elasticity than the surrounding tissue. The third phantom contains a cylindrical inclusion with a higher stiffness than the background. Real experimental datasets include data from a breast elastography phantom and three liver cancer patients.

\subsubsection{Four-layer Phantom}
A phantom consisting of four tissue layers with elastic moduli of $20$~kPa, $40$~kPa, $80$~kPa, and $20$~kPa was compressed by $4\%$ using closed-form equations delineated in the Supplementary Material of~\cite{guest}. Pre- and post-compressed RF frames were simulated using Field II ~\cite{field2} setting the center and sampling frequencies to $7.27$~MHz and $40$~MHz, respectively. The number of active elements, transducer height, and width were set to $64$, $5$~mm, and $0.2$~mm, respectively. To emulate the real ultrasound data collection scenario, random Gaussian noise with $20$~dB peak signal-to-noise ratio (PSNR) was added to the simulated RF data.

\subsubsection{Thin-layer Phantom}    
A homogeneous phantom containing a thin stiff layer of height 4~mm placed at 30~mm depth was compressed by $4\%$ using closed-form equations. The target and background elastic moduli were set to $40$~kPa and $20$~kPa, respectively. RF data were simulated using Field II with the same parameter setting as the four-layer phantom. RF data were corrupted with random Gaussian noise of 20 dB PSNR to imitate the real acquisition environment.

\subsubsection{Hard-inclusion Phantom}
A homogeneous phantom with a hard cylindrical inclusion was simulated where the background and inclusion elasticity moduli were set to $4$~kPa and $40$~kPa, respectively. The phantom was compressed by $1\%$ of its height using ABAQUS, a finite element (FEM) package (Providence, RI). The center frequency and temporal sampling rates were set to $7.27$~MHz and $100$~MHz during RF data simulation with Field II.

\subsubsection{Experimental Breast Phantom}
Free-hand compression was applied to an experimental breast elastography phantom (Model 059, CIRS: Tissue Simulation \& Phantom Technology, Norfolk, VA) with a background elasticity modulus of $20 \pm 5$~kPa. The elasticity modulus of the inclusion was at least twice as large as that of the background. RF datasets were acquired from the phantom during its deformation using an Alpinion E-Cube R12 research ultrasound machine with an L3-12H linear array probe setting the center and sampling frequencies to $10$~MHz and $40$~MHz, respectively.

\subsubsection{\textit{In vivo} Liver Cancer Datasets}
The \textit{in vivo} datasets were acquired from three liver cancer patients before open surgical RF thermal ablation at the Johns Hopkins Hospital (Baltimore, MD). Free-hand compressions were made by pushing a VF 10-5 linear array probe against the liver. RF data were acquired using an Antares Siemens research ultrasound system where the center frequency and the sampling rate were set to $6.67$~MHz and $40$~MHz, respectively. The Institutional Review Board approved all procedures related to this experiment, and all three patients provided written consent for this study. Interested individuals are suggested to visit~\cite{DPAM} for further details of this data.

\subsection{Quantitative Metrics}
We demonstrate the potency of $L1$-SOUL by comparing it with GLUE~\cite{glue}, OVERWIND~\cite{overwind}, and SOUL~\cite{soul}, three recently published TDE algorithms. In addition to visual assessment, we analyze the strain image quality using edge-resolution, mean Structural SIMilarity index (Mean SSIM)~\cite{Zhou_2004}, $L2$-error, edge-spread function (ESF), signal-to-noise ratio (SNR), contrast-to-noise ratio (CNR), and strain ratio (SR). We define edge-resolution as the depth of strain transition between different elastic regions. Mean SSIM is defined as~\cite{Zhou_2004}:

\begin{equation}
\textrm{Mean SSIM}(S_{g}, S_{e})=\frac{1}{W}\sum\limits_{w=1}^{W}\textrm{SSIM}(s_{g,w},s_{e,w}) 
\end{equation}

\noindent
where $S_{g}$ and $S_{e}$ denote the ground truth and estimated strain images, respectively. $s_{g,w}$ and $s_{e,w}$ stand for the $w$-th local windows on the ground truth and estimated strain images, respectively. $W$ indicates the total number of windows considered to calculate the SSIM map~\cite{Zhou_2004}. $L2$-error is defined as:

	\begin{equation}
	L2-\textrm{error}=\sqrt{\sum\limits_{j=1}^{n}\sum\limits_{i=1}^{m}[S_{e}(i, j) - S_{g}(i, j)]^{2}} 
	\end{equation}

Elastographic SNR, CNR, and SR are defined as follows.

\begin{equation}
\textrm{CNR}=\frac{C}{N}=\sqrt{\frac{2(\bar{{s_{b}}}-\bar{{s_{t}}})^2}{{\sigma_b}^2+{\sigma_t}^2}}, \textrm{SNR}=\frac{\bar{s_{b}}}{\sigma_{b}},     \textrm{SR}=\frac{\bar{{s_{t}}}}{\bar{{s_{b}}}}
\end{equation}            

\noindent
where $\bar{s_{b}}$, $\sigma_b$ and $\bar{s_{t}}$, $\sigma_t$ denote the mean and standard deviation of the background and target strain windows, respectively. The parameters were tuned to produce optimal strain results for all four techniques.

\section{Results}
The tunable parameters associated with all four techniques were carefully chosen to ensure the optimality of the strain images’ visual appearance. All algorithms’ parameter sets for every validation experiment were optimized individually to accomplish a fair comparison. All of these values are reported in the Supplementary Material.

It is common to use a large window, often of size 51 or more RF samples, to differentiate the displacement field and obtain the strain field since differentiation amplifies noise. Although this process masks samples with incorrect displacement estimates, it further blurs the strain image. Therefore, we use the smallest possible window size of 3 to differentiate the displacement field in all results.

\subsection{Simulated Layer Phantoms}
The ground truth and the axial strain images for the four- and thin-layer simulated phantoms are shown in Fig.~\ref{layer}. GLUE obtains the noisiest strain image. OVERWIND exhibits undesired variability in the uniform stiff tissue region. Although SOUL yields better noise suppression than GLUE and OVERWIND, it suffers from edge-thickening. In addition, for the thin-layer phantom, both GLUE and SOUL fail in shallow and deep tissue regions. $L1$-SOUL yields sharp edges and smooth backgrounds, which is substantiated by the ESFs shown in Fig.~\ref{esf}. The edge-resolutions reported in Table~\ref{table_edge} agree with our visual judgement. Finally, the Mean SSIM (Table~\ref{table_mssim}) and $L2$-error values (Table~\ref{table_l2error}) indicate that the $L1$-SOUL strain shows the highest resemblance to the ground truth.

\begin{figure*}
	\centering
	\subfigure[B-mode]{{\includegraphics[width=.2\textwidth, height=.19\textwidth]{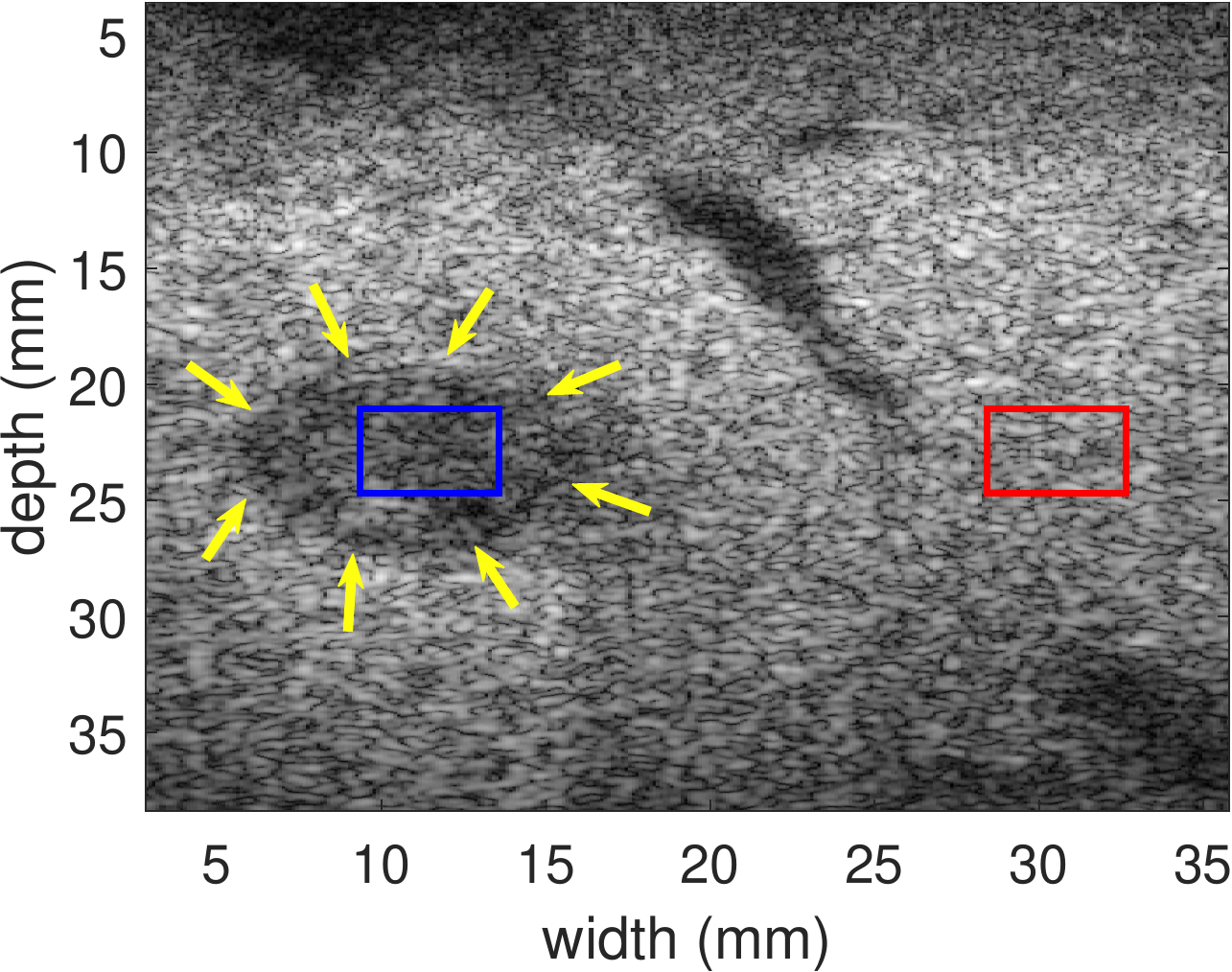}}}%
	\subfigure[GLUE]{{\includegraphics[width=.2\textwidth, height=.19\textwidth]{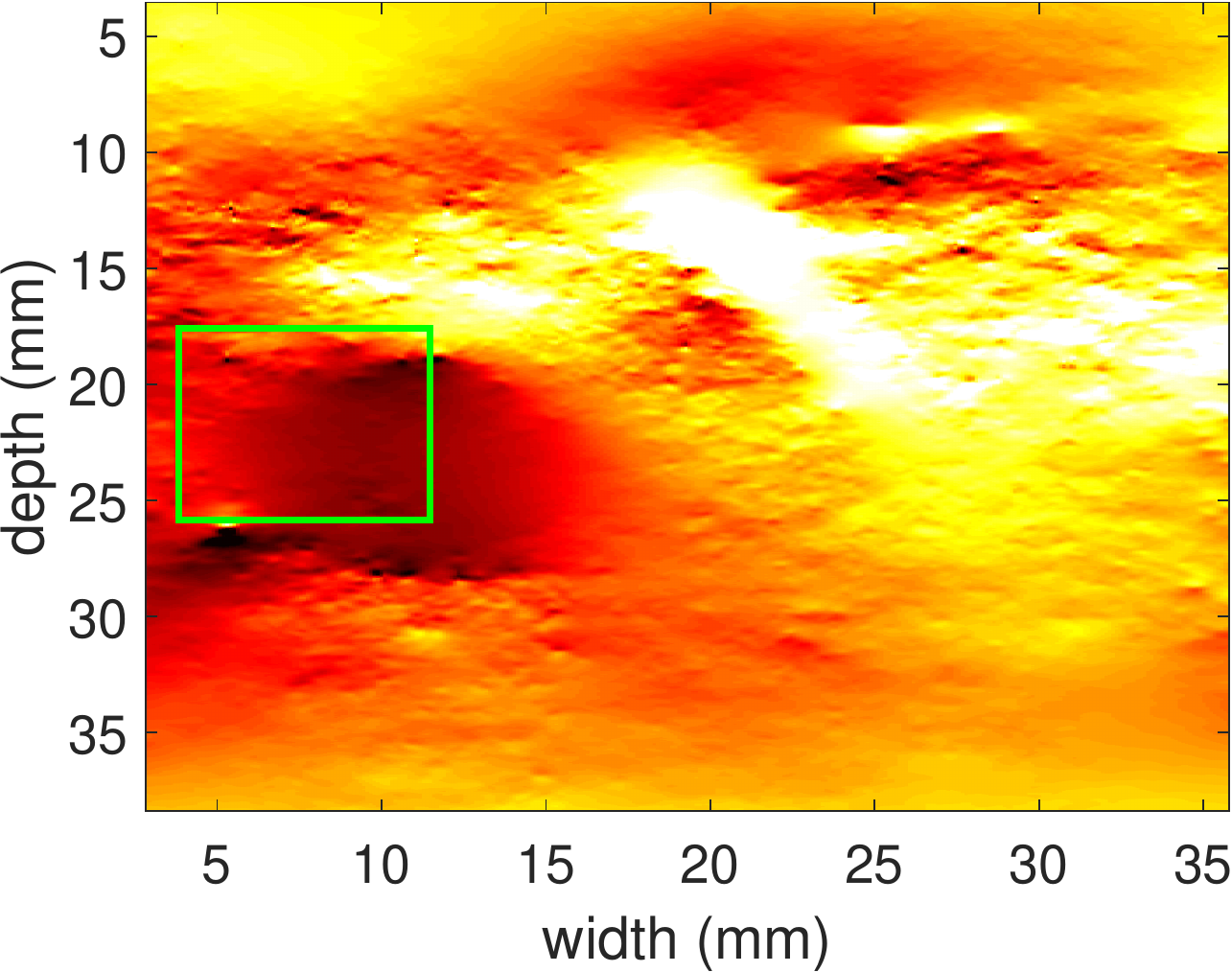}}}%
	\subfigure[OVERWIND]{{\includegraphics[width=.2\textwidth, height=.19\textwidth]{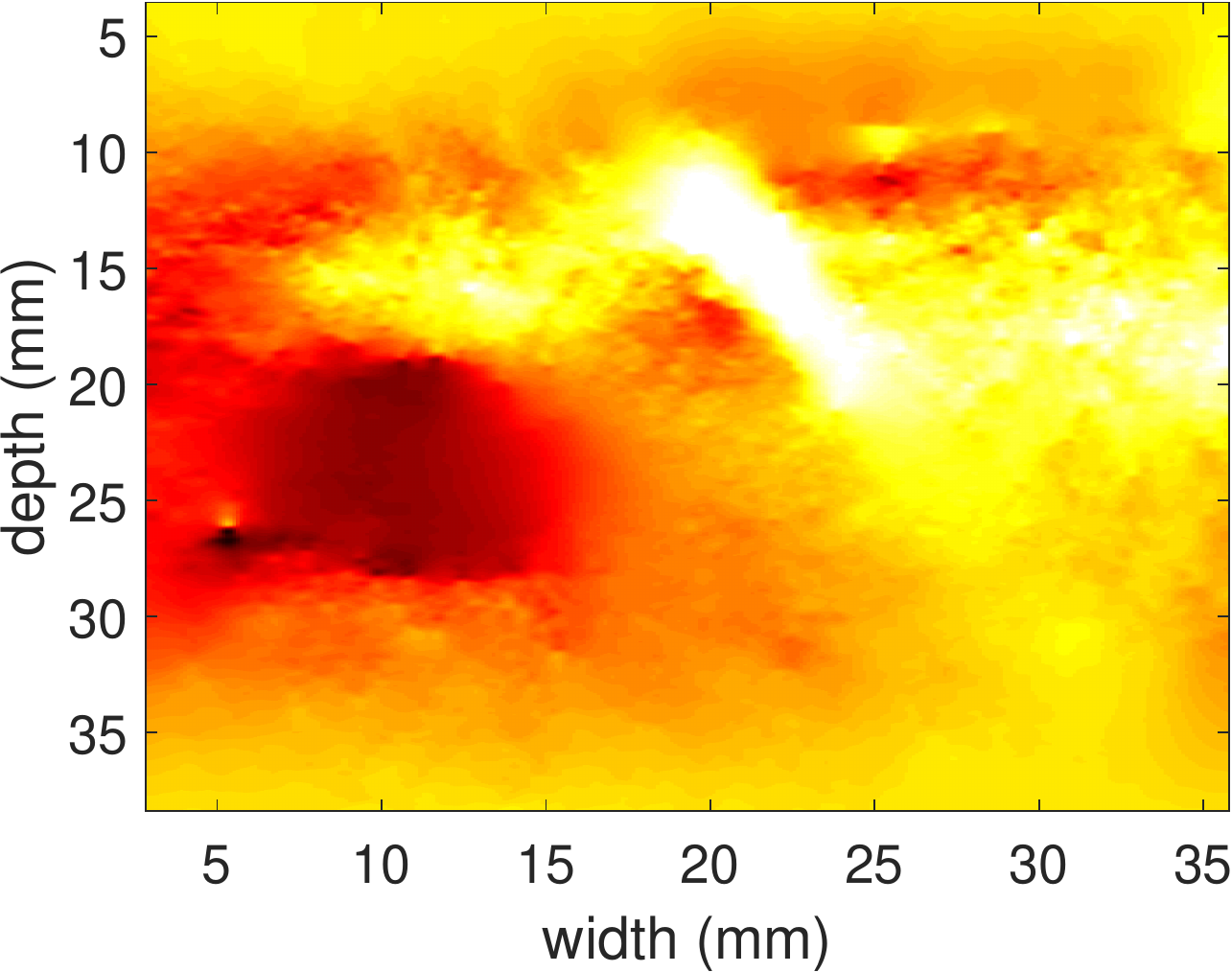} }}%
	\subfigure[SOUL]{{\includegraphics[width=.2\textwidth, height=.19\textwidth]{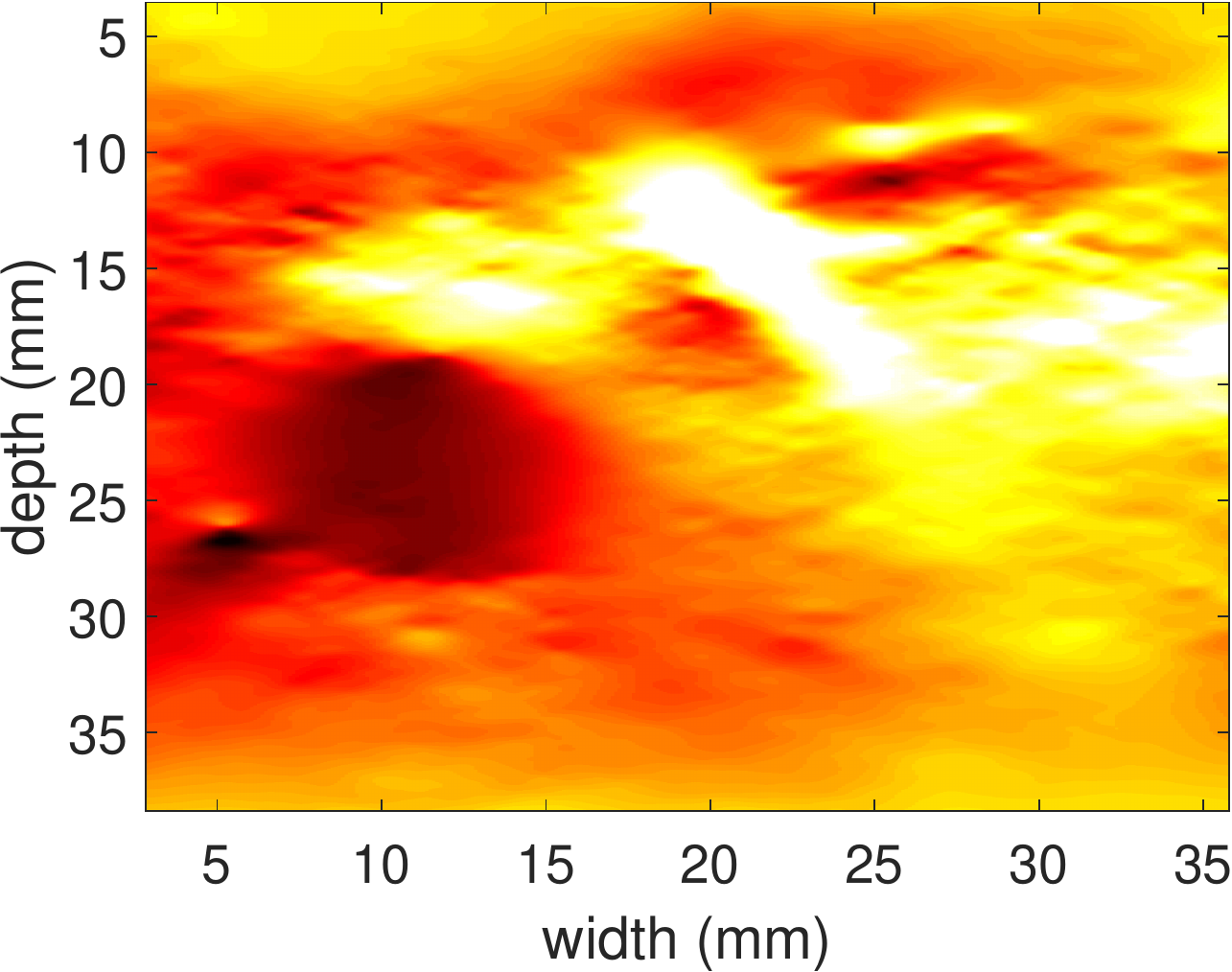} }}%
	\subfigure[$L1$-SOUL]{{\includegraphics[width=.2\textwidth, height=.19\textwidth]{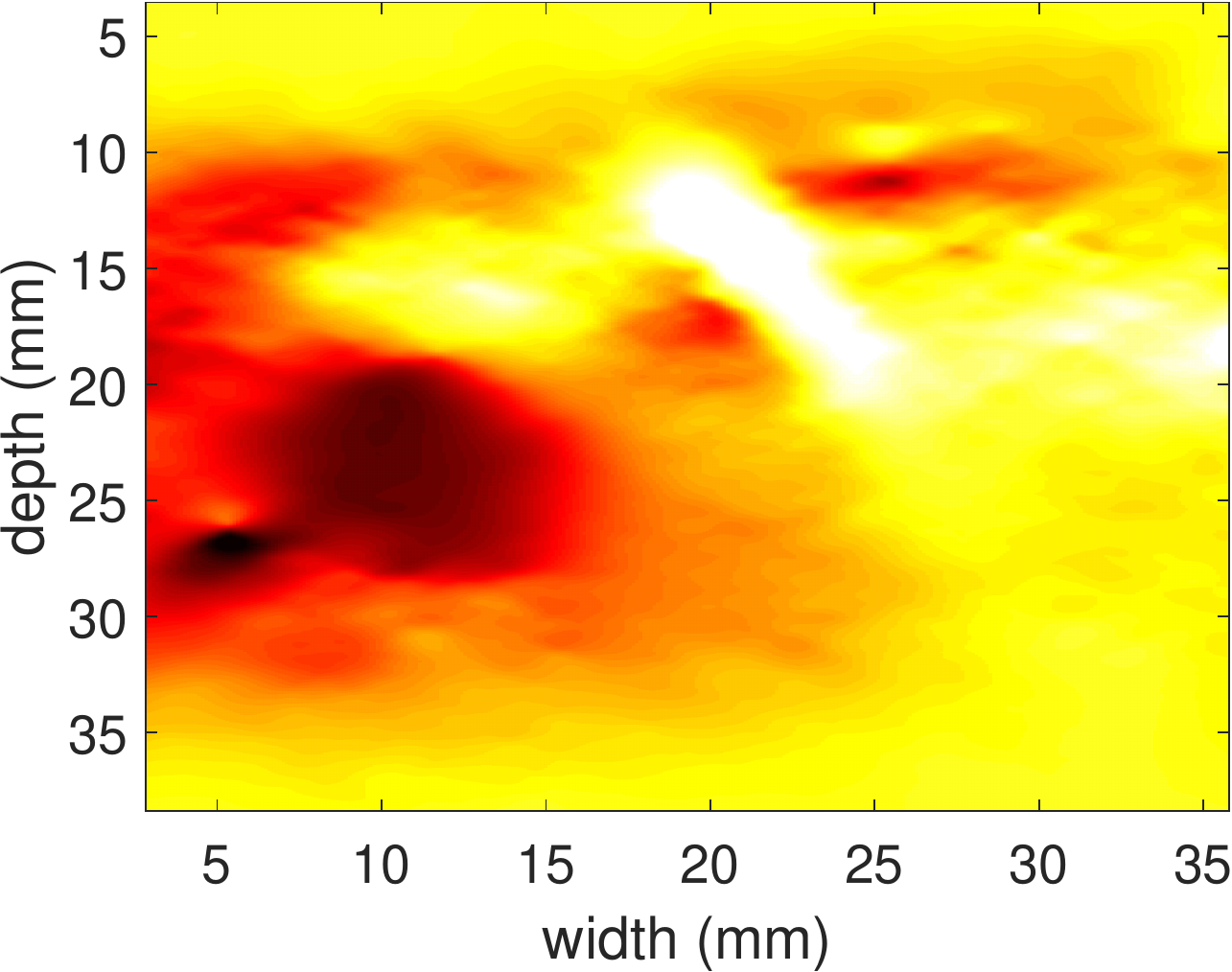} }}
	\subfigure[B-mode]{{\includegraphics[width=.2\textwidth, height=.19\textwidth]{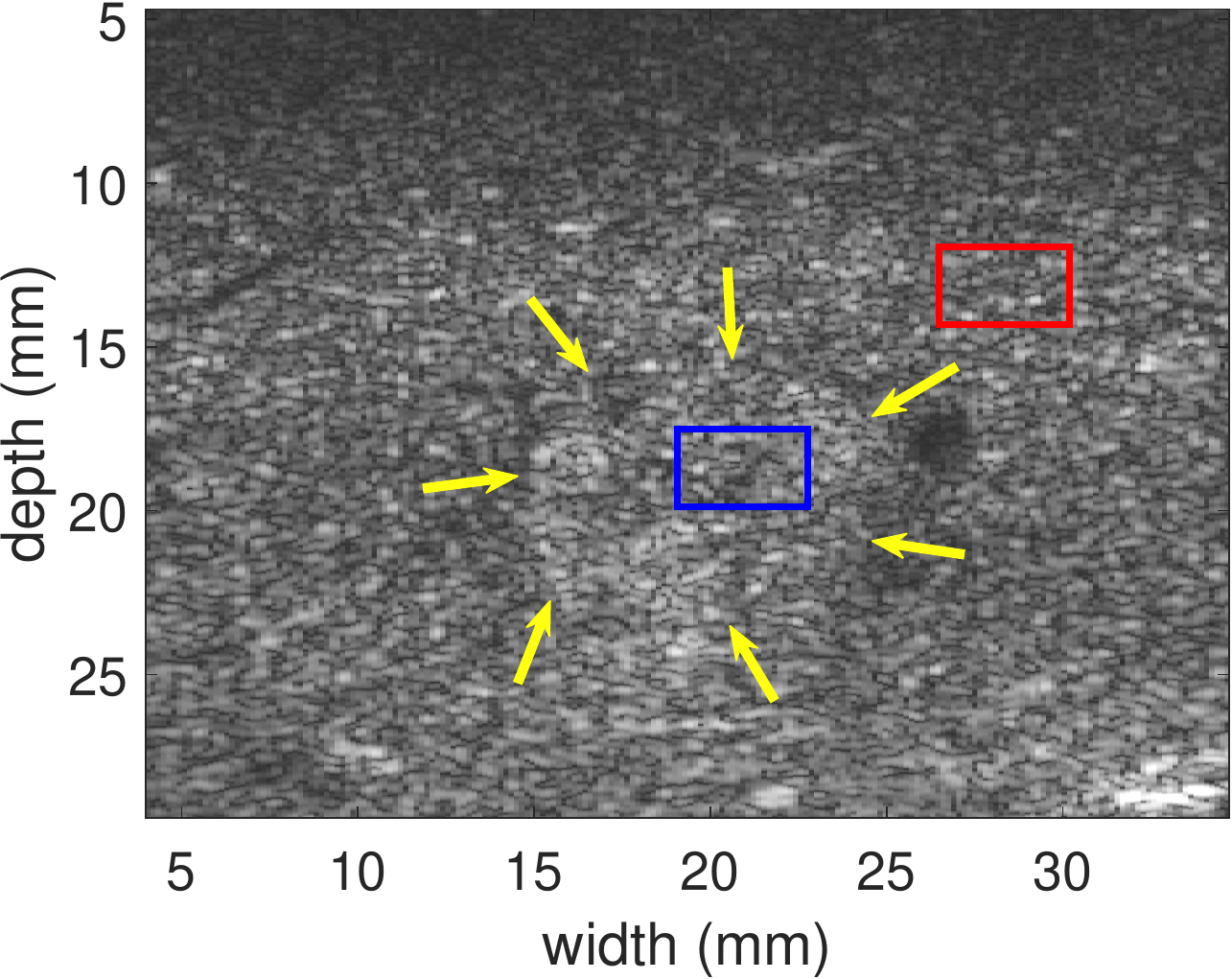}}}%
	\subfigure[GLUE]{{\includegraphics[width=.2\textwidth, height=.19\textwidth]{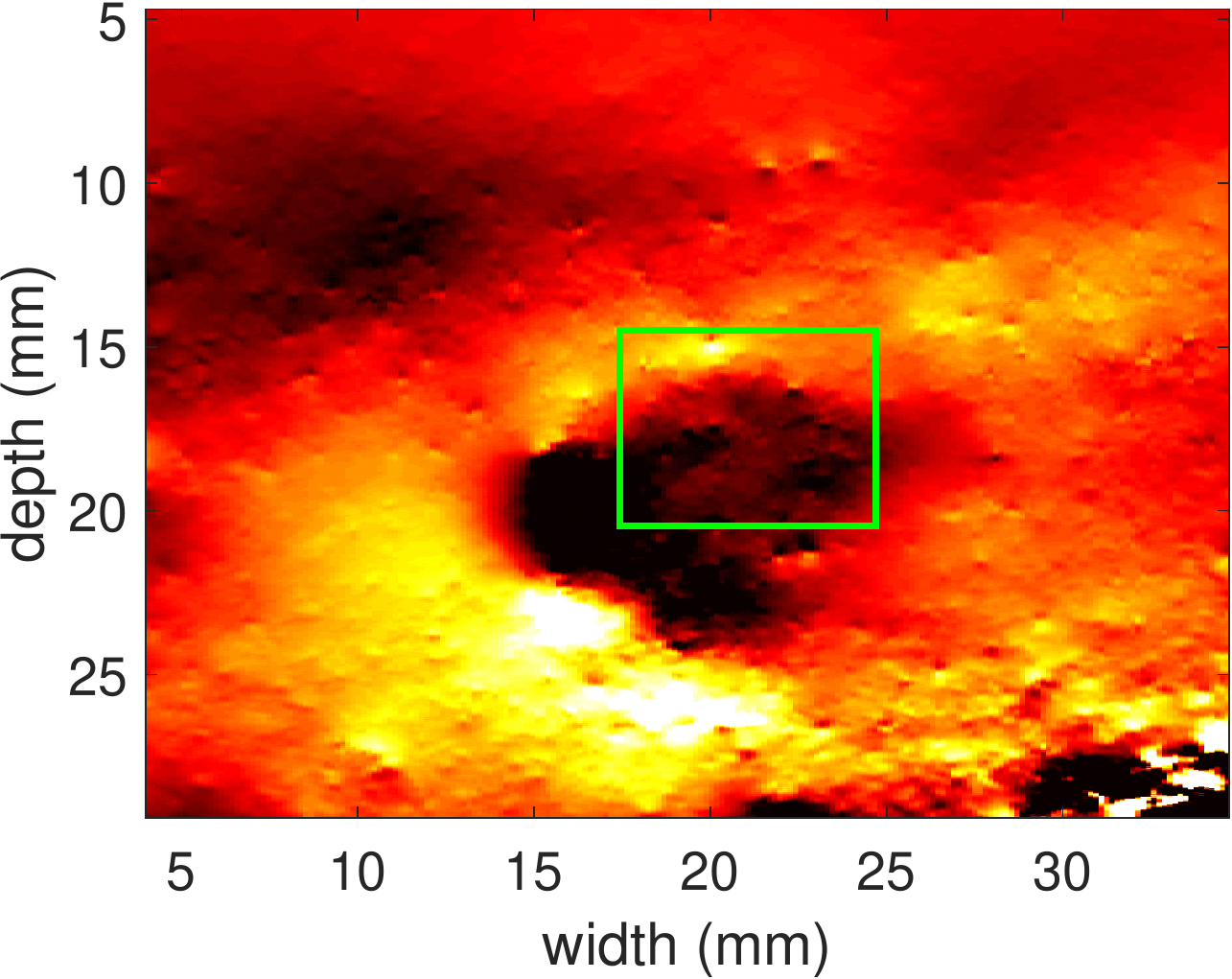}}}%
	\subfigure[OVERWIND]{{\includegraphics[width=.2\textwidth, height=.19\textwidth]{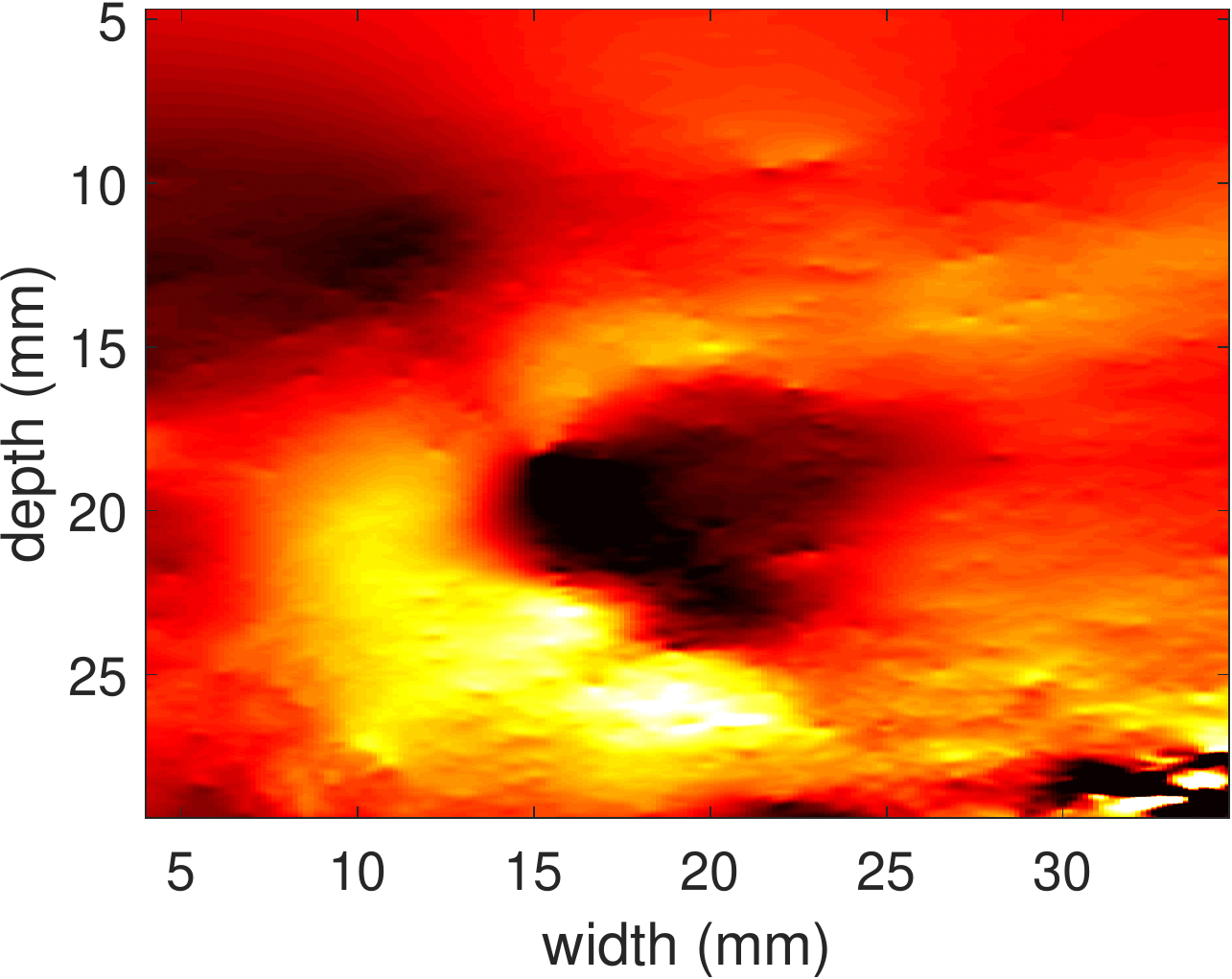} }}%
	\subfigure[SOUL]{{\includegraphics[width=.2\textwidth, height=.19\textwidth]{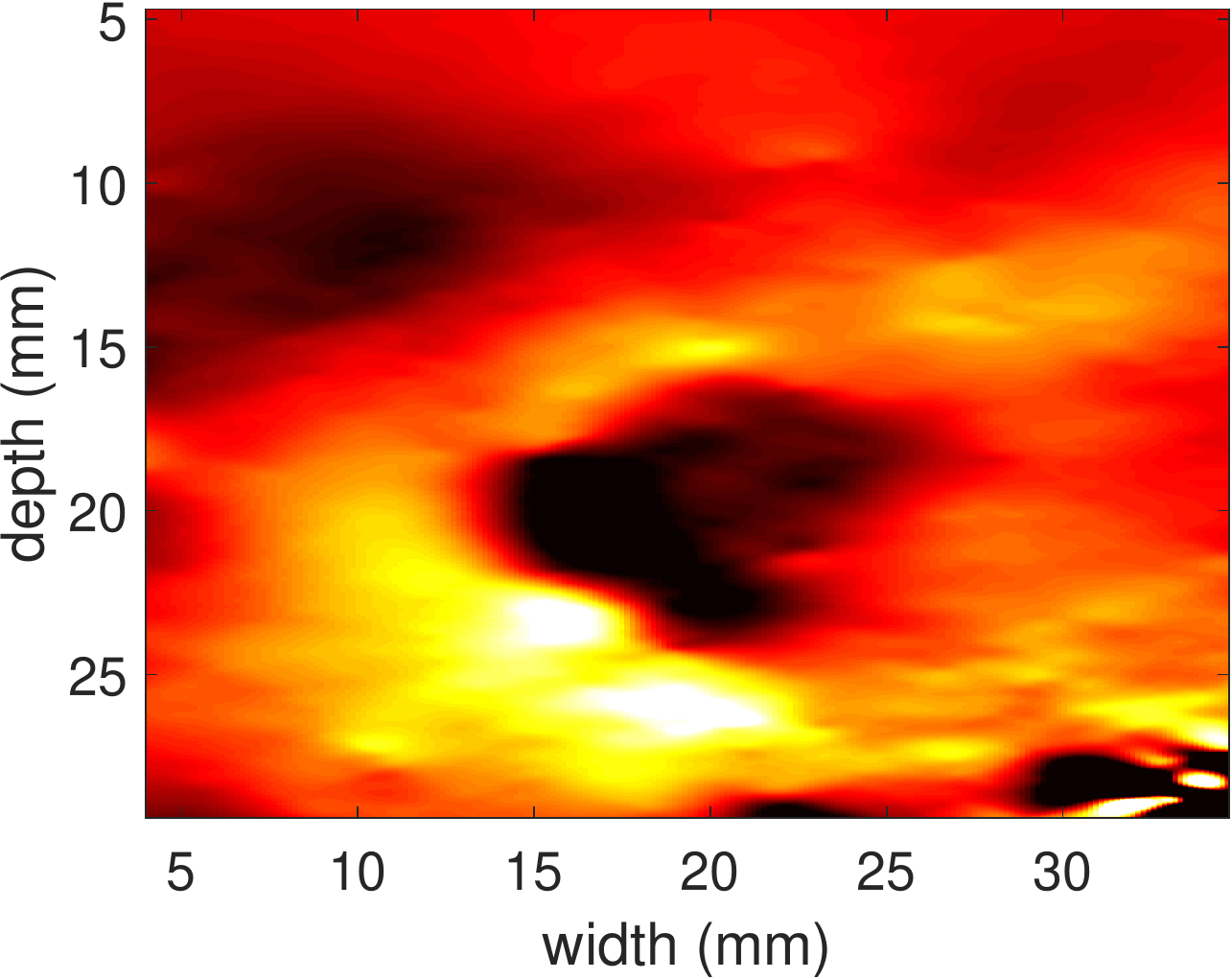} }}%
	\subfigure[$L1$-SOUL]{{\includegraphics[width=.2\textwidth, height=.19\textwidth]{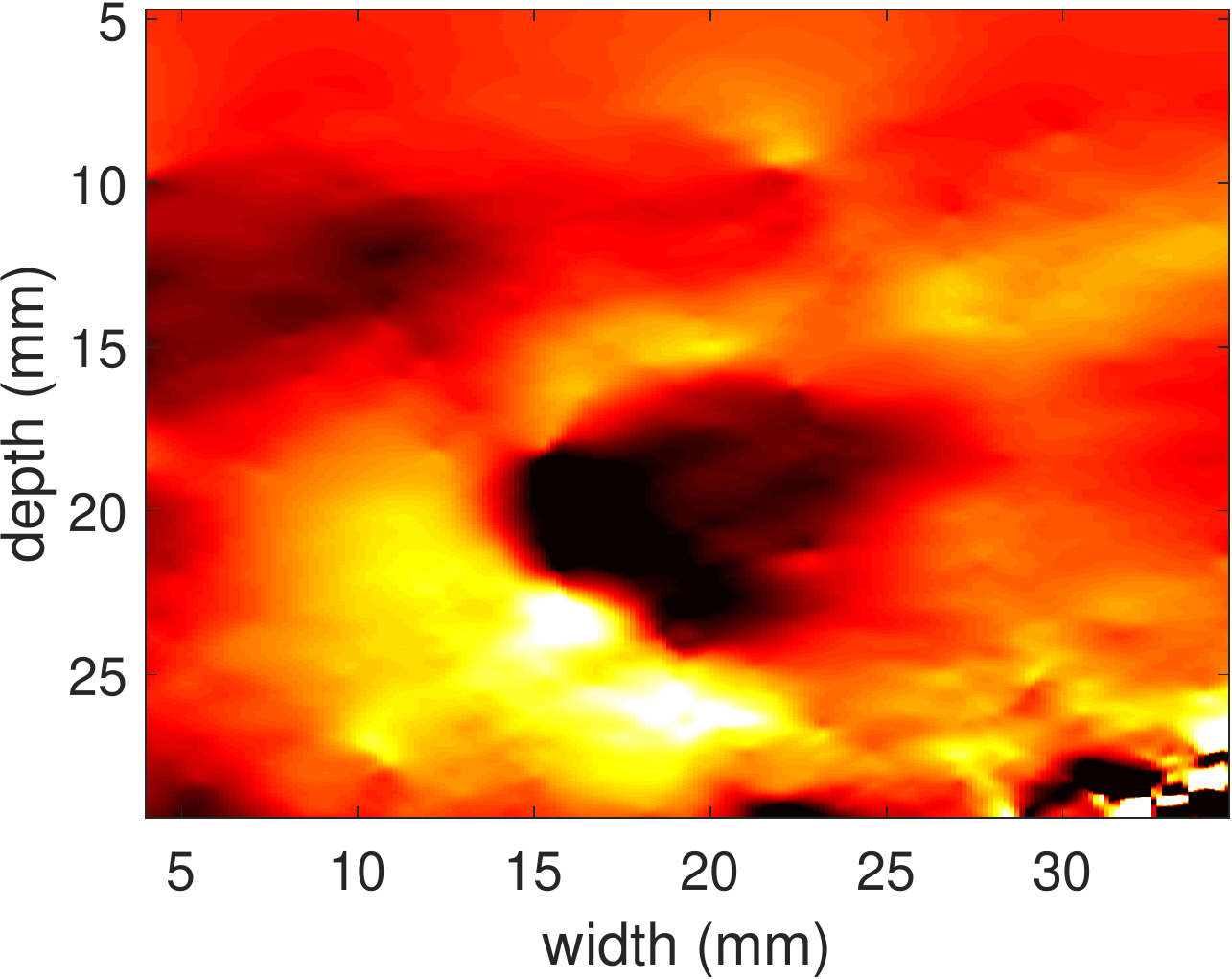} }}
	\subfigure[Strain, patient 1]{{\includegraphics[width=.48\textwidth]{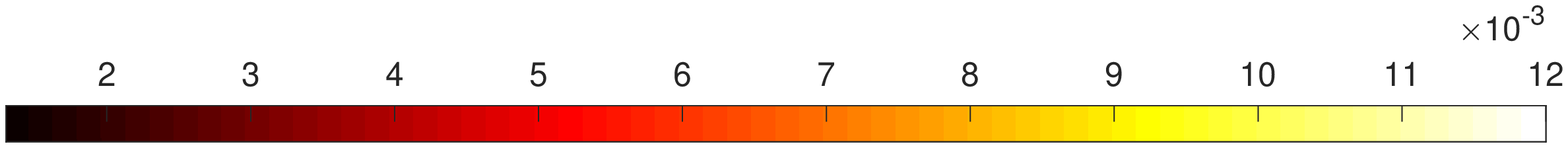}}}%
	\subfigure[Strain, patient 2]{{\includegraphics[width=.48\textwidth]{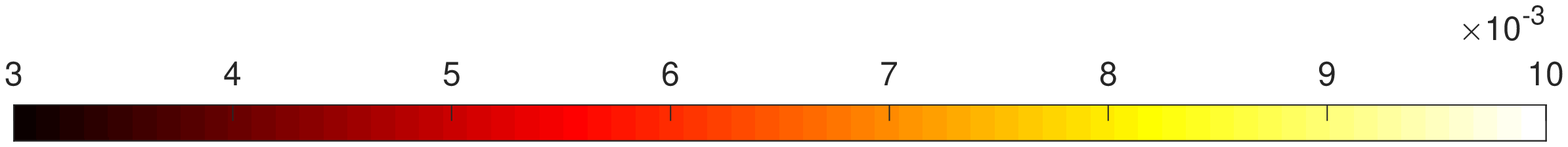}}}
	\caption{\textit{In vivo} axial strain results obtained from liver patients 1 and 2. Rows 1 and 2 correspond to patients 1 and 2, respectively, whereas columns 1 to 5 correspond to the B-mode and axial strain images obtained from GLUE, OVERWIND, SOUL, and $L1$-SOUL, respectively. The kernel length is set to 3 samples in all strain images for differentiating the displacement images.}
	\label{liver}
\end{figure*}

\begin{figure*}
	\centering
	\subfigure[GLUE]{{\includegraphics[width=.23\textwidth, height=.23\textwidth]{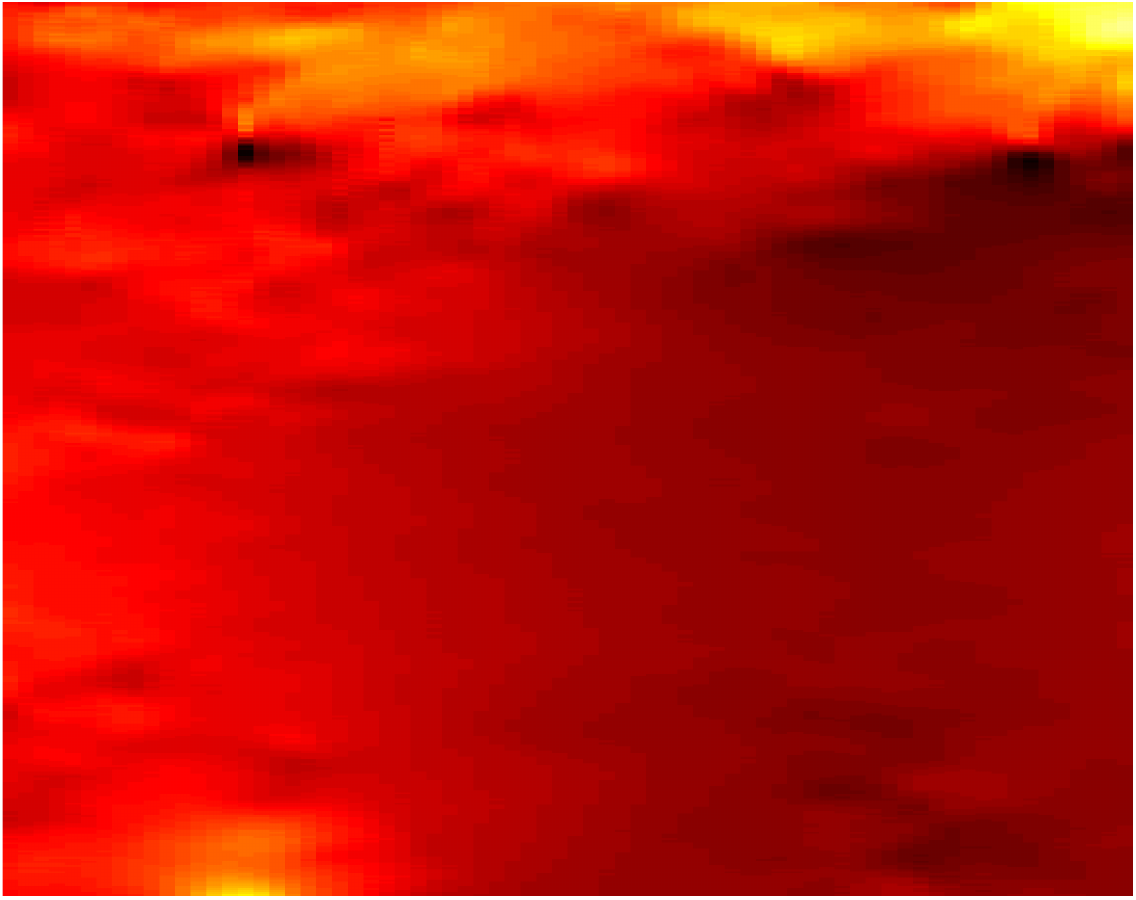}}}%
	\quad
	\subfigure[OVERWIND]{{\includegraphics[width=.23\textwidth, height=.23\textwidth]{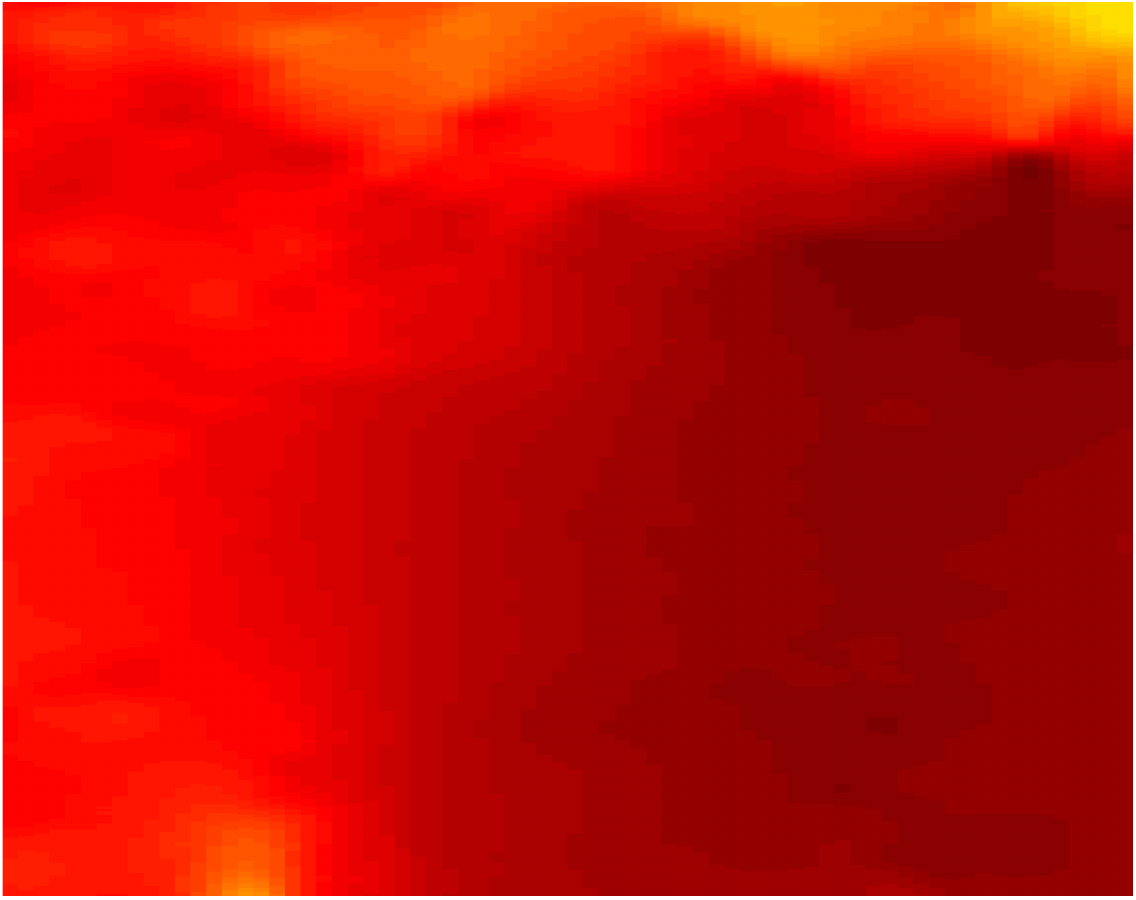}}}
	\quad
	\subfigure[SOUL]{{\includegraphics[width=.23\textwidth, height=.23\textwidth]{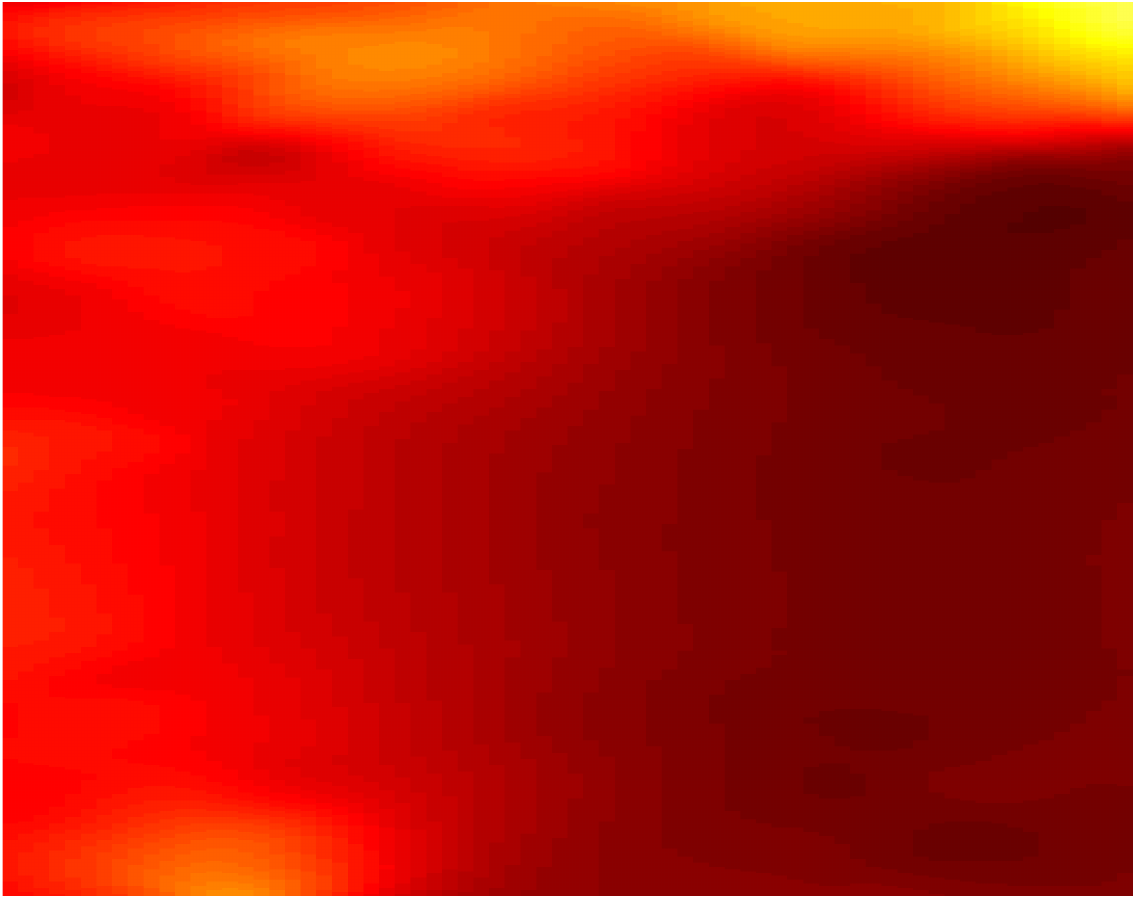}}}%
	\quad
	\subfigure[$L1$-SOUL]{{\includegraphics[width=.23\textwidth, height=.23\textwidth]{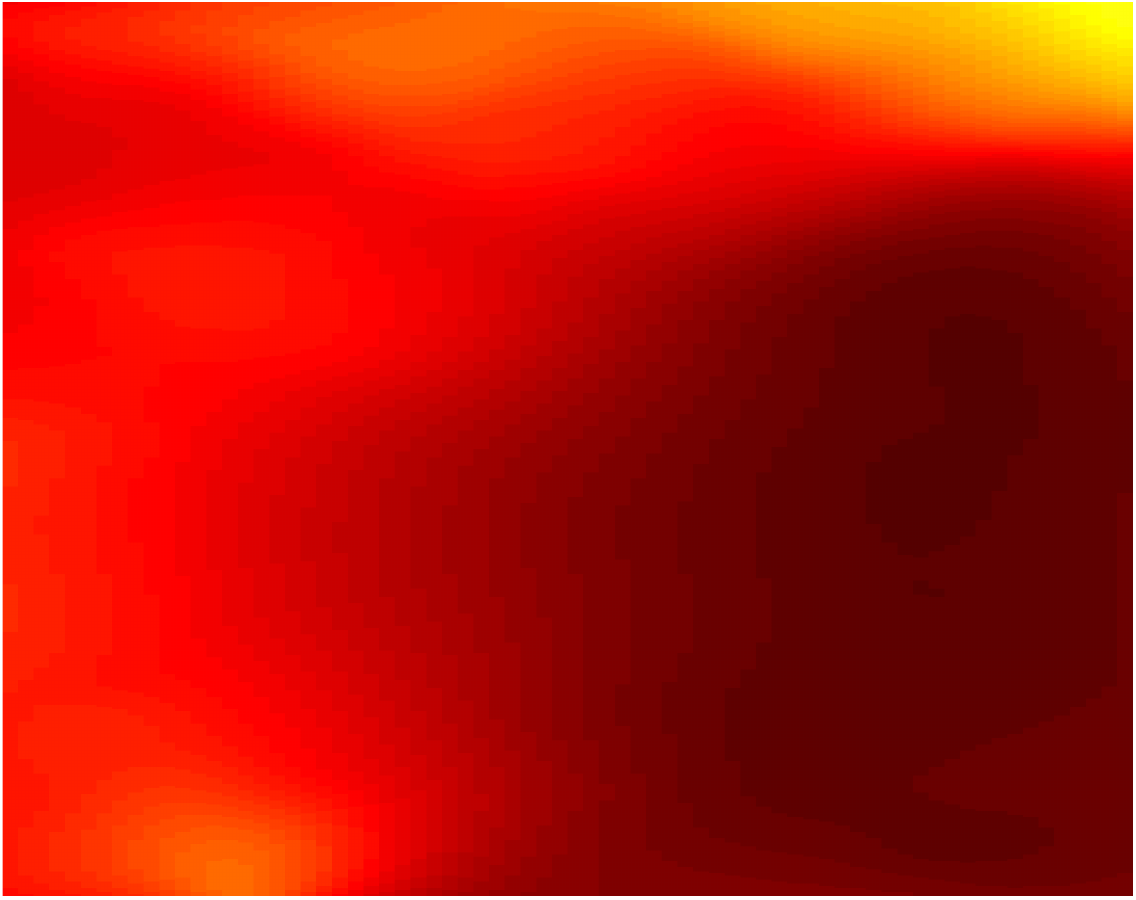}}}%
	\quad
	\subfigure[GLUE]{{\includegraphics[width=.23\textwidth, height=.23\textwidth]{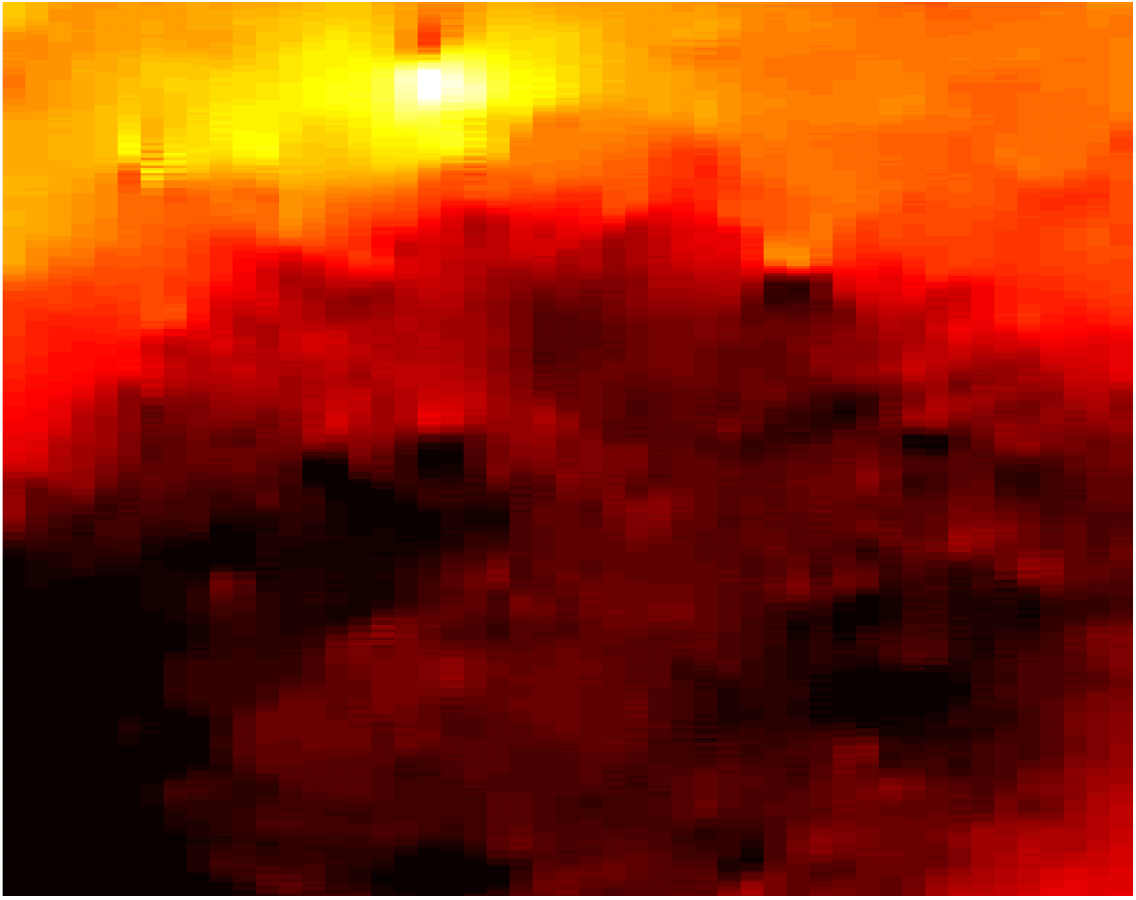}}}%
	\quad
	\subfigure[OVERWIND]{{\includegraphics[width=.23\textwidth, height=.23\textwidth]{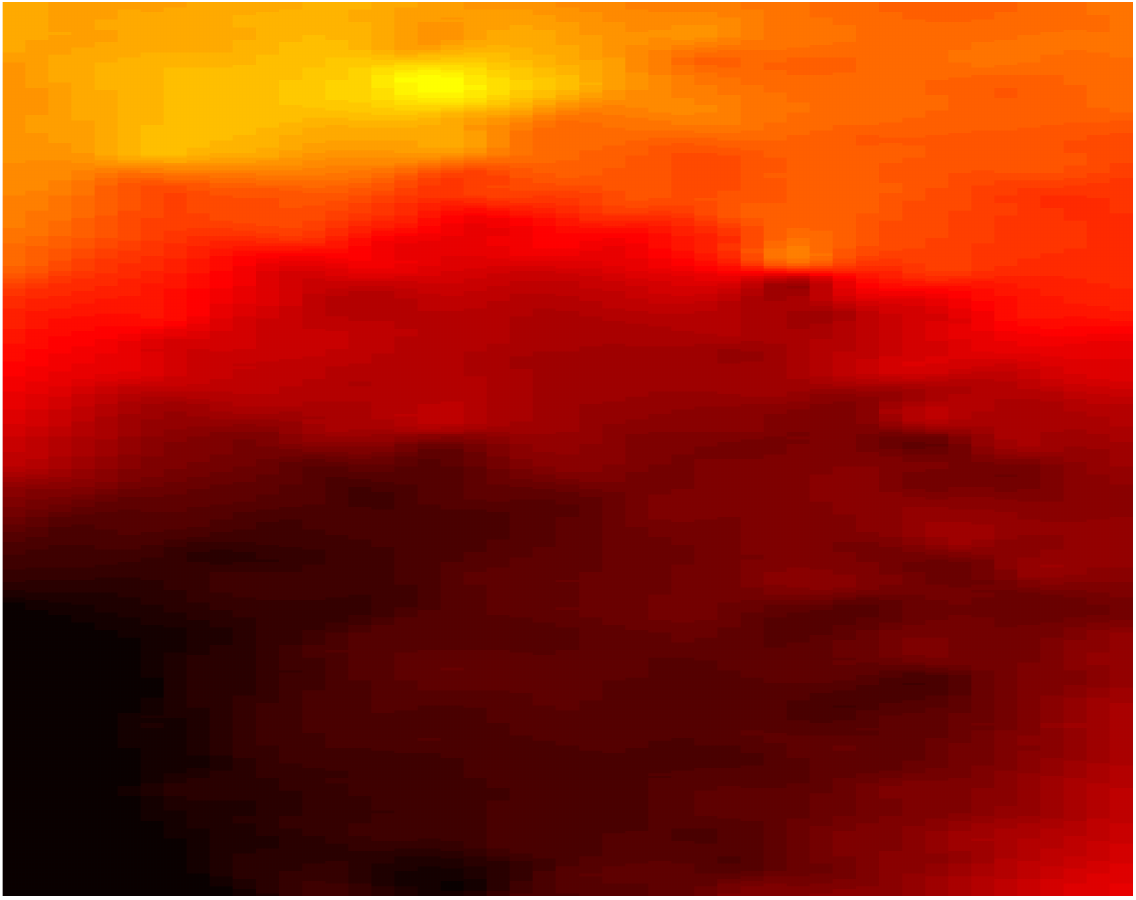}}}
	\quad
	\subfigure[SOUL]{{\includegraphics[width=.23\textwidth, height=.23\textwidth]{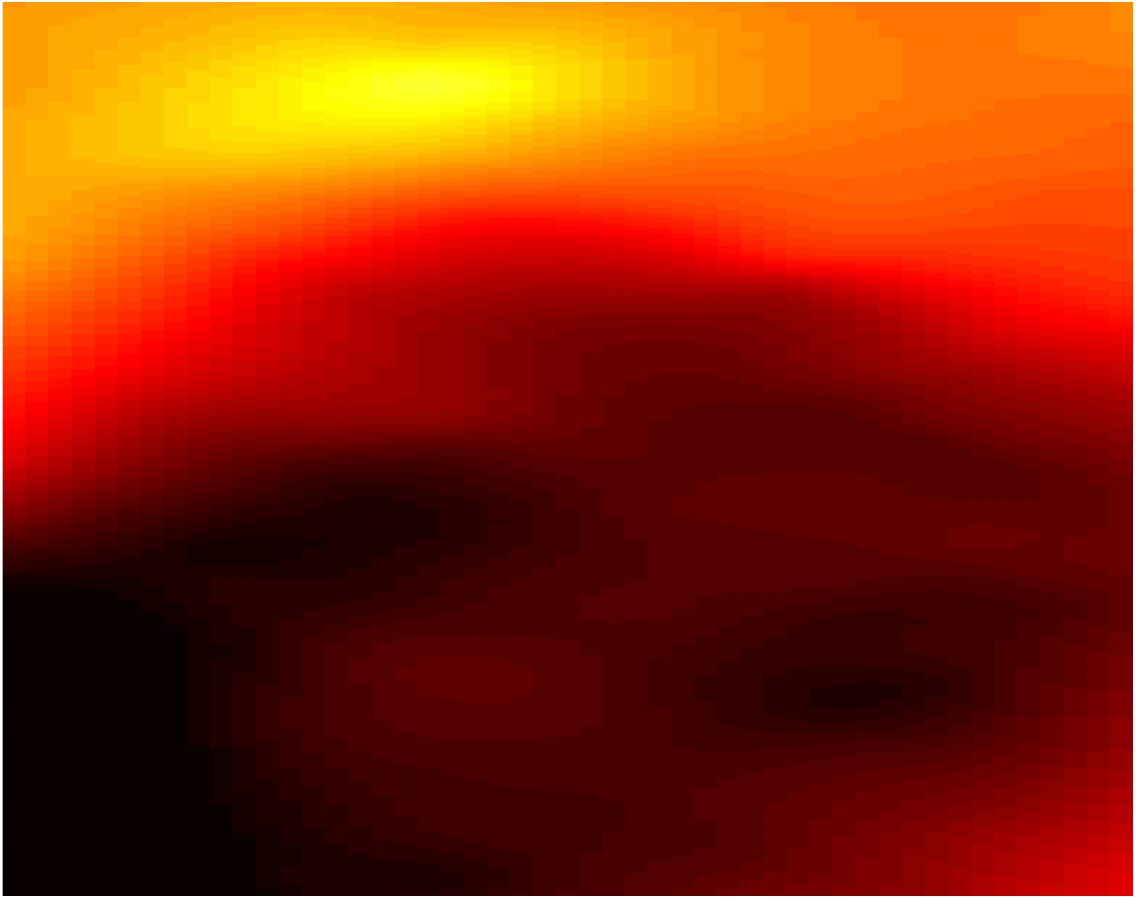}}}%
	\quad
	\subfigure[$L1$-SOUL]{{\includegraphics[width=.23\textwidth, height=.23\textwidth]{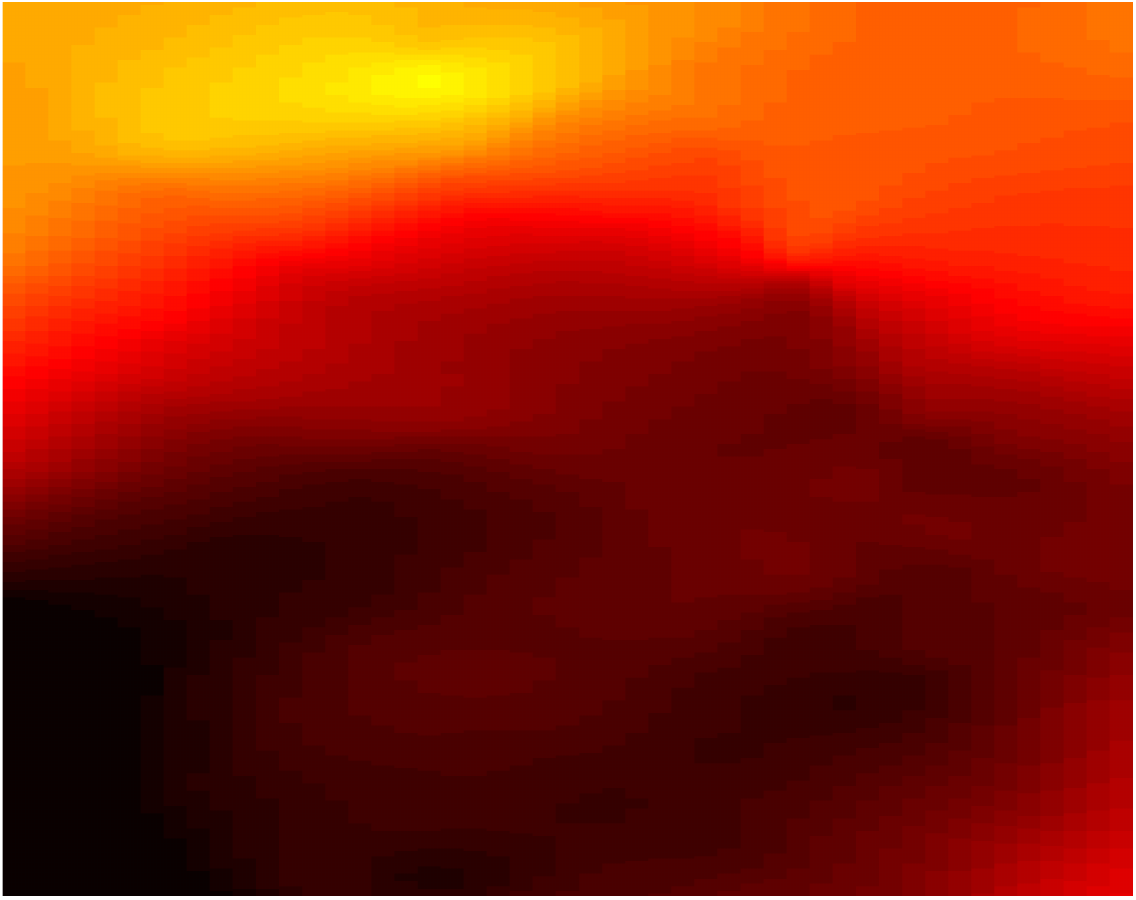}}}%
	\caption{Strain results for liver Patients 1 and 2 in the green windows shown in Figs.~\ref{liver}(b) and \ref{liver}(g), respectively. Rows 1 and 2 correspond to patients 1 and 2, whereas columns 1-4 correspond to GLUE, OVERWIND, SOUL, and $L1$-SOUL, respectively.}
	\label{liver12_zoomed}
\end{figure*}

\begin{figure*}
	\centering
	\subfigure[B-mode]{{\includegraphics[width=.33\textwidth, height=.36\textwidth]{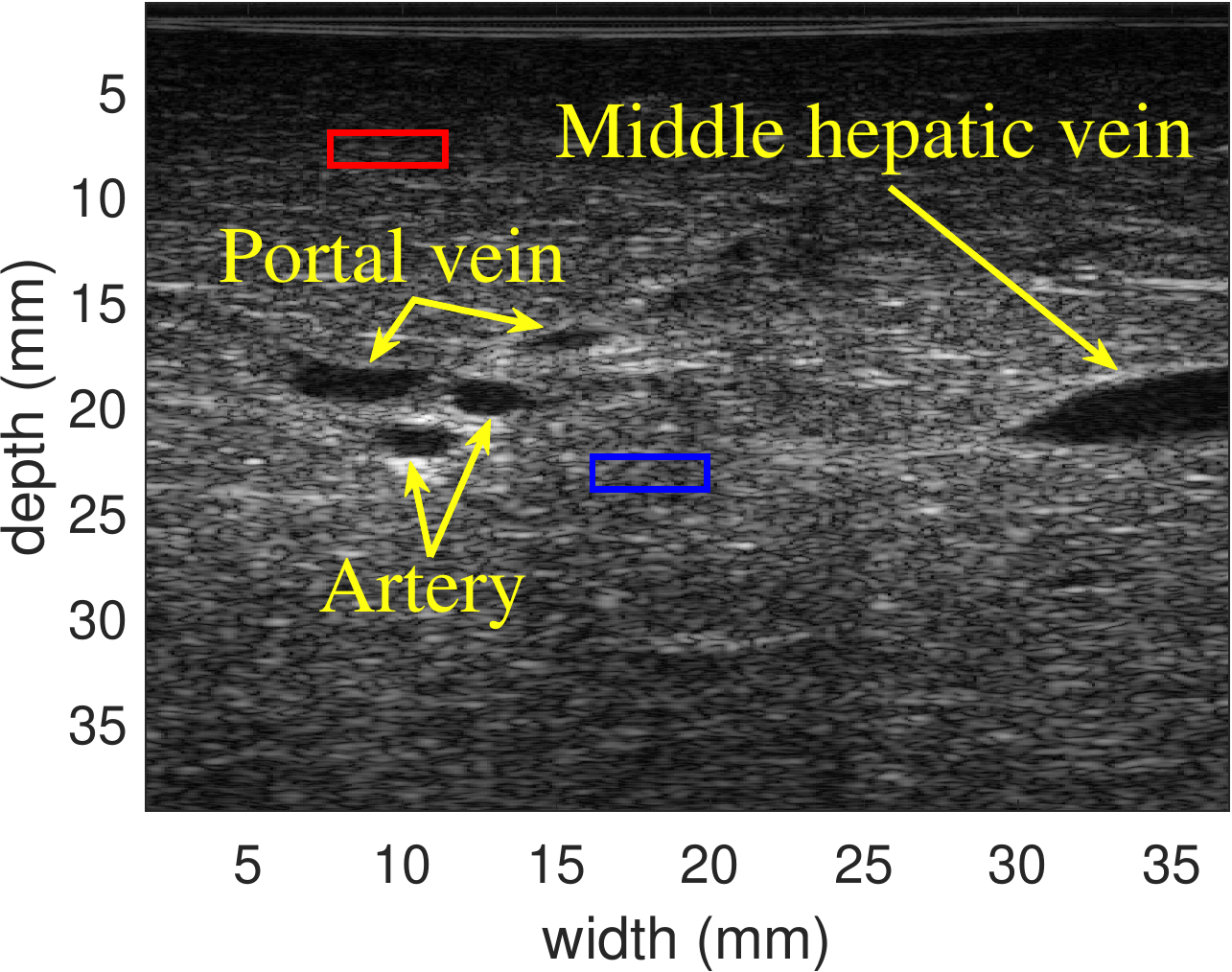}}}\\
	\subfigure[GLUE]{{\includegraphics[width=.48\textwidth, height=.3\textwidth]{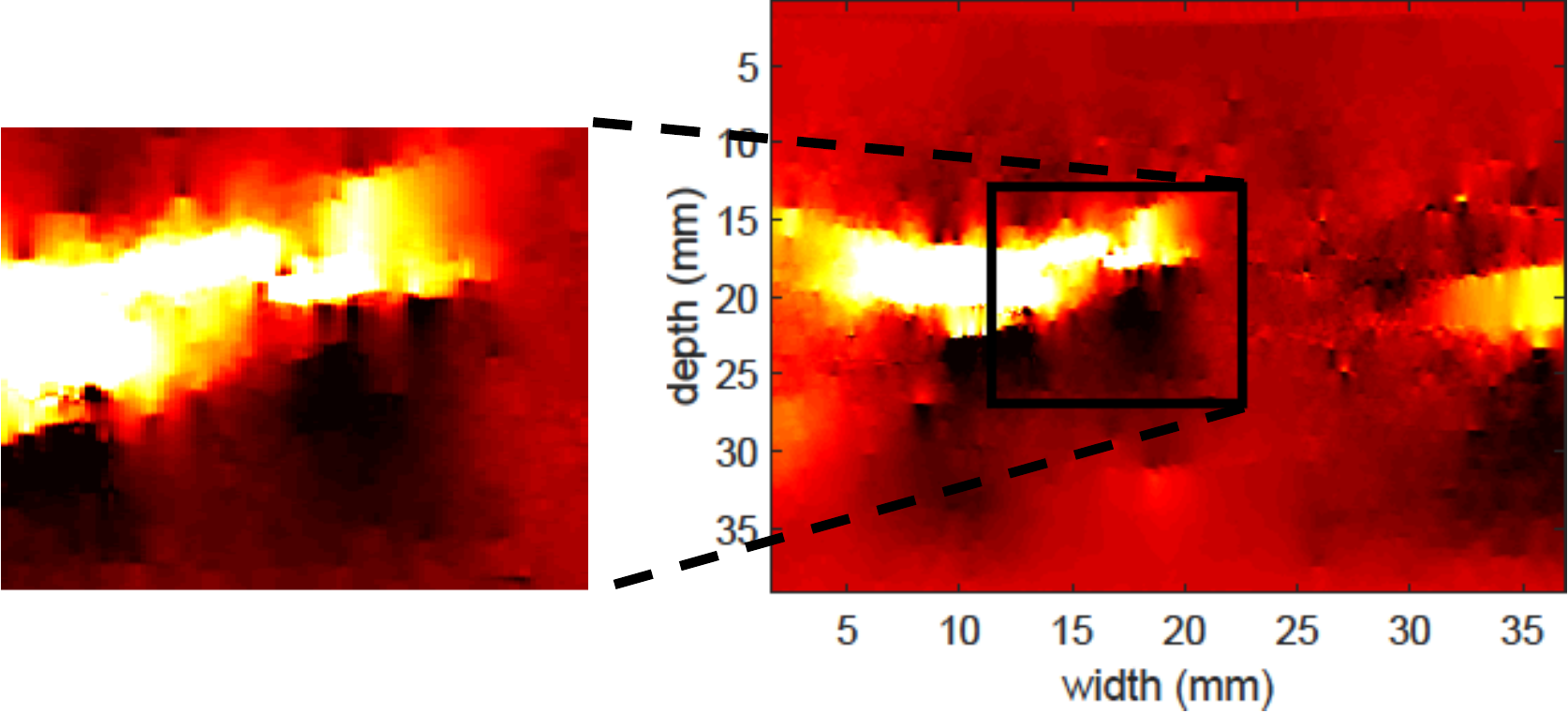}}}%
	\quad
	\subfigure[OVERWIND]{{\includegraphics[width=.48\textwidth, height=.3\textwidth]{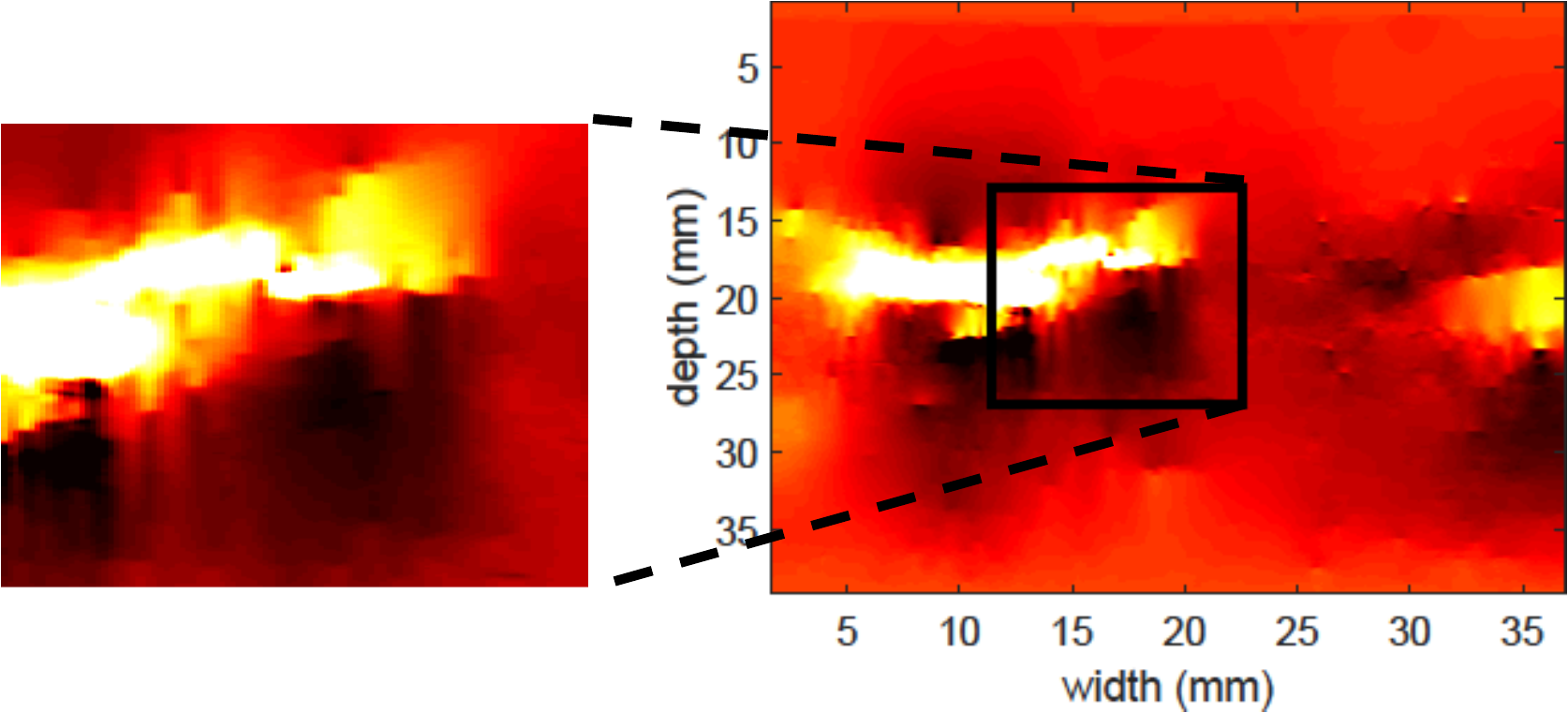} }}
	\quad
	\subfigure[SOUL]{{\includegraphics[width=.48\textwidth, height=.3\textwidth]{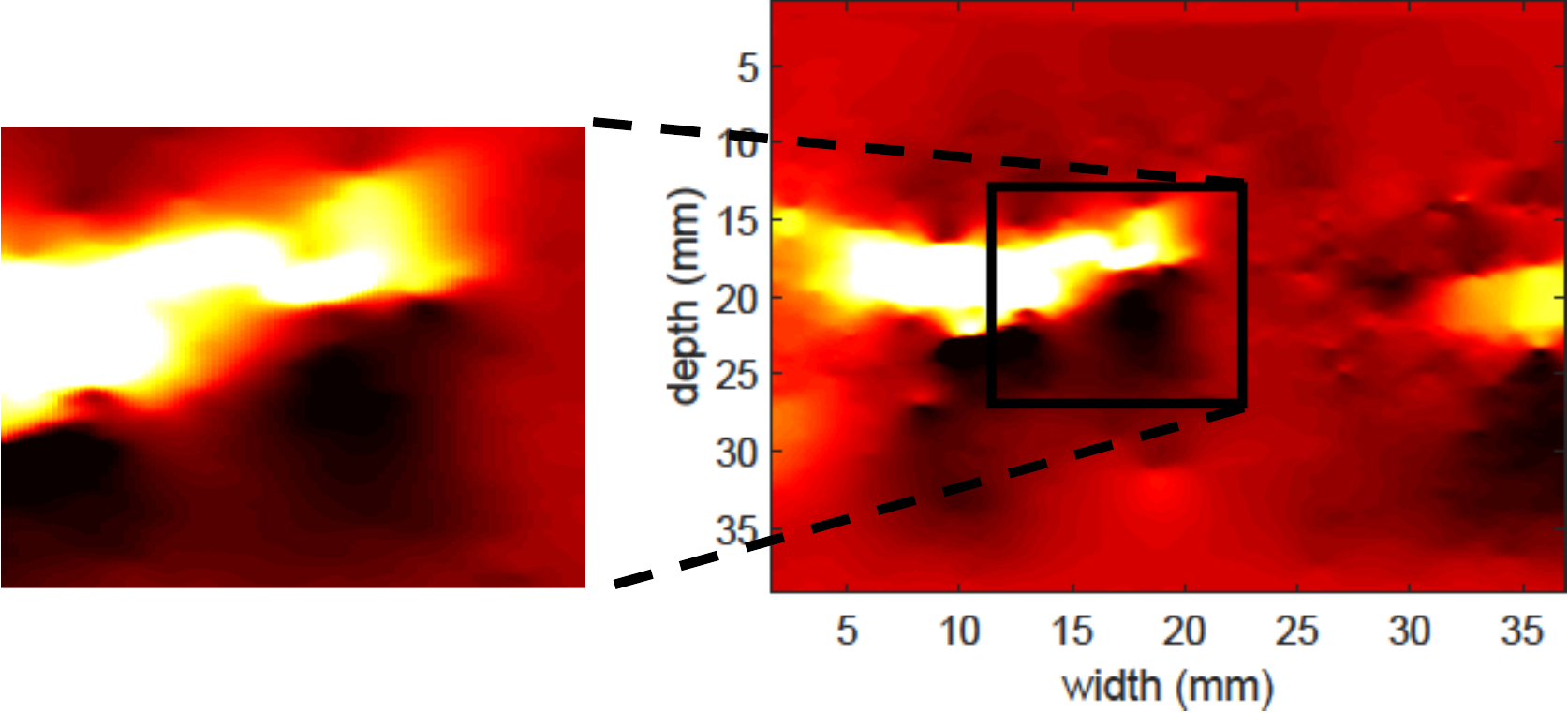} }}%
	\quad
	\subfigure[$L1$-SOUL]{{\includegraphics[width=.48\textwidth, height=.3\textwidth]{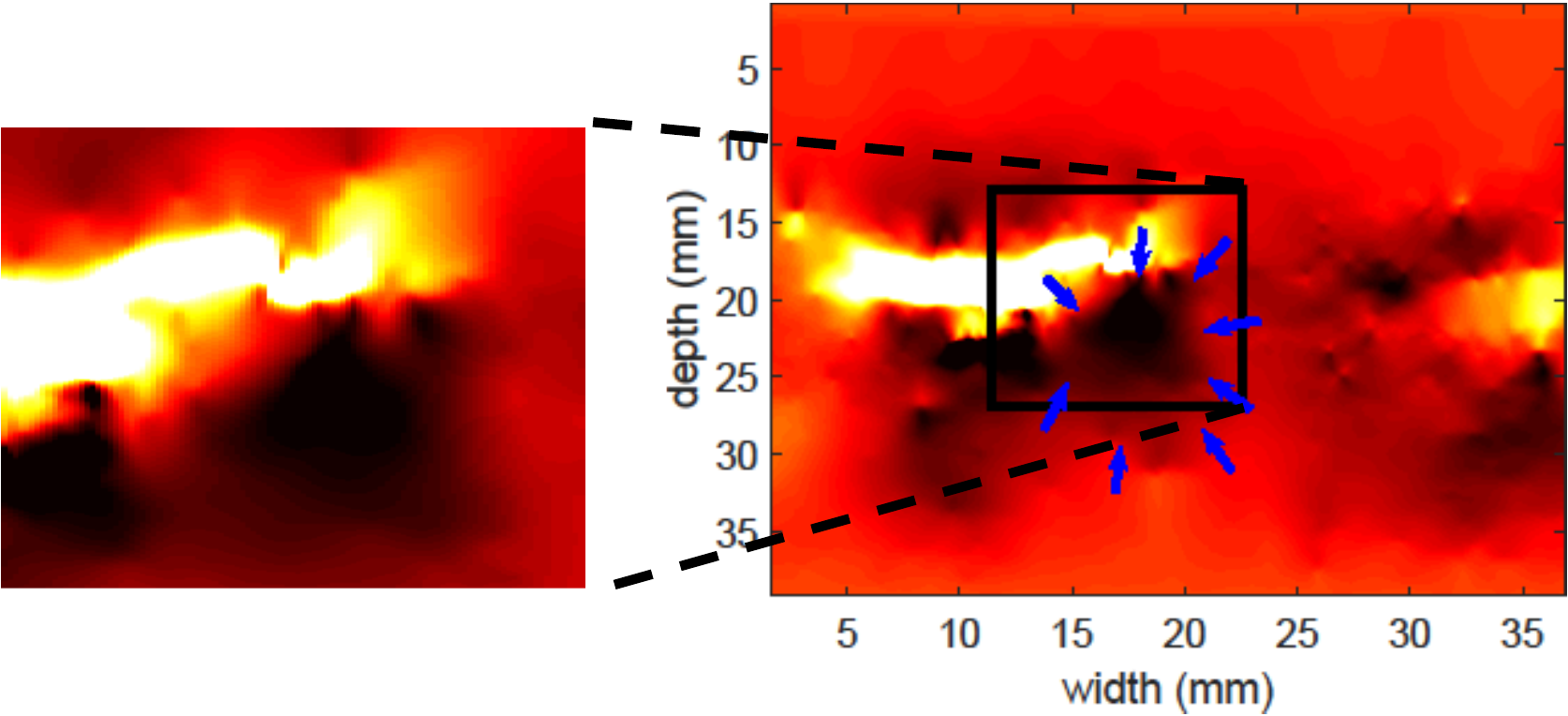} }}
	\quad
	\subfigure[Strain]{{\includegraphics[width=.6\textwidth]{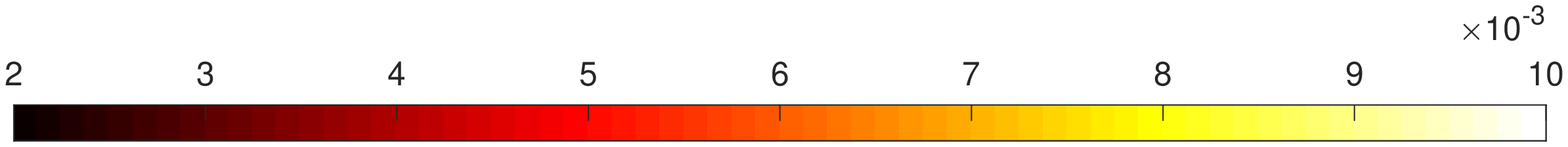}}}
	\caption{\textit{In vivo} strain results for liver patient 3. (a) shows the B-mode image, whereas (b)-(e) correspond to the axial strain images obtained from GLUE, OVERWIND, SOUL, and $L1$-SOUL, respectively. The kernel length is set to 3 samples in all strain images for differentiating the displacement images.}
	\label{liver3}
\end{figure*}

\begin{table}[!t]
	\renewcommand{\arraystretch}{1.3}
	\caption{Average strain in the tumor region of liver patient 1. Average strain values are calculated on the blue colored rectangular window shown in Fig.~\ref{liver}(a).}
	\label{table_liver1_tumor}
	\centering
		\begin{tabular}[c]{lllr}
			\hline		
			&	& Strain\\
			\hline
			&	GLUE   & $3.7 \times 10^{-3}$\\
			&	OVERWIND & $3.7 \times 10^{-3}$\\
			&	SOUL  & $3.2 \times 10^{-3}$\\
			&	$L1$-SOUL & $\mathbf{3 \times 10^{-3}}$\\
			\hline			
		\end{tabular}
\end{table} 

\begin{table*}[tb]  
	\centering
	\caption{SNR, CNR, and SR of the strain images pertaining to patients 1, 2, and 3. CNR and SR are calculated using the blue colored target and red colored background windows depicted in Figs.~\ref{liver}(a), \ref{liver}(f), and \ref{liver3}(a). SNR values are obtained from the background windows.}
	\label{table_vivo}
	\begin{tabular}{c c c c c c c c c c c c} 
		\hline
		\multicolumn{1}{c}{} &
		\multicolumn{3}{c}{Patient 1} &
		\multicolumn{1}{c}{} &
		\multicolumn{3}{c}{Patient 2} &
		\multicolumn{1}{c}{} &
		\multicolumn{3}{c}{Patient 3}\\
		\cline{2-4} 
		\cline{6-8}
		\cline{10-12}
		$ $  $ $&    SNR & CNR & SR $ $  $ $&$ $  $ $ &$ $  $ $ SNR & CNR & SR$ $  $ $&$ $  $ $ &$ $  $ $ SNR & CNR & SR\\
		\hline
		GLUE &  15.27 &  12.21 & 0.39 && 15.75 & 8.81 & 0.52 && 22.90 & 8.07 & 0.62\\
		OVERWIND & 23.23 &  15.46 & 0.40 && 25.90 & 12.62 & 0.56 && 56.92 & 16.43 & 0.56\\
		SOUL & 18.76 &  15.56 & 0.35 && 22.60 & 13.63 & \textbf{0.51} && 25.31 & 8.53 & 0.62\\
		$L1$-SOUL &  \textbf{32.60}  & \textbf{20.69} & \textbf{0.31}  && \textbf{28.81} & \textbf{15.95} & 0.52 && \textbf{66.36} & \textbf{18.54} & \textbf{0.46}\\
		\hline
	\end{tabular}
\end{table*}

\subsection{Simulated Phantom with Hard Inclusion}
Fig.~\ref{hard_simu} shows the strain images for the hard-inclusion simulated phantom obtained from FEM and the four techniques under consideration. GLUE strain is pixelated in the inclusion and noisy in the background. SOUL stretches out the strain in the stiff tissue regions. Although OVERWIND overcomes some of these limitations, its result still lacks visual contrast between the target and the background. $L1$-SOUL substantially outperforms all other techniques by ensuring the highest visual contrast and a sharp inclusion edge which is corroborated by the SNR, CNR, and SR values reported in Table~\ref{table_hard}.          

To provide a quantitative assessment of overall image quality, we calculate 120 CNR values from 6 target and 20 background windows. The CNR histogram (see Fig.~\ref{cnr_histograms}(a)) demonstrates the superiority of $L1$-SOUL over the other techniques by showing that it occupies most of the high CNR values. The mean CNRs corresponding to GLUE, OVERWIND, SOUL, and $L1$-SOUL, respectively, are 12.30, 28.96, 18.20, and 32.98. In addition, a statistical analysis using paired \textit{t}-test confirms that $L1$-SOUL is significantly better than GLUE, OVERWIND, and SOUL with $p$-values of $7.55 \times 10^{-28}$, $4.4 \times 10^{-3}$, and $4.97 \times 10^{-23}$, respectively. 

\subsection{Breast Phantom Results}
The B-mode and the axial strain images for the experimental breast elastography phantom are presented in Fig.~\ref{perform_phan}. Both GLUE and SOUL distort the inclusion shape at the bottom. In addition, OVERWIND exhibits low contrast in deep regions. $L1$-SOUL obtains the highest quality strain image by preserving the inclusion shape and maximizing the visual contrast. The quantitative values shown in Table~\ref{table_phan} support our visual interpretation.

Similar to the simulated dataset, we plot the histogram of 120 CNR values corresponding to 6 target and 20 background strain windows. It is evident from Fig.~\ref{cnr_histograms}(b) that $L1$-SOUL exhibits the highest frequency in cases of high CNR values. The average CNR values associated with GLUE, OVERWIND, SOUL, and $L1$-SOUL are 9.18, 16.63, 13.73, and 19.29, respectively. An additional statistical analysis using paired  \textit{t}-test reassures that $L1$-SOUL is significantly better than GLUE, OVERWIND, and SOUL with $p$-values of $1.19 \times 10^{-18}$, $3.98 \times 10^{-5}$, and $8.97 \times 10^{-12}$, respectively.

\subsection{\textit{In vivo} Liver Cancer Results}
Figs.~\ref{liver} and \ref{liver3} show the B-mode and the strain images for the \textit{in vivo} liver cancer datasets. The yellow arrow marks in Figs.~\ref{liver}(a) and \ref{liver}(f) indicate the tumors of patients of 1 and 2. In the case of patient 3, since the B-mode image does not show any echogenic difference between the tumor and the background, the tumor is indicated in the strain image (Fig.~\ref{liver3}(e)) by blue arrow marks. While all four techniques visualize the tumor successfully in the case of patient 1, GLUE obtains the noisiest strain image. Although OVERWIND and SOUL better handle the noise, they exhibit irregular strain fluctuation along the tumor edge (see Fig.~\ref{liver12_zoomed}) and underestimate the background strain. In addition, the tumor regions in GLUE, OVERWIND, and SOUL strain images are not sufficiently dark (see Table~\ref{table_liver1_tumor}). $L1$-SOUL outperforms the other techniques by showing a darker tumor, sharper edge, and high target-background contrast. For patient 2, SOUL and $L1$-SOUL outperform GLUE and OVERWIND in terms of background smoothness. In addition, the tumor edges in GLUE and OVERWIND strain images suffer from irregular strain fluctuations (see Fig.~\ref{liver12_zoomed}). $L1$-SOUL fully resolves the earlier techniques' limitations by generating sharp edges at the borders of the tumor. In the case of patient 3, GLUE and SOUL strain images suffer from two major drawbacks. First, due to over-smoothing, they stretch the branches of the portal vein axially. Second, they yield dark backgrounds due to the underestimation of strain. Although OVERWIND overcomes these limitations, it suffers from discontinuities in the vessel wall and the background. In addition, the tumor region shows a low contrast in the OVERWIND strain. $L1$-SOUL substantially outperforms all three techniques and produces a high-quality strain image with a sharp and continuous vessel wall, dark tumor, and bright and smooth background. The quantitative values of SNR, CNR, and SR reported in Table~\ref{table_vivo} align with our visual inference.                    

\begin{figure*}
	\centering
	\subfigure[Hard-inclusion simulated phantom]{{\includegraphics[width=.2\textwidth]{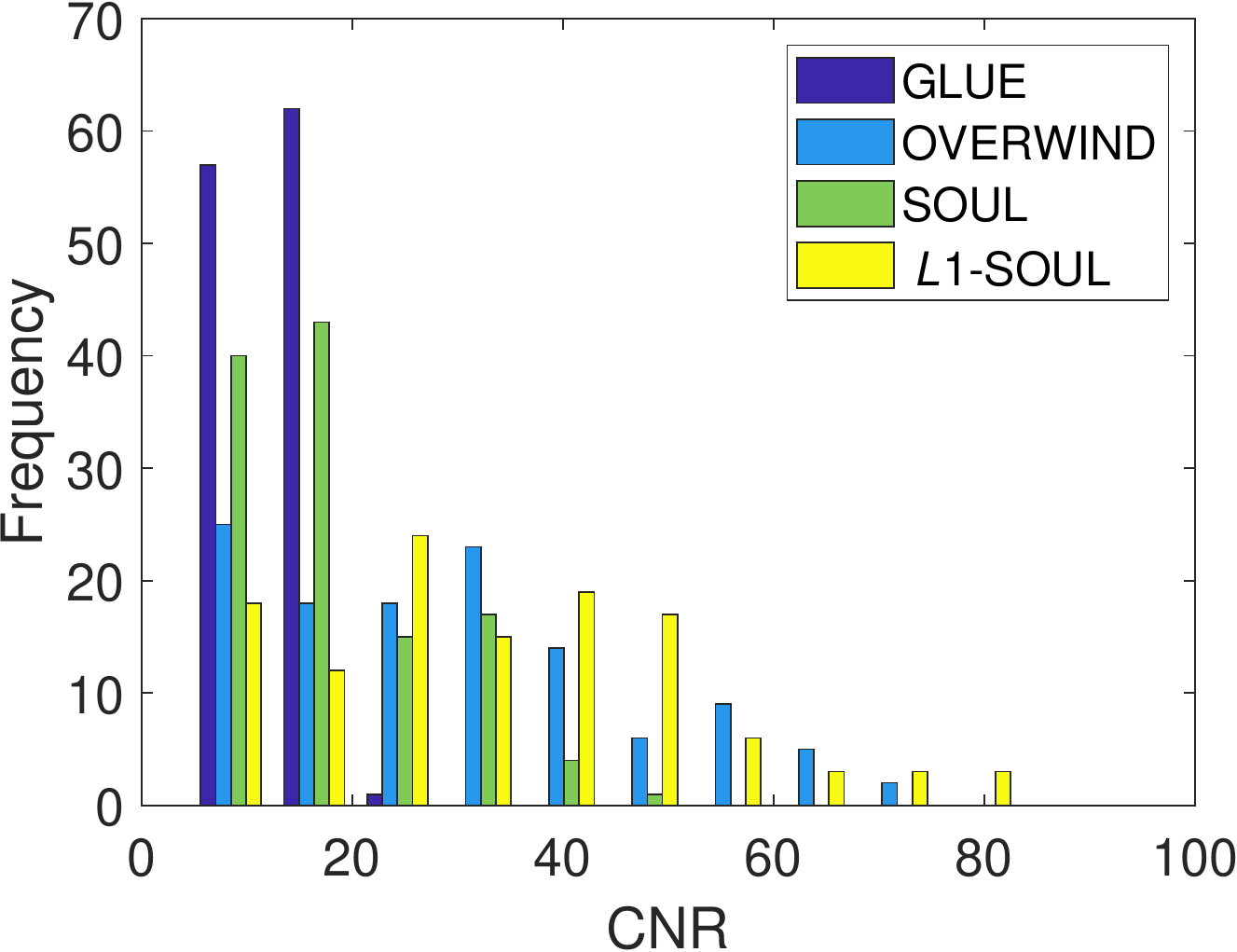}}}%
	\subfigure[Breast phantom]{{\includegraphics[width=.2\textwidth]{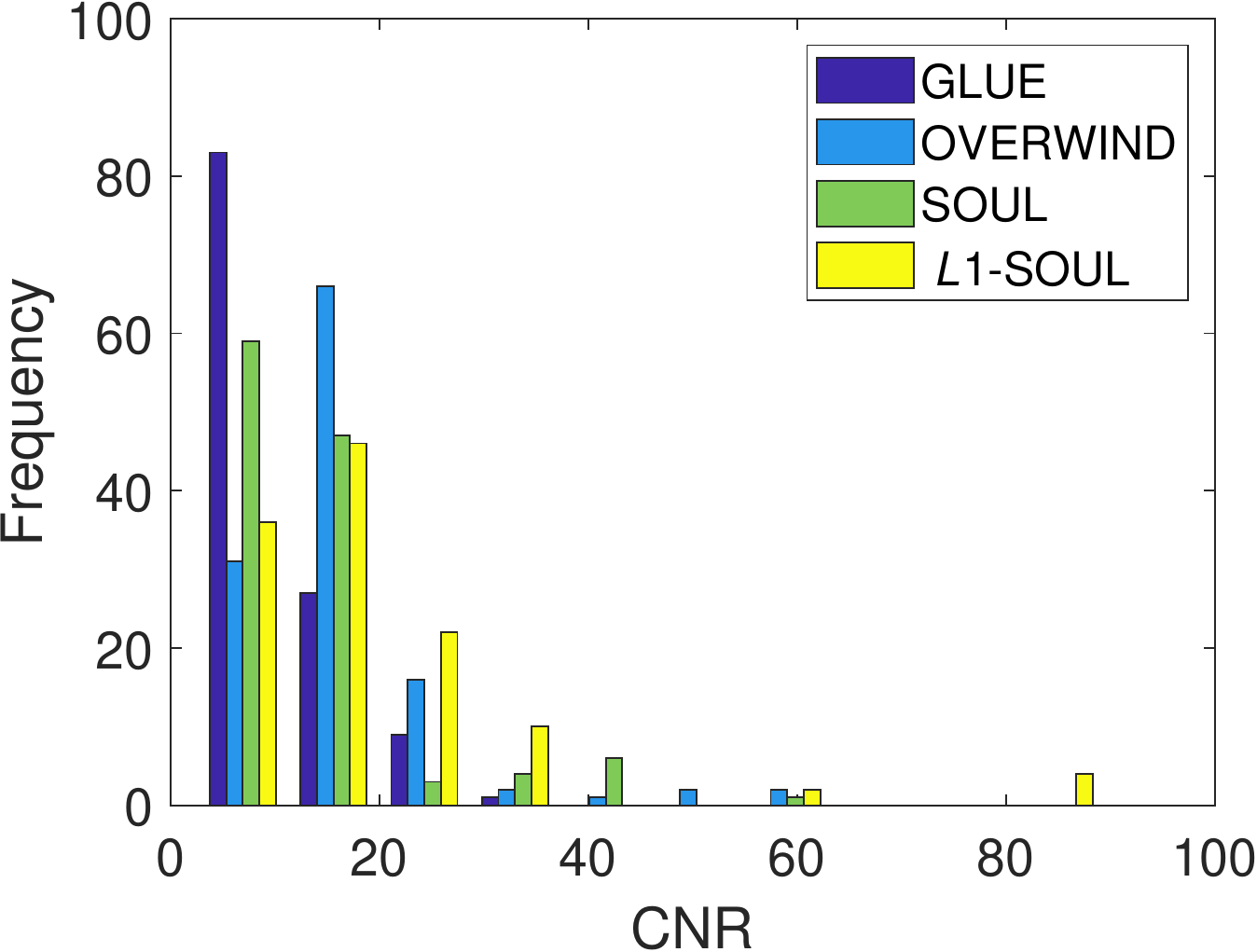}}}%
	\subfigure[Liver patient 1]{{\includegraphics[width=.2\textwidth]{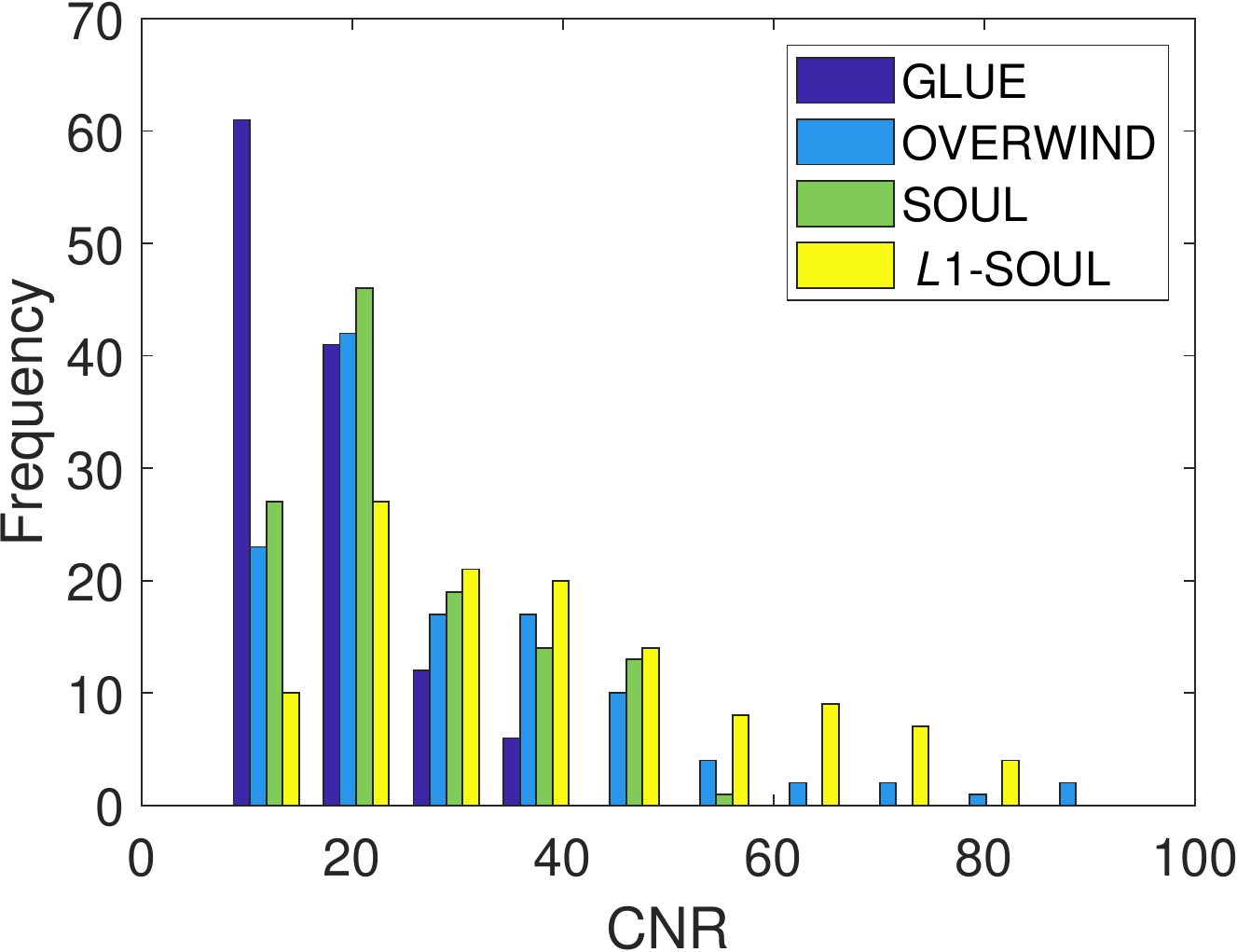}}}%
	\subfigure[Liver patient 2]{{\includegraphics[width=.2\textwidth]{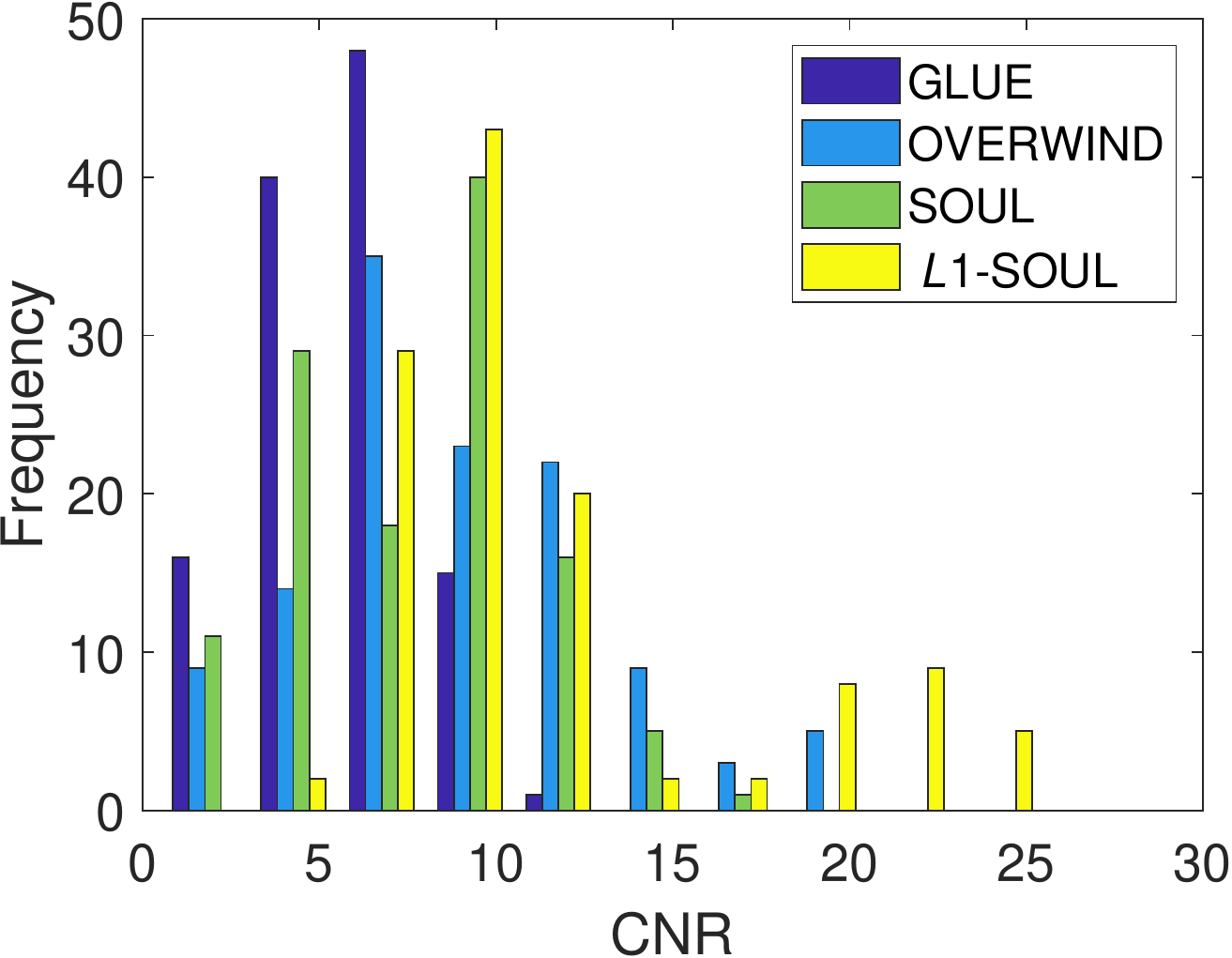}}}%
	\subfigure[Liver patient 3]{{\includegraphics[width=.2\textwidth]{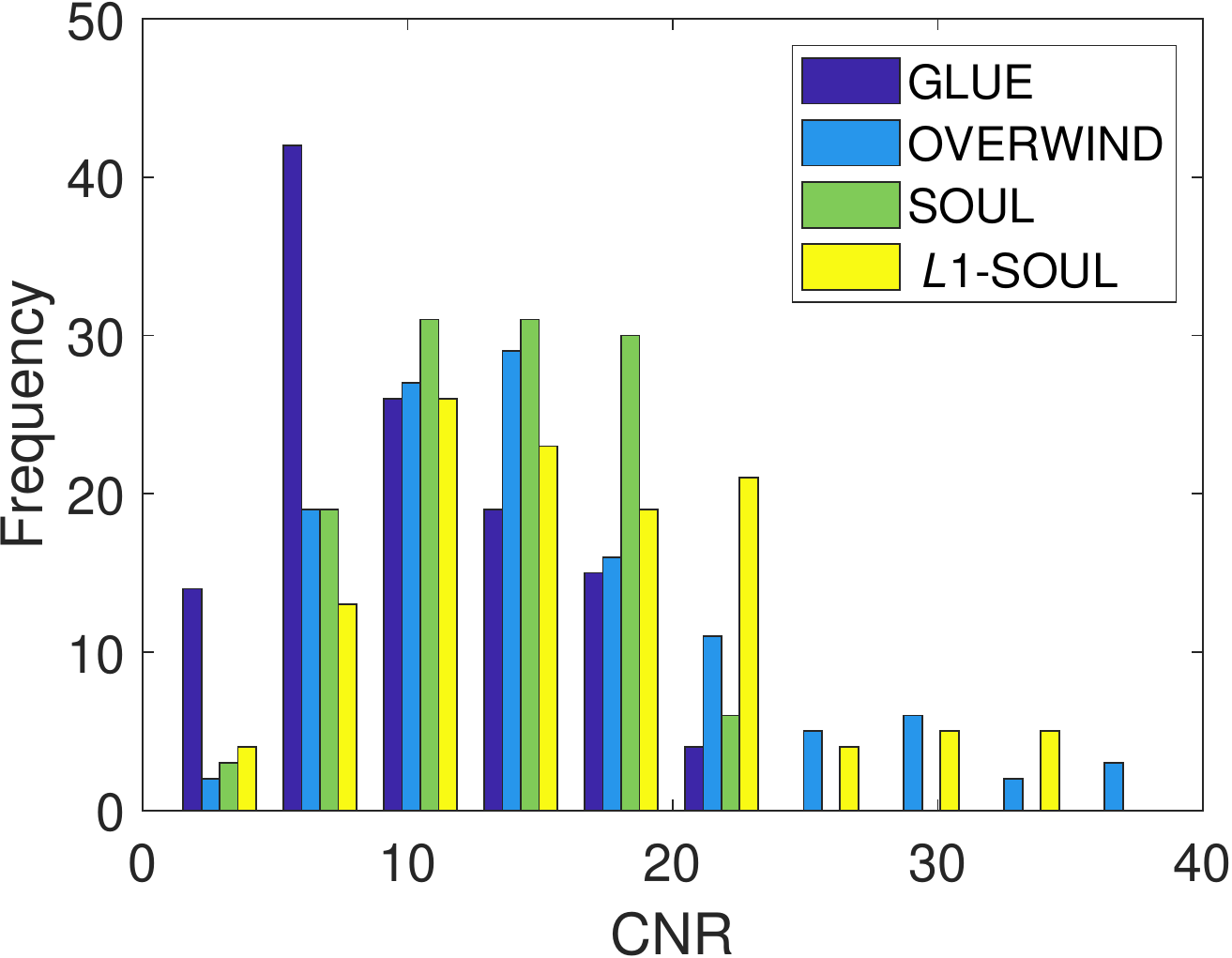}}}%
	\caption{CNR histograms obtained from 120 target-background combinations. Columns 1 to 5 correspond to the hard-inclusion simulated phantom, experimental breast phantom, and \textit{in vivo} liver datasets from Patients 1, 2, and 3, respectively.}
	\label{cnr_histograms}
\end{figure*}

The CNR histograms (Figs.~\ref{cnr_histograms}(c)-(e)) illustrate that $L1$-SOUL achieves the high CNR values most frequently for all three patients. GLUE, OVERWIND, SOUL, and $L1$-SOUL, respectively, obtain average CNR values of 17.43, 29.19, 24.99, and 37.37 for patient 1; 5.61, 8.79, 7.71, and 11.60 for patient 2 and 10.15, 15.38, 12.99, and 16.16 for patient 3. Moreover, paired  \textit{t}-tests confirm that $L1$-SOUL is significantly better than GLUE, OVERWIND, and SOUL with $p$-values of $7.2 \times 10^{-32}$, $7.28 \times 10^{-16}$, and $1.08 \times 10^{-24}$, respectively for patient 1; $4.15 \times 10^{-24}$, $2.81 \times 10^{-11}$, and $2.56 \times 10^{-16}$, respectively for patient 2; $5.07 \times 10^{-35}$, $4.7 \times 10^{-3}$, and $4.54 \times 10^{-12}$, respectively for patient 3.

\subsection{Execution Time}
All four techniques have been implemented on MATLAB R2018b using an $8^{th}$ generation Intel Core-i7 Computer. Table~\ref{table_time} shows the runtimes of GLUE, OVERWIND, SOUL, and $L1$-SOUL for computing the displacement field between two $1000 \times 100$-sized RF frames. We have reported the average computation time of ten independent runs. GLUE exhibits the lowest execution time since it optimizes the simplest penalty function. SOUL requires additional runtime due to the insertion of second-order continuity constraints. Since OVERWIND and $L1$-SOUL iteratively optimize their cost functions, they yield higher execution times than the other two techniques.    

\begin{table}[!t]
	\renewcommand{\arraystretch}{1.3}
	\caption{Time required for calculating the displacement field between RF frames of size $1000 \times 100$.}
	\label{table_time}
	\centering
	\begin{tabular}[c]{lllr}
		\hline		
		&	& Time (seconds)\\
		\hline
		&	GLUE   & \textbf{0.49}\\
		&	OVERWIND  & 2.79\\
		&	SOUL  & 1.01\\
		&	$L1$-SOUL & 5.33\\
		\hline			
	\end{tabular}
\end{table}  

\begin{figure}
	\centering
	\subfigure[B-mode]{{\includegraphics[width=.24\textwidth, height=.33\textwidth]{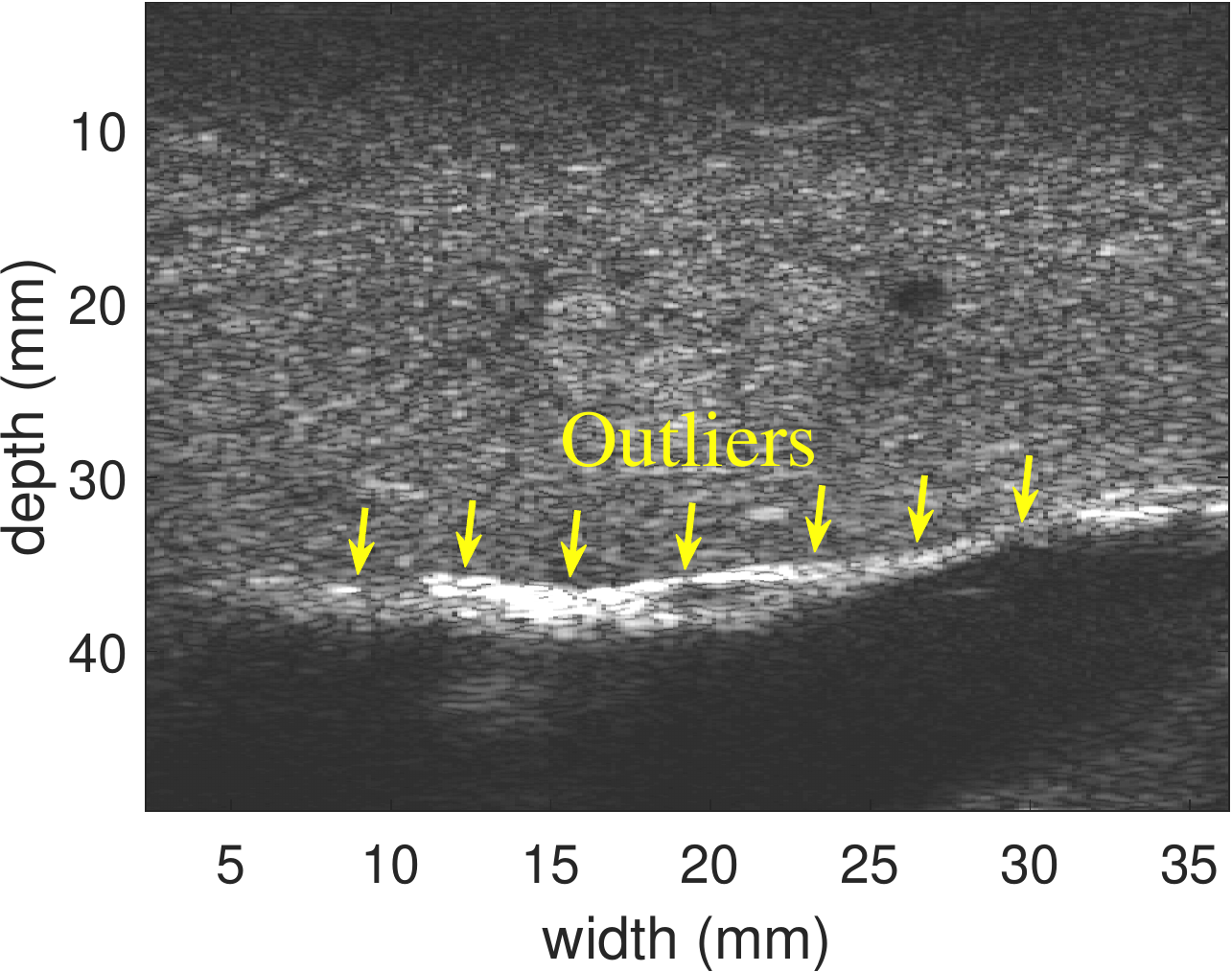}}}%
	\subfigure[Data residual]{{\includegraphics[width=.24\textwidth, height=.33\textwidth]{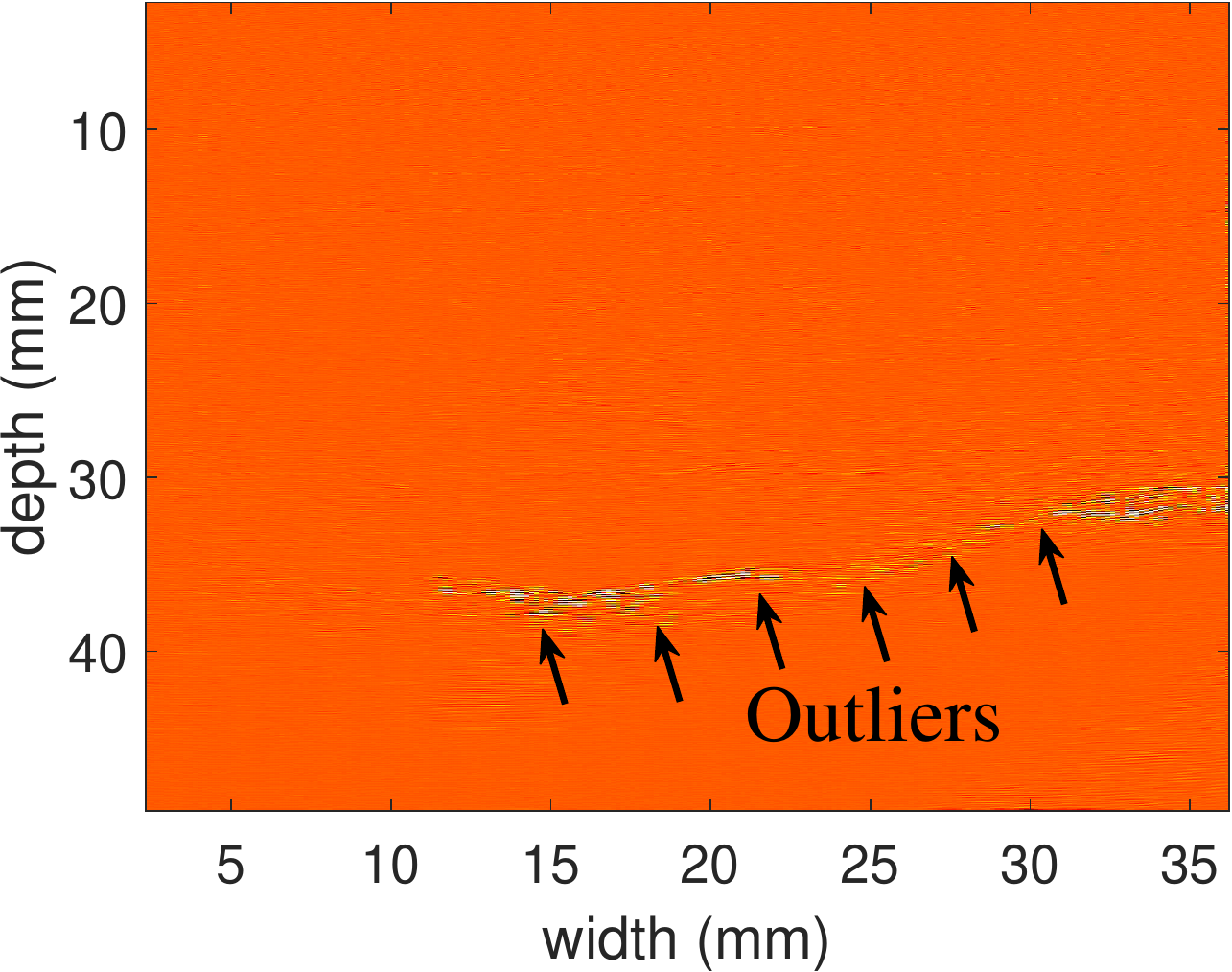}}}
	\subfigure[$L1$-SOUL, $L2$-norm data fidelity]{{\includegraphics[width=.24\textwidth, height=.33\textwidth]{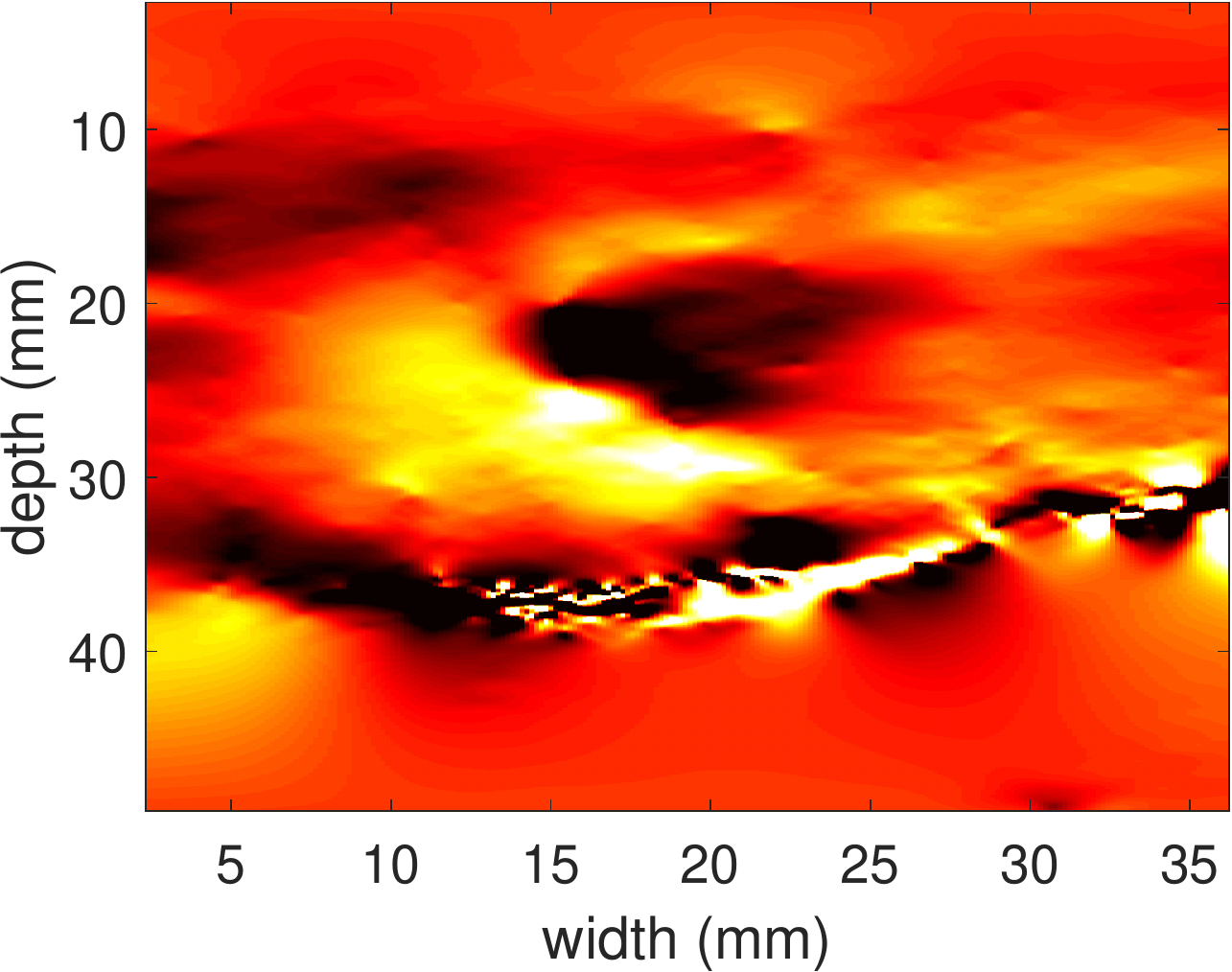}}}%
	\subfigure[$L1$-SOUL, $L1$-norm data fidelity]{{\includegraphics[width=.24\textwidth, height=.33\textwidth]{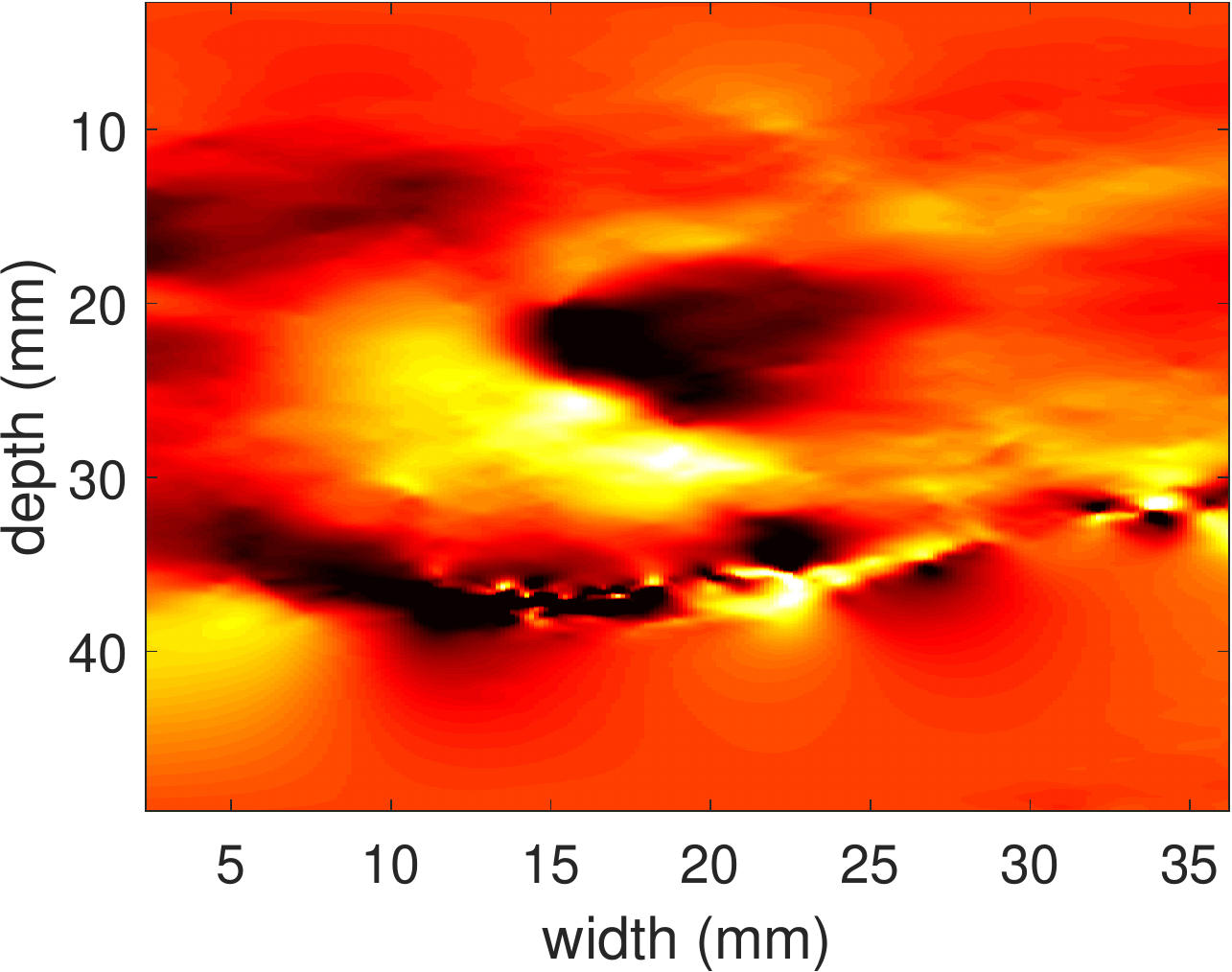}}}
	\caption{Strain results for liver Patient 2. (a) and (b) show the B-mode image and the data residual image, respectively, whereas (c) and (d) show the strain images obtained by $L1$-SOUL with $L2$- and $L1$-norm data fidelity terms, respectively.}
	\label{outliers}
\end{figure}

\begin{figure}
	\centering
	\subfigure[GLUE, 1 iteration]{{\includegraphics[width=0.24\textwidth,height=0.372\textwidth]{Results/small_layer_simu/new/glue_small}}}
	\subfigure[GLUE, 5 iterations]{{\includegraphics[width=0.24\textwidth,height=0.372\textwidth]{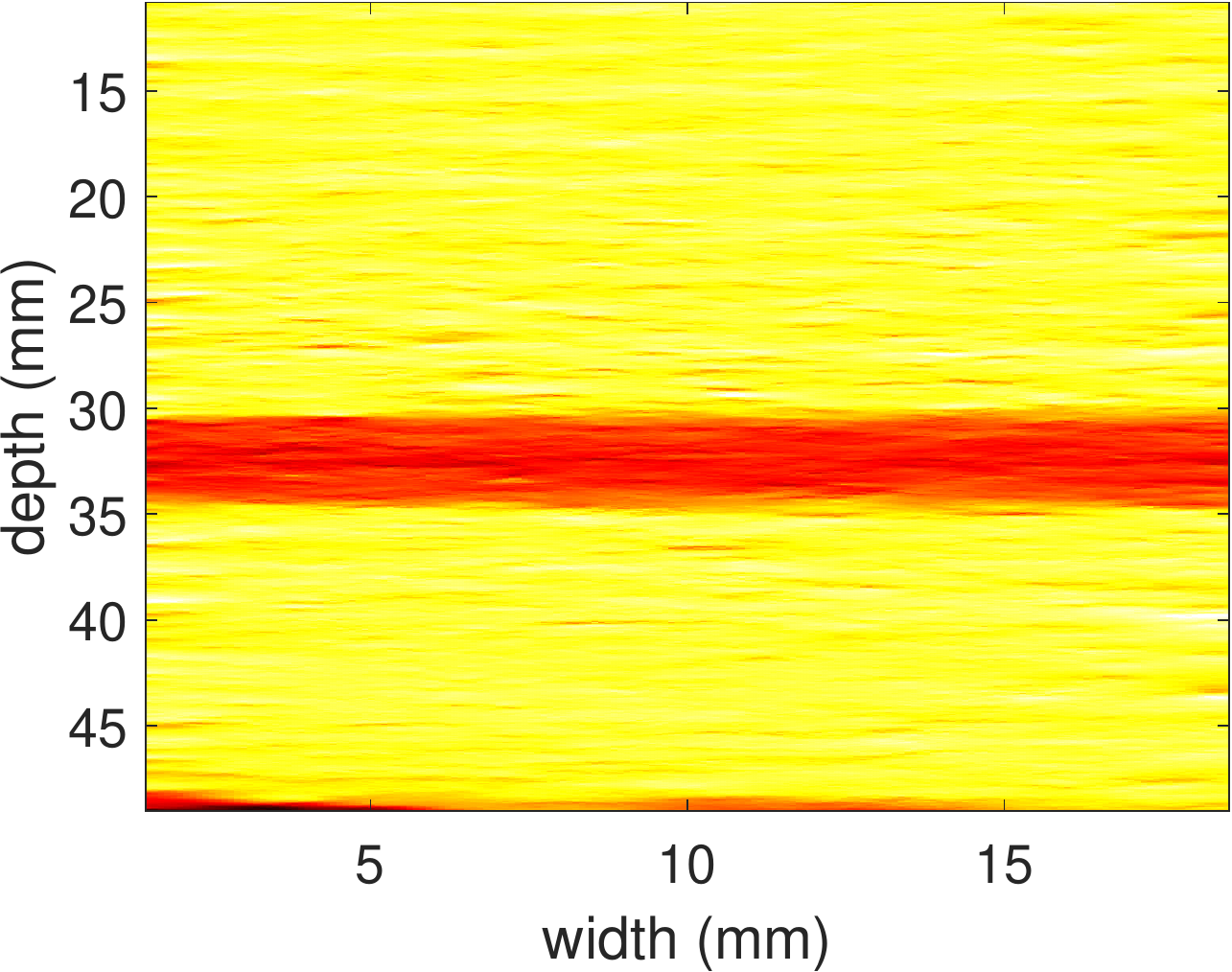}}}
	\subfigure[SOUL, 1 iteration]{{\includegraphics[width=0.24\textwidth,height=0.372\textwidth]{Results/small_layer_simu/new/soul_small}}}
	\subfigure[SOUL, 5 iterations]{{\includegraphics[width=0.24\textwidth,height=0.372\textwidth]{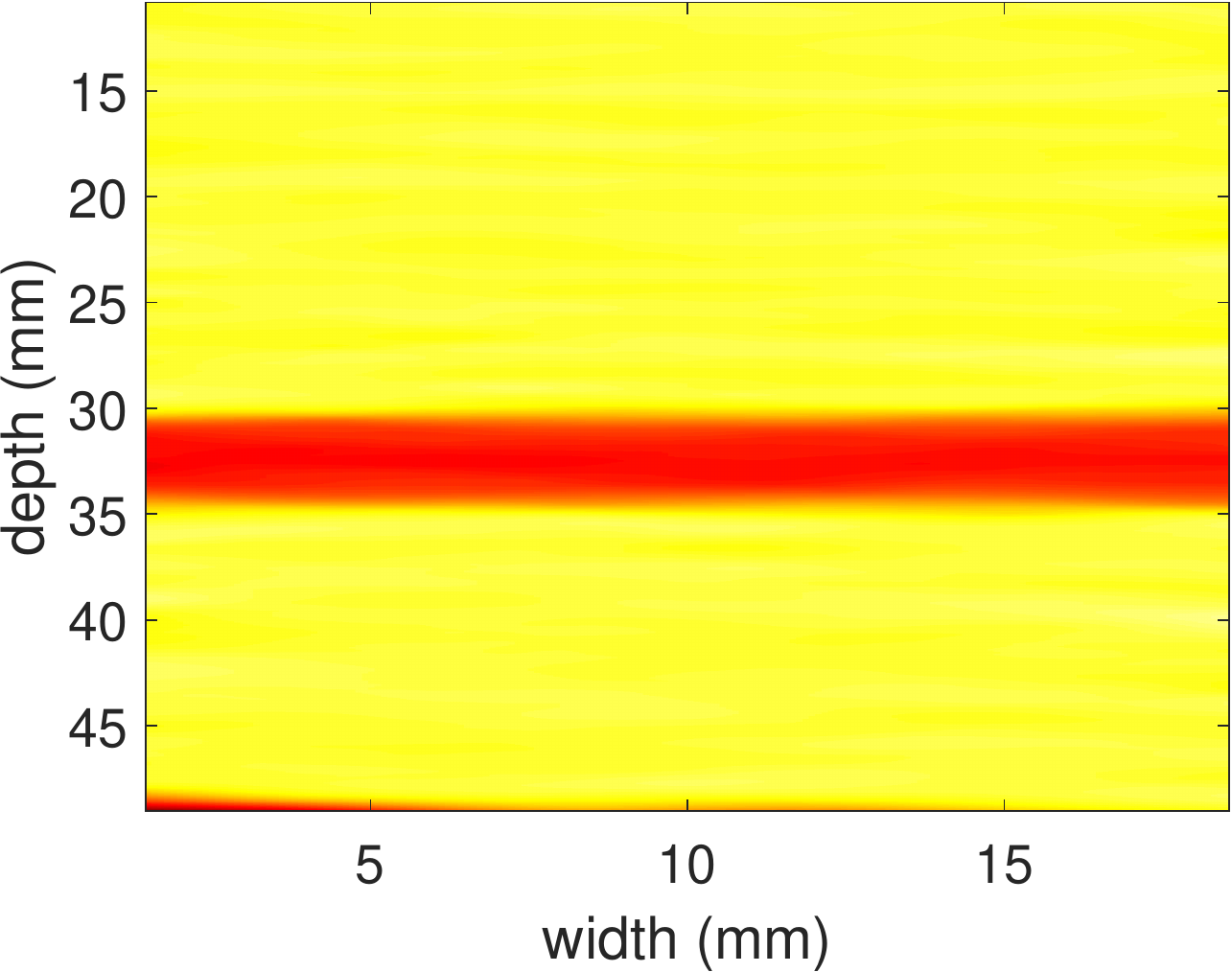}}}
	\caption{Results for the thin-layer simulated phantom with multiple iterations. Rows 1 and 2 correspond to GLUE and SOUL, respectively, whereas columns 1 and 2 correspond to 1 and 5 iterations, respectively.}
	\label{iters}
\end{figure}

\begin{figure}
	\centering
	\subfigure[Optimal regularization]{{\includegraphics[width=.2\textwidth, height=.23\textwidth]{Results/hard_simu/soulmate_hard_corr}}}%
	\subfigure[$50\%$ stronger regularization]{{\includegraphics[width=.2\textwidth, height=.23\textwidth]{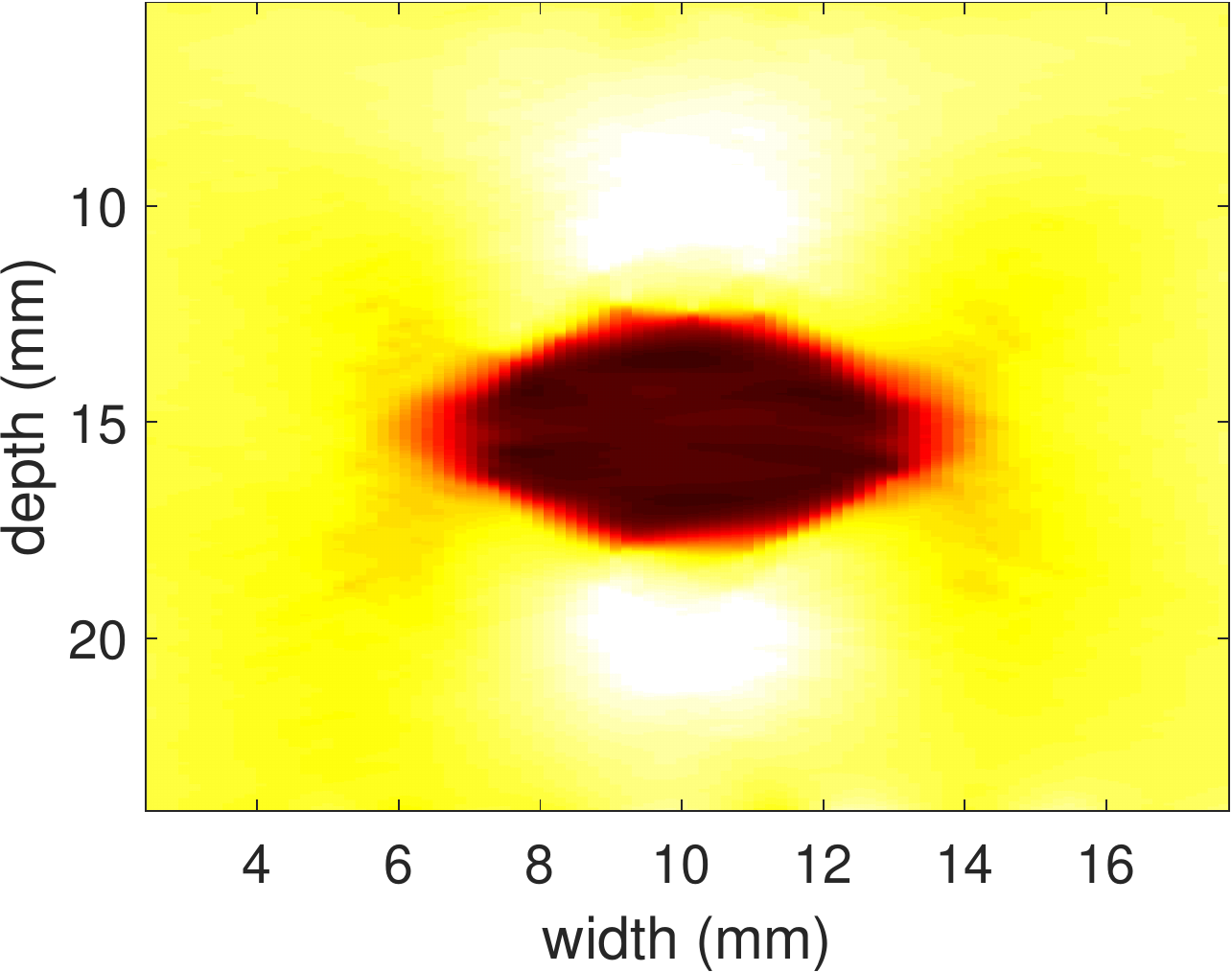}}}%
	\caption{Strain results for the hard-inclusion simulated phantom obtained by $L1$-SOUL. Columns 1 and 2 correspond to optimal and $50\%$ stronger regularization weights, respectively.}
	\label{regu_param}
\end{figure}

\section{Discussion}
First- and second-order continuity constraints together take the true tissue deformation physics into account, and as such, denoise the displacement and strain images accurately. The $L1$-norm regularization preserves the edges and maximizes the contrast by allowing sharp transitions at borders between different tissue types. $L1$-SOUL takes the advantage of both the combination of first- and second-order smoothness and the $L1$-norm to produce sharp and high-contrast strain images.

This work considers $L1$-norm for imposing only the regularization constraints. The other term in the cost function, i.e. the data term, uses $L2$-norm of the amplitude residuals. RF data usually contains additive Gaussian noise, and therefore, the $L2$-norm of the amplitude residuals provides the maximum likelihood estimator. Therefore, we employ $L2$-norm instead of $L1$-norm for the data term. Our experiments corroborated this choice. However, the $L1$-norm may be advantageous even in the data term if large regions of the image contain outlier data. To test this, we selected an image where large regions are corrupted by shadows from a specular bone surface. The arrows in Fig.~\ref{outliers} point to the rib bones under the liver, which create a large shadow. The $L1$-norm for the data term outperforms $L2$-norm in the region created by specular reflection from the bone, which is hard to track, as evidenced by the large residual in part (b). This might happen due to the fact that the derivative of $2\eta \sqrt{\eta^{2} + residual^{2}}$-like formulation of data $L1$-norm adaptively weights the data samples in an iterative manner. In case of a large residual for a data sample, the next iteration reduces the data weight for that sample and removes the outlier from the strain image. Nevertheless, we encountered such a case only once and therefore did not use the $L1$-norm in the data term.

The dark bands in the shallow and deep tissue regions of Figs.~\ref{layer}(g) and \ref{layer}(i) stem from the added Gaussian noise and the TDE techniques' boundary conditions. OVERWIND and $L1$-SOUL do not suffer from this issue due to two reasons. First, these techniques are inherently iterative. Second, the $L1$-norm regularization scheme guides them to converge to a more accurate solution. Fig.~\ref{iters} shows that GLUE and SOUL partially ameliorate the aforementioned issue when multiple iterations are taken into consideration. However, in our experience, multiple iterations of GLUE and SOUL slightly reduce the visual contrast.

In Fig.~\ref{esf}, a consistent shift of the estimated ESFs to the right of the ground truth ESF is observed. Fig. 1 of the Supplementary Material demonstrates that this shifting issue is not associated with the strain imaging techniques. Instead, it originates from the spatial gap between the actual location of the stiff layer and the one simulated by Field II.

In Figs.~\ref{liver}(g) and \ref{liver}(i), GLUE and OVERWIND exhibit dark-band artifacts in shallow tissue regions which have been partially removed by OVERWIND and $L1$-SOUL. This issue of strain underestimation often stems from over-penalizing the displacement derivative, especially in the regions of low echogenicity. In such a region, the regularization term dominates the data term~\cite{soul} and leads to a substantial reduction in average strain. However, since we do not have the ground truth strain for \textit{in vivo} datasets, a rigorous conclusion cannot be drawn.

We smoothen the sharp corners of the absolute value functions used in $L1$-norm to make them differentiable. The derivatives of these approximation functions introduce non-linearity to the equations, which is handled by using an iterative scheme. Therefore, both the $L1$-norm formulation as well as the optimization framework used in this work are inexact and have areas of improvement. Advanced techniques such as alternating direction method of multipliers (ADMM)- and duality-based~\cite{Zach_2007} approaches can be adapted for an improved $L1$-norm optimization. However, since these methods are beyond the scope of this work, they are interesting areas of future research.\\

Since $L1$-SOUL sharpens the strain image, one concern is that it might introduce over-sharpening and produce a noisy strain background. However, the simulation, phantom, and \textit{in vivo} results presented in this study demonstrate that $L1$-SOUL makes a proper balance between smoothness and sharpness to generate an edge-preserving and high-contrast strain map.

This work optimizes all techniques' regularization parameters for each dataset individually and uses the same set of weights for all RF samples in a particular ultrasound frame. In the future, an adaptive parameter selection strategy can be implemented where spatial structure and noise distribution are taken into account to make discrete decision on the continuity weight associated with each sample. Moreover, the regularization parameters are manually tuned based on the visual appearance of the strain images, which can make the procedure user-dependent. However, it is worth mentioning that the regularization parameters of $L1$-SOUL are not sensitive to moderate changes and lead to visually similar strain results even with a $50\%$ alteration in their values (see Fig.~\ref{regu_param}). To fully automatize the strain imaging task, the optimal set of regularization weights can be obtained from L-curve~\cite{curve_2001}, axial and lateral sampling rates, tissue properties, and parameters related to imaging physics. Since this automatic parameter estimation strategy demands extensive investigation, we postpone it as future work.

The fundamental assumption, mathematical formulation, and the validation results indicate that $L1$-SOUL is a substantially superior technique than GLUE, OVERWIND, and SOUL. However, to facilitate both second-order continuity and $L1$-norm, $L1$-SOUL requires higher running time than the other three algorithms. Therefore, the current unoptimized implementation of $L1$-SOUL might not be suitable for applications where high temporal resolution is important. This limitation of the proposed technique can be resolved by optimizing its implementation on a modern GPU.

All four techniques, namely GLUE, OVERWIND, SOUL, and $L1$-SOUL handle matrices of size $2mn \times 2mn$, which may lead to a memory footprint of around 300 GB for standard ultrasound frames of size $1000 \times 100$. However, majority of the entries are zero and therefore, we formulate the matrices as sparse block matrices. This efficient representation restricts the memory requirement to less than 100 MB.

\section{Conclusion}	
This work proposes $L1$-SOUL, a novel ultrasonic strain estimation technique. The proposed technique optimizes a non-linear energy functional consisting of data similarity and first- and second-order spatial continuity constraints to obtain the TDE between two time-series RF frames. The principal contribution of $L1$-SOUL is the incorporation of $L1$-norms of both first- and second-order displacement derivatives instead of $L2$-norms to obtain an edge-preserving and high-contrast strain estimate. Extensive validation against simulation, phantom, and \textit{in vivo} datasets demonstrate the superiority of $L1$-SOUL over the existing ultrasound elastography algorithms.                    

\section*{Acknowledgment}
This work is supported in part by Natural Sciences and Engineering Research Council of Canada (NSERC) RGPIN-2020-04612. Md Ashikuzzaman is a recipient of PBEEE and B2X Doctoral Research Fellowships provided by the Fonds de Recherche du Québec - Nature et Technologies (FRQNT). We thank Drs. E. Boctor, M. Choti, and G. Hager for allowing us to use the liver data. Authors thank Dr. M. Mirzaei for his help optimizing the results of OVERWIND and for providing the hard-inclusion simulated phantom. We also thank the anonymous reviewers for their constructive comments.

\FloatBarrier
\balance
\bibliographystyle{IEEEtran}
\bibliography{ref}

\end{document}


\title{Second-Order Ultrasound Elastography with $L1$-norm Spatial Regularization}
%
%
\author{Md~Ashikuzzaman, \textit{Student Member}, \textit{IEEE}
        and~Hassan~Rivaz, \textit{Senior Member}, \textit{IEEE}
}

%

\maketitle
\setlength{\abovedisplayskip}{1pt}
\setlength{\belowdisplayskip}{1pt}

\IEEEpeerreviewmaketitle
                 
In this Supplementary Material, we report the optimal parameter values associated with different techniques used in this work. In addition, we investigate the origin of the rightward shifts of ESFs in Fig. 3 of the paper.\\ \\ \\ \\ \\ \\ \\

\section{Results}
Time-delay estimation (TDE) using all methods entails some parameters that should be set before running the algorithm. For window-based methods, those parameters are search range, window sizes in the axial and lateral directions, minimum correlation threshold, and overlap in axial and lateral directions. These parameters can be stored beforehand for different imaging settings and for imaging different organs. It is important to note that ultrasound imaging itself also entails several tunable parameters, which are loaded when the sonographer selects the probe and the target anatomy (e.g. breast, thyroid, etc). In case of the layer phantoms, the first-order parameter set \{$\alpha_{1}$, $\alpha_{2}$, $\beta_{1}$, $\beta_{2}$\} was set to \{10, 2, 10, 2\}, \{40, 8, 20, 4\}, \{5, 1, 5, 1\}, and \{10, 2, 10, 2\} for GLUE, OVERWIND, SOUL, and $L1$-SOUL, respectively. For the simulated phantom with a hard inclusion, the first-order parameters corresponding to GLUE, OVERWIND, SOUL, and $L1$-SOUL were set to \{40, 0.1, 20, 0.05\}, \{40, 0.5, 20, 0.25\}, \{40, 0.05, 20, 0.025\}, and \{60, 0.05, 30, 0.025\}, respectively. In case of the breast phantom, the first-order continuity weights were set to \{40, 2, 20, 1\}, \{80, 1, 40, 0.5\}, \{80, 0.2, 40, 0.1\}, and \{120, 0.1, 60, 0.05\} for GLUE, OVERWIND, SOUL, and $L1$-SOUL, respectively. Best results for liver Patient 1 were obtained by setting \{$\alpha_{1}$, $\alpha_{2}$, $\beta_{1}$, $\beta_{2}$\} to \{20, 1, 20, 1\}, \{100, 1, 50, 0.5\}, \{20, 0.3, 10, 0.15\}, and \{80, 0.5, 40, 0.25\} for GLUE, OVERWIND, SOUL, and $L1$-SOUL, respectively. Optimal strain images for liver Patient 2 were found by setting \{$\alpha_{1}$, $\alpha_{2}$, $\beta_{1}$, $\beta_{2}$\} to \{0.25, 0.003, 0.2, 0.002\}, \{0.5, 0.025, 0.25, 0.0125\}, \{0.25, 0.003, 0.2, 0.002\}, and \{0.35, 0.0035, 0.3, 0.003\} for GLUE, OVERWIND, SOUL, and $L1$-SOUL, respectively. The first-order parameters for liver Patient 3 were set to \{10, 0.02, 5, 0.01\}, \{20, 0.16, 10, 0.08\}, \{10, 0.02, 5, 0.01\}, and \{20, 0.02, 10, 0.01\} for GLUE, OVERWIND, SOUL, and $L1$-SOUL, respectively. The second-order continuity weights \{$\theta_{1}$, $\theta_{2}$, $\lambda_{1}$, $\lambda_{2}$\} were set to the optimal multiple of \{$\alpha_{1}$, $\alpha_{2}$, $\beta_{1}$, $\beta_{2}$\}. For both SOUL and $L1$-SOUL, the multiplier corresponding to four-layer simulated phantom, thin-layer simulated phantom, hard-inclusion simulated phantom, breast phantom, liver patients 1, 2, and 3, respectively, was 1000, 1000, 50, 100, 100, 100, and 100. $\gamma$ was set to 0.1 for the simulated and phantom datasets, whereas it was set to 0 for the \textit{in vivo} experiments. $L1$-SOUL's smoothness controlling parameters \{$\eta_{0}$, $\eta_{1}$, $\eta_{2}$\} were set to \{0.0001, 0.0001, 0.0001\}, \{0.0001, 0.0001, 0.0001\}, \{0.0001, 0.001, 0.0005\}, \{0.0001, 0.0006, 0.0001\}, \{0, 0.003, 0.003\}, \{0, 0.02, 0.0003\}, and \{0, 0.002, 0.0005\} for four-layer simulated phantom, thin-layer simulated phantom, hard-inclusion simulated phantom, breast phantom, liver patients 1, 2, and 3, respectively.

We simulated a pre-deformed RF frame using Field II with the same parameter settings as the four- and thin-layer phantoms. A rectangular band (6 mm height) of this pre-deformed frame starting at 12 mm depth was axially shifted by 10 samples ($\approx$ 0.19231 mm). We performed this shifting using two different methods. First, we used closed form equation to deform the phantom and then simulated the post-deformed frame with Field II. Second, we shifted the aforementioned band with MATLAB (i.e. the second image is generated from the first image using image processing as opposed to Field II simulation) and used the shifted version of the pre-deformed frame as the post-deformed frame.

\begin{figure*}
	\centering
	\subfigure[Ground truth]{{\includegraphics[width=.5\textwidth,height=.45\textwidth]{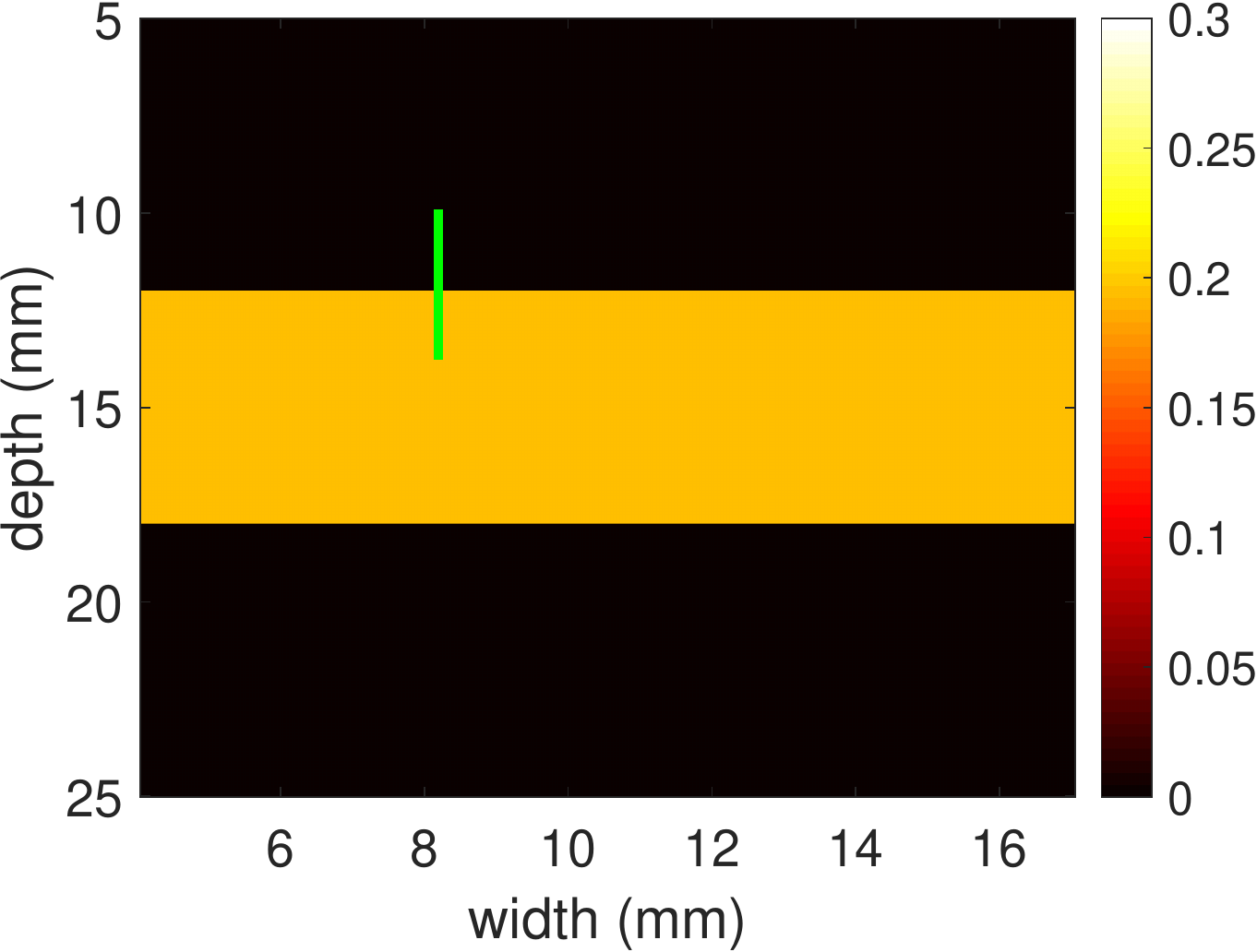}}}\\
	\subfigure[ESF, deformation with Field II]{{\includegraphics[width=.45\textwidth]{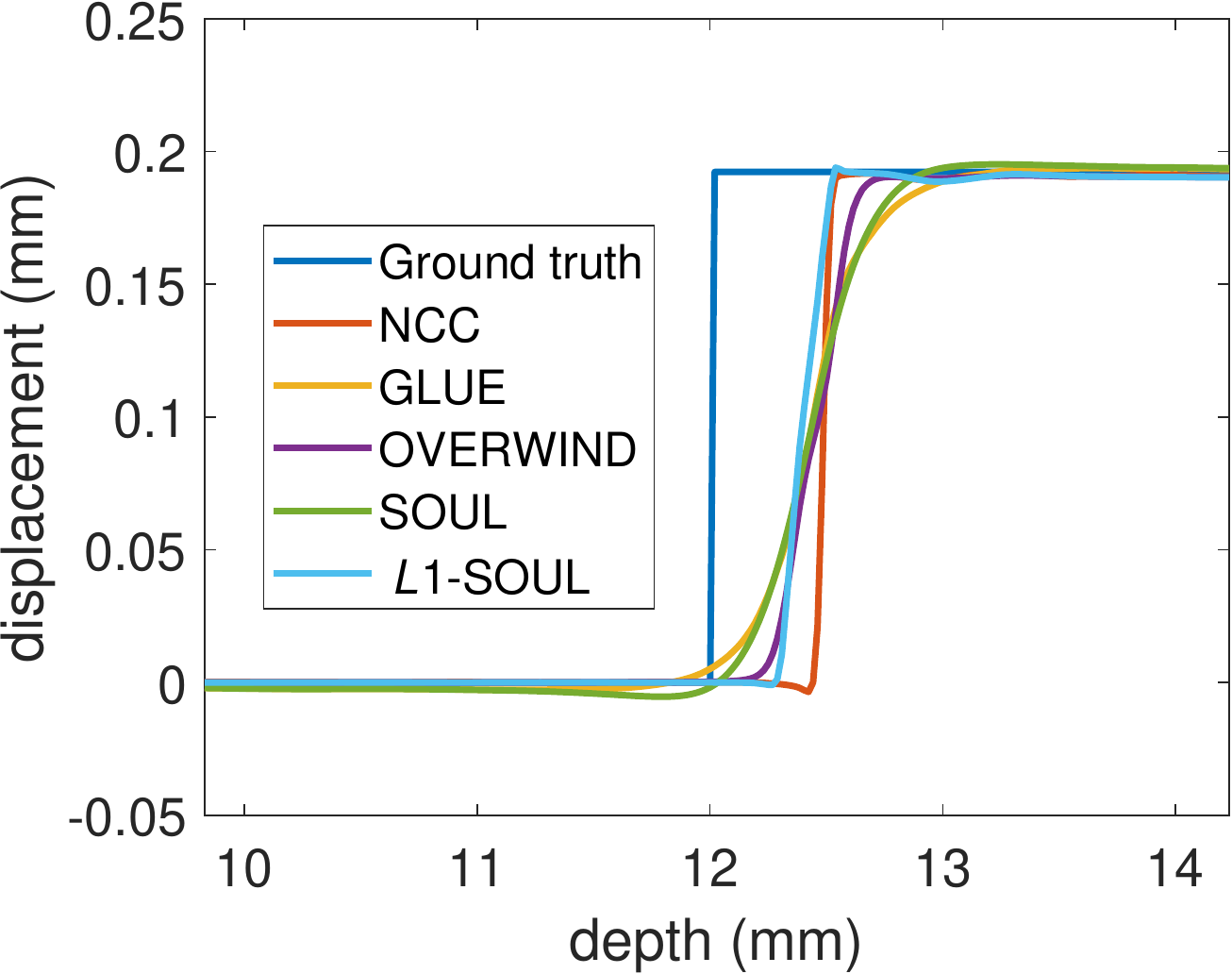}}}
	\subfigure[ESF, deformation with MATLAB]{{\includegraphics[width=.45\textwidth]{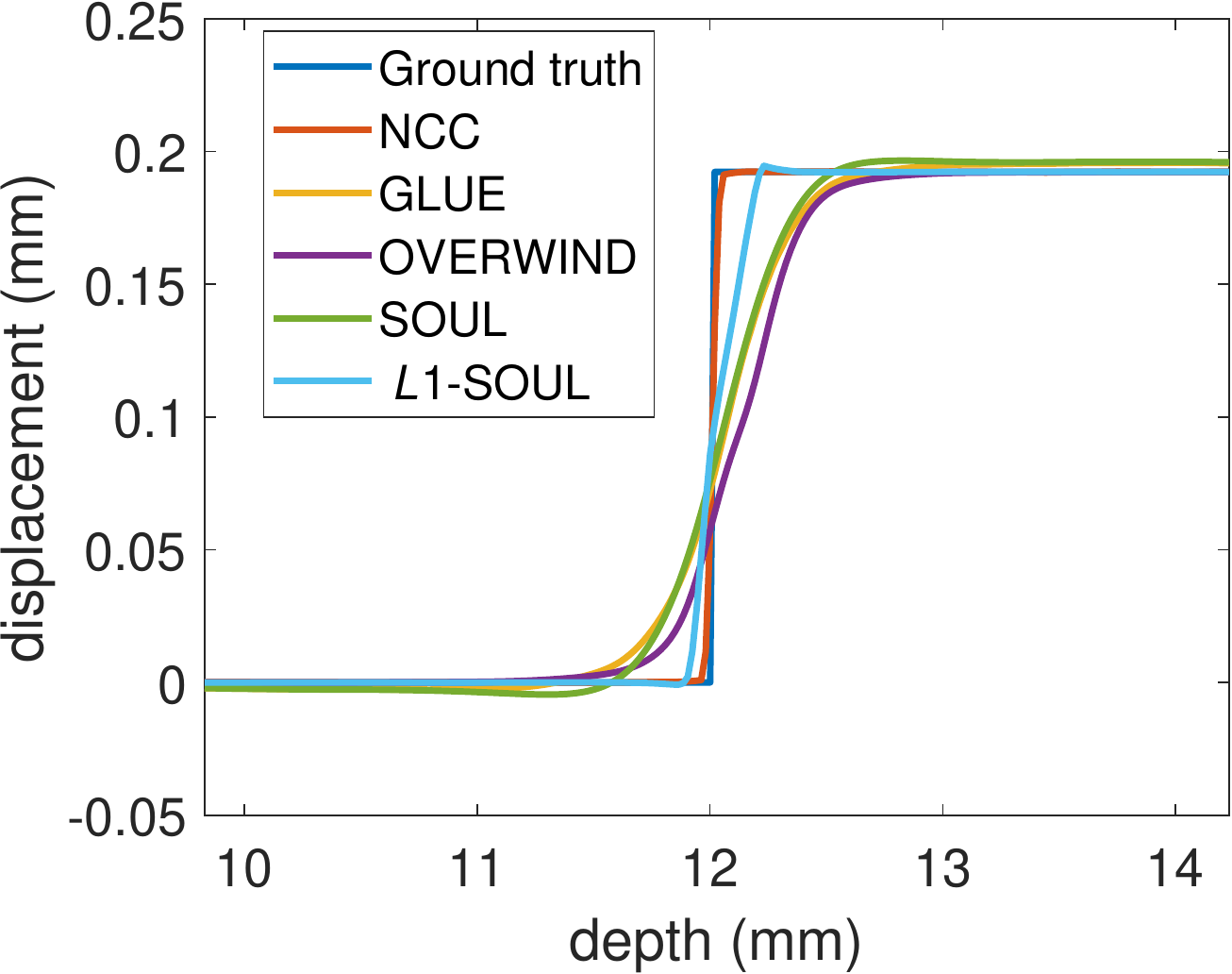}}}
	\caption{Comparison between the deformation fields generated using Field II and MATLAB. (a)-(c) correspond to the ground truth displacement field, ESFs for the deformations using Field II and MATLAB, respectively. The ESFs are plotted over the vertical green line shown in (a).}
	\label{field_simple}
\end{figure*}

Fig.~\ref{field_simple} shows the ground truth axial deformation field and the displacement ESFs. In the case of deformation with Field II, the ESFs are slightly shifted to the right of the ground truth. No such shift is observed when the rectangular band is deformed using a MATLAB command. Therefore, it is evident that the shifting issue is not associated with the TDE techniques. Instead, there is a spatial gap between the actual location of the deformed band and the one simulated by Field II.  

